\documentclass[galaxies,article,submit,moreauthors]{mdpi}

\firstpage{1} 
\pubvolume{xx}
\issuenum{1}
\articlenumber{5}
\pubyear{2019}
\copyrightyear{2019}
\externaleditor{Academic Editors: }
\history{Received: date; Accepted: date; Published: date}

\pdfoutput=1
\preto{\abstractkeywords}{\nolinenumbers}

\usepackage{color}
\usepackage{xcolor}
\usepackage[normalem]{ulem}
\usepackage{cancel} 
\definecolor{refkey}{rgb}{1,1,1}
\definecolor{labelkey}{rgb}{1,0,0}	

\usepackage{amssymb}	
\usepackage{wasysym}
\usepackage{upgreek}
\usepackage{enotez}
\usepackage{changepage}

\def\Rm{\rm Rm}

\def\Rey{{\rm Re}}
\def\Pm{\rm Pm}
\newcommand{\sigmafd}{\sigma_{\rm FD}}
\newcommand{\xx}{\mbox{\boldmath $x$}{}}

\newcommand{\FF}{\mbox{\boldmath $F$} {}}
\newcommand{\cmn}{\,{\rm cm^{-3}}}
\newcommand{\gcm}{\,{\rm gm\,cm^{-3}}}

\newcommand{\kpc}{\,{\rm kpc}}

\newcommand{\kms}{\,{\rm km\,s^{-1}}}

\newcommand{\nG}{\,{\rm nG}}

\newcommand{\radm}{\,{\rm rad\,m^{-2}}}

\newcommand{\ghz}{\,{\rm GHz}}
\def\vec#1{\ensuremath{\mathchoice{\mbox{\boldmath$\displaystyle#1$}}
		{\mbox{\boldmath$\textstyle#1$}}
		{\mbox{\boldmath$\scriptstyle#1$}}
		{\mbox{\boldmath$\scriptscriptstyle#1$}}}}

\newcommand{\bra}[1]{\langle #1\rangle}

\newcommand{\BB}{{\vec{B}}}

\newcommand{\Isync}{{I_{\rm sync}}}
\newcommand{\pmean}{{\langle p \rangle}}
\newcommand{\lf}{{l_{\rm f}}}
\newcommand{\EQ}{\begin{equation}}
\newcommand{\EN}{\end{equation}}
\newcommand{\EQA}{\begin{eqnarray}}
\newcommand{\ENA}{\end{eqnarray}}

\newcommand\aap{Astron.\ Astrophys.}
\newcommand\aapr{Astron.\ Astrophys.\ Rev.}
\newcommand\aj{Astron.\ J.}
\newcommand\apj{Astrophys.\ J.}
\newcommand\apjs{Astrophys.\ J.\ S.}
\newcommand\apjl{Astrophys.\ J.\ L.}
\newcommand\araa{Ann.\ Rev.\ Astron.\ Astrophys.}

\newcommand\mnras{MNRAS}
\newcommand\pasj{PASJ}
\newcommand\physrep{Phys.\ Rep.}
\newcommand\ssr{Space Sci. Rev.}

\Title{Properties of Polarized Synchrotron Emission from Fluctuation Dynamo Action -- II. 
Effects of Turbulence Driving in the ICM and Beam Smoothing}

\Author{Aritra Basu$^{1,\dagger}$*\orcidA,\ Sharanya Sur
$^{2,\dagger,}$*\orcidB}
\AuthorNames{Aritra Basu and, Sharanya Sur}
\address{%
$^{1}$ \quad Th\"{u}ringer Landessternwarte, Sternwarte 5, D-07778 Tautenburg, Germany; abasu@tls-tautenburg.de\\
$^{2}$ \quad Indian Institute of Astrophysics, 2nd Block, Koramangala, Bangalore 560034, India; sharanya.sur@iiap.res.in\\
}

\corres{Correspondence: abasu@tls-tautenburg.de, sharanya.sur@iiap.res.in}

\firstnote{These authors contributed equally to this work.}

\abstract{Polarized synchrotron emission from the radio halos of diffuse
intracluster medium (ICM) in galaxy clusters are yet to be observed. To
investigate the expected polarization in the ICM, we use high resolution
($1$\,kpc) magnetohydrodynamic simulations of fluctuation dynamos, which
produces intermittent magnetic field structures, for varying scales of
turbulent driving ($l_{\rm f}$) to generate synthetic observations of the
polarized emission.  We focus on how the inferred diffuse polarized emission
for different $l_{\rm f}$ is affected due to smoothing by a finite telescope
resolution.  The mean fractional polarization $\langle p\rangle$ vary as
$\langle p \rangle \propto l_{\rm f}^{1/2}$ with $\langle p \rangle > 20\%$ for
$l_{\rm f} \gtrsim 60$\,kpc, at frequencies $\nu > 4\,{\rm GHz}$. Faraday
depolarization at $\nu < 3$\,GHz leads to deviation from this relation, and in
combination with beam depolarization, filamentary polarized structures are
completely erased, reducing $\langle p \rangle$ to below 5\% level at $\nu
\lesssim1$\,GHz. Smoothing on scales up to $30$\,kpc reduces $\langle p
\rangle$ above $4$\,GHz by at most a factor of 2 compared to that expected at
$1$\,kpc resolution of the simulations, especially for $l_{\rm f} \gtrsim
100$\,kpc, while at $\nu < 3$\,GHz, $\langle p \rangle$ is reduced by a factor
of more than 5 for $l_{\rm f} \gtrsim 100$\,kpc, and by more than 10 for
$l_{\rm f} \lesssim 100$\,kpc.  Our results suggest that observational
estimates of, or constrain on, $\langle p \rangle$ at $\nu \gtrsim 4$\,GHz
could be used as an indicator of the turbulent driving scale in the ICM.}


\begin{document}

\section{\label{Intro}Introduction}

Galaxy clusters are the largest known gravitationally bound systems which
provide important clues on how structures were formed in the Universe.  Besides
gravity, magnetic fields are also believed to play an important role in the
evolution of the intracluster medium (ICM) of galaxy clusters which emit in the
radio and X-ray wavebands. A large fraction of massive merging clusters show
diffuse radio emission (radio halo) originating from relativistic electrons,
possibly accelerated by turbulence in the ICM, illuminating the cluster
magnetic fields via synchrotron radiation \citep{Fer+12, weere19}. This
emission is expected to be partially polarized, and its measurement provides
insights into the statistical properties of magnetic field structure in the ICM
\citep{Vazza+18,Domin+19,sur21}. So far, microgauss ($\upmu$G) strength magnetic
fields ordered on several kpc scales in halos have been indirectly inferred via
Faraday rotation measure (RM) estimated towards polarized sources located in
the background, or from depolarization studies of radio relics, remnants of
cluster collisions \citep{CKB01, Bonafede+09, bonaf10, Kierdorf+17}.  In the
absence of large-scale rotation of the ICM, such field strengths can naturally
arise from {\it fluctuation dynamo} action where dynamically insignificant seed
magnetic fields are amplified by random stretching of the field by turbulent
eddies \citep{SSH06,CR09,BS13,PJR15,Vazza+18,SBS18,Sur19,sur21}.  Understanding
the structural and coherence properties of these fields is important as they
contribute to pressure balance, control the acceleration and propagation of
relativistic particles \citep{BJ14}, and possibly play an important role in
governing microphysical processes such as thermal conduction, spatial mixing of
gas and kinetic viscosity \citep{Kunz+11,Komarov+14,Roberg+16}.  A direct
detection of polarized halo emission remains elusive and is a major science
driver for the Square Kilometre Array (SKA) later this decade,
\mbox{e.g., \citep{bonaf15}.}

Tentative detection of polarized emission for only three clusters have
been reported, namely, for Abell\,2255 \citep{Govoni+05}, MACS\,0717.5+3745
\citep{Bonafede+09}, and Abell\,523 \citep{girardi16}. It is likely that for
all the three cases the polarized signal is related to a radio relic seen in
projection rather than the halo emission itself \citep{P+11, rajpurohit21}. It
is noteworthy that all three reports of polarized halo emission are based on
observations at 1.4\,GHz, a frequency at which many radio relics show polarized
emission \citep{wittor19}. Due to the large Faraday dispersion in the ICM,
\mbox{\citet{bohringer16}} found that the presence of a cluster medium increases the
Faraday depth dispersion ($\sigmafd$) by about 60\,rad\,m$^{-2}$, and
therefore, polarized halo emission is expected to be highly depolarized at
1.4\,GHz. Hence, no clean detection of diffuse polarized halo emission has been
made so far.

Most of the previous attempts to constrain magnetic fields in the ICM by using
RM measured towards polarized background sources and/or depolarization of
cluster radio sources are based on the assumption of Gaussian random fields, e.g., 
\citep{bonaf10, Vacca+10}.  However, in stark contrast, numerical
magnetohydrodynamic (MHD) simulations reveal that fluctuation dynamo-generated
fields are spatially intermittent \citep{Schek04, BS05, Vazza+18, Seta+20,
sur21} with the field components exhibiting non-Gaussian distributions. In
order to maximize the chance to confidently detect polarized emission from
cluster halos using current and/or future radio telescope facilities, it is
therefore imperative to investigate the expected polarized signal directly from
fluctuation dynamo-generated fields obtained in MHD simulations. To this end,
to gain information on the properties of polarized synchrotron emission from
intermittent magnetic fields amplified by fluctuation dynamo, we have recently
performed realistic broad-bandwidth, synthetic observations using MHD
simulations of ICM see \mbox{\citet{sur21}}, for details. Here, to mimic cluster
merger-driven turbulence, also the same process which produces the radio halo,
incompressible turbulence is solenoidally forced on 256\,kpc for a simulation
box-size of 512\,kpc giving rise to magnetic fields ($B$) being correlated on
$\sim$110$\kpc$ (also indicated by cosmological simulations; \cite{Domin+19}).
This results in polarized intensity, ${\rm PI}\propto\int B^2\,{d}l$,
correlated on $\sim$200 kpc scales. However, due to strong Faraday
depolarization at frequencies below $\sim$3 $\ghz$, polarized structures on
significantly smaller scales are produced. These simulations roughly correspond
to the central regions of massive Coma-like galaxy-clusters. The main outcome
from the work of \citet{sur21} is that, one needs high frequency ($\nu \gtrsim
5\ghz$) observations to detect polarization in cluster radio halos, and with
high spatial resolution ($\approx$1 $\kpc$) to infer its structural properties.

In this paper, we expand the scope of our work and present detailed
investigation of the effects of turbulent driving at different scales on the
properties of the polarized synchrotron emission when smoothed by a telescope
beam at three representative frequencies of 0.6, 1.2 and 5~GHz. These
frequencies represent the typical frequencies at which galaxy clusters are
observed, and also corresponds to Band 1 (covering 0.35--1.05\,GHz),
Band 2 (covering 0.95--1.76\,GHz), and Band 5 (covering 4.6--15.3\,GHz) of the
SKA's mid-frequency component, SKA1-MID. A key question which we seek to
address using the synthetic observations concerns whether the mean fractional
polarization can inform us about the turbulent driving scale in the ICM. To
investigate the effects different scales of turbulent forcing have on the
polarized emission in the ICM, in this work, we forced turbulence on scales of
$256, 102$ and $64\kpc$ which gives rise to magnetic field structures
correlated on widely different scales. For a representative cluster like Coma,
these roughly correspond to scales ranging from the core radius down to scales
of pressure scale-height. Magnetic fields generated by fluctuation dynamos can
be ordered at the most on the scale of turbulent motions. Consequently,
turbulent driving on the aforementioned scales enables us to probe the
statistical properties of the polarized emission due to random magnetic fields
with varying correlation lengths. This paper is structured as follows: in
Section~\ref{sec:method}, we present in brief the details on the numerical
simulations and methodology of obtaining synthetic observations from the
simulation data. In Section~\ref{sec:results}, we discuss the results by first
focusing on the power spectra of the kinetic and the magnetic energies, and
that of the Faraday depth (FD).  Next, we analyze the statistical properties of
the total and polarized emission in radio halos when synthetic observations are
smoothed over various scales to study the impact of telescope beam. We focus on
the dependence of the mean fractional polarization of the diffuse ICM on the
turbulent driving scale. Finally, in Section~\ref{sec:discuss} we conclude
with a summary of the main results and discuss the implications of our work on
the properties of polarized emission in the ICM.

\section{Methodology of Numerical Calculations} \label{sec:method}

In this section, we present in brief the setup used for performing direct
numerical MHD simulations, and the methodology of computing the synthetic
observations using the simulation data. For further detail on these methods, we
refer an interested reader to \citet{Sur19, basu19b} and \citet{sur21}.

\subsection{Summary of Magnetohydrodynamic (MHD) Simulations}\label{sec:mhd}

In this work, we make use of numerical simulations of fluctuation dynamos
performed with the FLASH code \citep{Fry+00} to address the aforementioned
goals of this work.  We focus on three simulations performed with an isothermal
equation of state where turbulence is driven non-helically at three different
forcing wave numbers, $k_{\rm f}L/2\pi = 2, 5$ and $8$, referred to as run A, B
and C, respectively, in Table~\ref{tab:sumsim}.  The setup of each of these
simulations is identical to the one presented in \citet{sur21}. We therefore
highlight only the essential features here. All the three simulations are
performed at magnetic Prandtl number $\Pm = \Rm/\Rey = 1$, where $\Rm$ and
$\Rey$ are the magnetic and fluid Reynolds numbers, respectively. Recent
cosmological simulations of hierarchical structure formation including the
formation of galaxy clusters show that turbulence in the cluster core is
dominated by solenoidal modes~\mbox{\citep{Miniati15, Vazza+17, Wittor+17, PPQ21}}.
Accordingly, and to maximize the efficiency of the fluctuation dynamo, we use
only solenoidal modes {(i.e., $\nabla\cdot\FF = 0$)} for the turbulent driving.
Here, $\FF$ is the forcing term in the MHD equations.  Furthermore, we adjust
the amplitude of the forcing such that the resulting root mean square (rms)
value of the Mach number $\mathcal{M} = u_{\rm rms}/c_{\rm s} \approx 0.18
\textrm{--} 0.19$, where $c_{\rm s}$ is the isothermal sound speed, and $u_{\rm
rms}$ is the rms velocity. The subsonic nature of these simulations imply that
density fluctuations ($\delta\rho/\rho \approx \mathcal{M}^2$) are negligible.
This is in accordance with observational results which show rms
turbulent velocity of $\approx$200--300~$\kms$ \citep{Hitomi18a}, and
density fluctuations, inferred from X-ray surface brightness fluctuations, are
at a $7 \textrm{--} 10\%$ level on scales of $\sim$500 $\kpc$
\citep{Churazov+12}, down to $\sim$4\% on scales of $30 \textrm{--} 50\kpc$
\citep{Zhurav+19}. Our simulations were initialized with weak seed magnetic
fields of the form $\BB = B_{0}[0,0,\sin(10\,\pi\,x)]$ where the amplitude
$B_{0}$ was adjusted to a value such that the initial plasma $\beta =
P_{\rm th}/P_{B}\sim 10^{6}$, where $P_{\rm th}$ and $P_{B}$ are the thermal
and magnetic pressures, respectively.  The divergence constraint of the
magnetic field in our simulations is satisfied by using the standard algorithms
available in FLASH see \citep{sur21} for details.  Each of these simulations
are run until we obtained many realizations of the saturated state of
the fluctuation dynamo for our further analyses.  Note that these realizations
span over many eddy turnover times in our simulations. Table~\ref{tab:sumsim}
highlights the important dimensionless parameters of the runs. 

\begin{table}
	\centering
\setlength{\tabcolsep}{7.0 pt}
\caption{Key parameters of the subsonic 
simulations used in this study. 
$N$ is the number of grid points in each dimension, $k_{\rm f}$ is 
the forcing wave number and $L$ is the size of the simulation domain.
$\mathcal{M}$ and $b_{\rm rms}$ are the average values of the rms Mach 
number and the magnetic field obtained in the steady state. Pm and 
Re are the magnetic Prandtl and fluid Reynolds numbers, respectively.}
\tabcolsep=0.506cm
\begin{tabular}{ccccccc} \toprule
\textbf{Run} & \boldmath{$N^{3}$} & \boldmath{$k_{\rm f}\,L/2\pi$} & \boldmath{$\mathcal{M}$} & \boldmath{$b_{\rm rms}$} & \boldmath{$\Pm$} & 
\boldmath{$\Rey = u\,l_{\rm f}/\nu$} \\ \midrule
A & $512^{3}$ & 2.0 & $\approx$0.18 & $\approx$0.08 & 1 & 1080 \\  \midrule      
B & $512^{3}$ & 5.0 & $\approx$0.19 & $\approx$0.12 & 1 & 1450 \\ \midrule 
C & $512^{3}$ & 8.0 & $\approx$0.19 & $\approx$0.13 & 1 & 1425 \\ \bottomrule
\end{tabular}
\label{tab:sumsim}
\end{table}

For our analysis, we generate three-dimensional (3-D) snapshots of the gas mass
density ($\rho$) and three components of the magnetic fields, $B_x$, $B_y$ and
$B_z$. FLASH outputs these physical variables in dimensionless units which are
then converted to physical units for computing the observable quantities, such
as, the synchrotron intensity ($\Isync$), the Stokes $Q, U$ parameters,
polarized intensity (${\rm PI}$) and the fractional polarization $p = {\rm
PI}/I$. To this effect, we first renormalize the length of the simulation
domain to $L = 512\kpc$ in each dimension which implies a resolution of $\Delta
x = \Delta y = \Delta z = 1\kpc$. Depending on $k_{\rm f} = 2, 5$ and $8$, the
scale of turbulent motions $l_{\rm f} = 2\pi/k_{\rm f} = 256, 102.4$ and
$64\kpc$, respectively. The local electron number number density ($n_{\rm e}$)
is computed from $\rho$ using $n_{\rm e}(\xx) = \rho(\xx)/\mu_{\rm e}\,m_{p}$.
Here `$\xx$' is the three dimensional position vector, $\mu_{\rm e}  = 1.18$ is
the mean molecular weight per free electron, and $m_{p}$ is the proton mass. We
assume $\langle n_{\rm e} \rangle = 10^{-3}\cmn$ and $c_{\rm s} =1 0^{3}\kms$
as typical values in the ICM \citep{Sar88} for all our runs. This implies a
mean gas mass density $\langle \rho\rangle = \langle n_{\rm e}\rangle\,\mu_{\rm
e}\,m_{p} \approx 1.97\times 10^{-27}\gcm$. For these values of the
density and the sound speed, an initial plasma $\beta \sim 10^{6}$ implies
$B_{0} \sim 22.2\nG$. The dimensionless values of the components of the
magnetic field are expressed in Gauss by scaling them with the unit of the
magnetic field strength $\sqrt{4\,\pi\,\rho\, c_{\rm s}^{2}} \approx 15.7~\upmu \text{G}$
in all the runs, while the equipartition field is given by $B_{\rm eq} =
\sqrt{4\,\pi\,\rho\, u_{\rm rms}^{2}}$. Thus, considering the chosen values of
the simulation domain, densities and sound speeds, our simulations can be
thought of as representative of the core regions of galaxy clusters.
Table~\ref{tab:pars} lists the values of $u_{\rm rms}$, $b_{\rm rms}$ and
$B_{\rm eq}$ for the different runs considered here. 

\begin{table}
	\centering
\caption{The rms values of the turbulent velocity $u_{\rm rms}$, the 
magnetic field $b_{\rm rms}$ and the equipartition field strength 
$B_{\rm eq}$ from simulations at three different turbulent driving scales. 
The simulation domain is $512\times512\times 512\kpc^{3}$ with a 
resolution of $1\times1\times 1\kpc^{3}$. The mean electron density 
$\langle n_{\rm e}\rangle = 10^{-3}\cmn$ and isothermal sound speed 
$c_{\rm s}= 10^{3}\kms$ across all the runs. 
}
\tabcolsep=0.3cm
\begin{tabular}{cccc} \toprule
\textbf{Parameter Name} & \boldmath{$l_{\rm f} = 256\kpc$} & \boldmath{$l_{\rm f} = 102.4\kpc$} & \boldmath{$l_{\rm f} = 64\kpc$} \\ \midrule 
Turbulent rms velocity ($u_{\rm rms}$) & $\approx$ 180 $\kms$ & $\approx$190 $\kms$ & $\approx$190 $\kms$ \\
rms field strength ($b_{\rm rms}$) & $\approx$1.3~$\upmu \text{G}$ & $\approx$1.57~$\upmu \text{G}$ & $\approx$1.7~$\upmu \text{G}$ \\
Equipartition field strength ($B_{\rm eq}$) & $\approx$2.8~$\upmu \text{G}$ & $\approx$3~$\upmu \text{G}$ & $\approx$3~$\upmu \text{G}$ \\ \bottomrule
\end{tabular}
\label{tab:pars}
\end{table}

\subsection{Synthetic Observations} \label{sec:cosmic}

The output of the MHD simulations, converted to physical dimensions, were used
as input for the \texttt{COSMIC} package \citep{basu19b} to compute
broad-bandwidth synthetic maps of the total and polarized synchrotron emission.
For our analysis, we chose the $x$- and $y$-axes to be in the plane of the sky,
and the $z$-axis to be parallel to the line of sight (LOS). Hence, $B_\parallel
\equiv B_z$ contributes to the Faraday rotation, while $B_\perp \equiv
\sqrt{B_x^2 + B_y^2}$ contributes to the polarized synchrotron emission. Our
MHD simulations are devoid of cosmic rays. Therefore, for computing the
synchrotron emissivity, we assume a uniform number density of cosmic ray
electrons ($n_{\rm CRE}$) in each mesh point, which follow a constant power-law
energy spectrum of the form $n_{\rm CRE}(E) = n_0\,E^\gamma$.  Here, the energy
index $\gamma=-3$ is constant for all mesh points and corresponds to spectral
index $\alpha = -1$ for the frequency spectrum of the synchrotron intensity
$\Isync(\nu) = I_0\,\nu^\alpha$ in the plane of the sky. The normalization
$n_0$ is chosen such that $I_0 = 1$\,Jy at 1\,GHz.\footnote{The
surface brightness presented in this work at any other $\nu_0$ can be scaled
depending on the choice of $I_0$ which is representative of a particular galaxy
cluster.} The choice of the value of the flux density normalization
would not affect the fractional polarization or Faraday depolarization, which
are the main focus of this study, unless the type of turbulence driving itself
is widely different for galaxy clusters at the extreme ends of luminosity
and/or mass function. Details of numerical calculations performed in
\texttt{COSMIC} are presented in \citet{basu19b} and \citet{sur21}. Note that,
for simplicity, we have not included any noise and/or systematic errors
introduced while observing with a telescope. Furthermore, to
investigate the effects of intermittent magnetic fields produced by the action
of fluctuation dynamo, we have assumed a constant $n_{\rm CRE}$. Although this
is an oversimplification, we believe that large turbulent diffusivity in the
ICM would mix the cosmic ray electrons efficiently, and thereby damp spatial
variations in $n_{\rm CRE}$ \citep{sur21}. In addition, spatial fluctuations in
$n_{\rm CRE}$ due to cooling, predominantly arising from inverse-Compton
scattering with the cosmic microwave background (CMB)
radiation,\footnote{Since the magnetic field strengths in the ICM is
$<$3.25\,$(1+z)^2\,\upmu \text{G}$ where $z$ is the redshift of a cluster, inverse-Compton
cooling due to scattering with the CMB photons would dominate over synchrotron
cooling.} is expected to be small. This is because, the CMB is smooth over the
scale of the cluster.

It was shown in \citet{sur21} that the magnetic fields in the non-linear
saturated state of the fluctuation dynamo at different times separated by at
least one eddy turnover time, $t_{\rm ed} = \lf/u_{\rm rms}$, are statistically
equivalent. Therefore, in this paper, we focus on results obtained for synthetic
observations performed at one snapshot in the saturated stage for each of the
three simulation runs listed in Table~\ref{tab:sumsim}, i.e., at $t/t_{\rm ed}
= 23, 20.2$ and $30.5$ for $k_{\rm f}=2, 5$ and $8$, respectively. Furthermore,
in order to study the effects of a telescope beam on the observed quantities,
the synthetic maps obtained at the native 1\,kpc resolution of the simulations
were convolved using unit-amplitude Gaussian kernels with full width at half
maximum (FWHM) of various sizes see \citep{sur21} for the smoothing
methodology.  In this paper, we present results for convolution on
scales ranging between $5$ and $80\kpc$ which roughly correspond to angular
resolution between $10$ and $160$\,arcsec at the distance of 100\,Mpc of the
Coma cluster.

\section{\label{sec:results}Results}


\begin{figure}
	\centering
\begin{tabular}{cc}
\includegraphics[width=0.45\columnwidth]{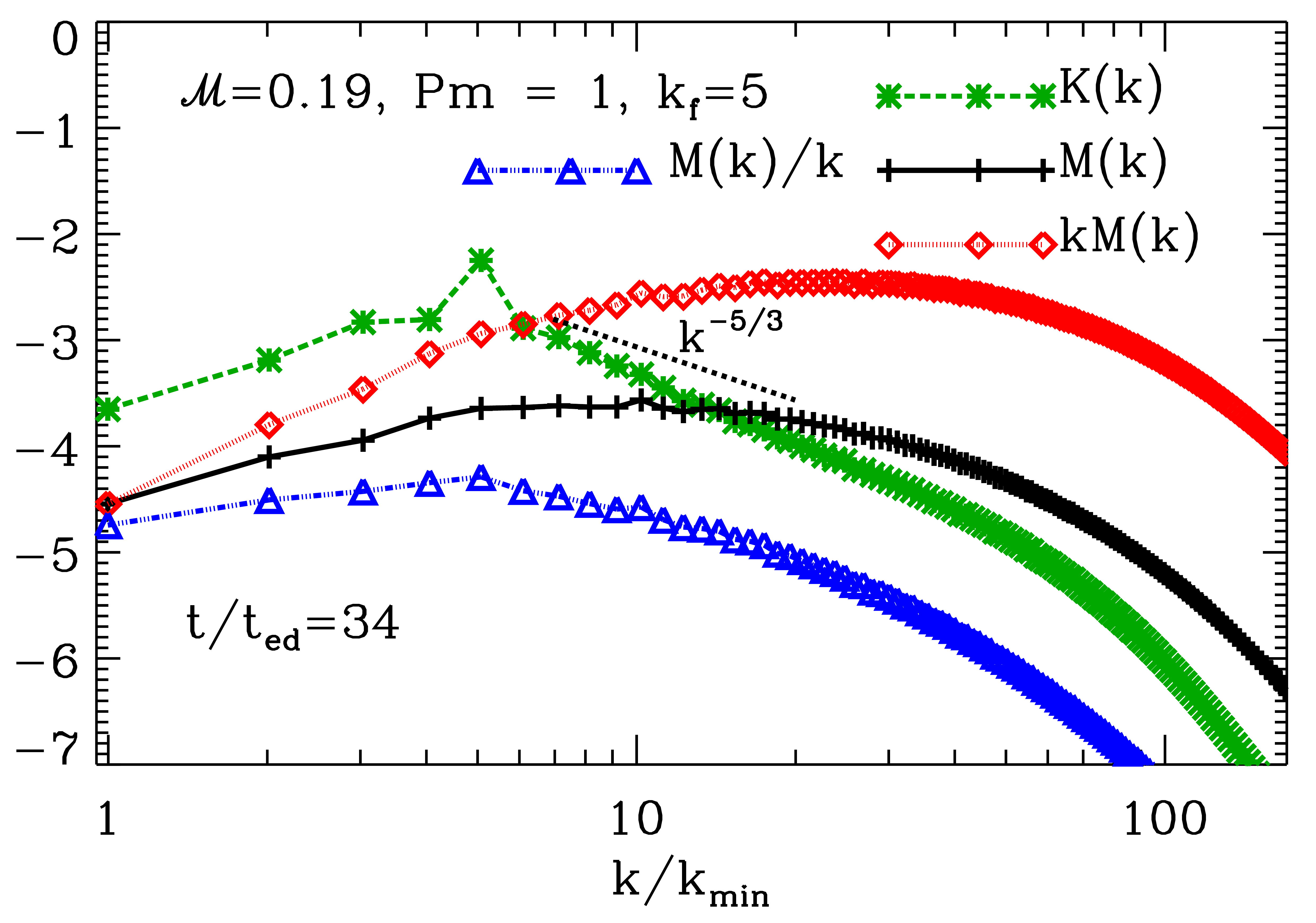} &
\includegraphics[width=0.45\columnwidth]{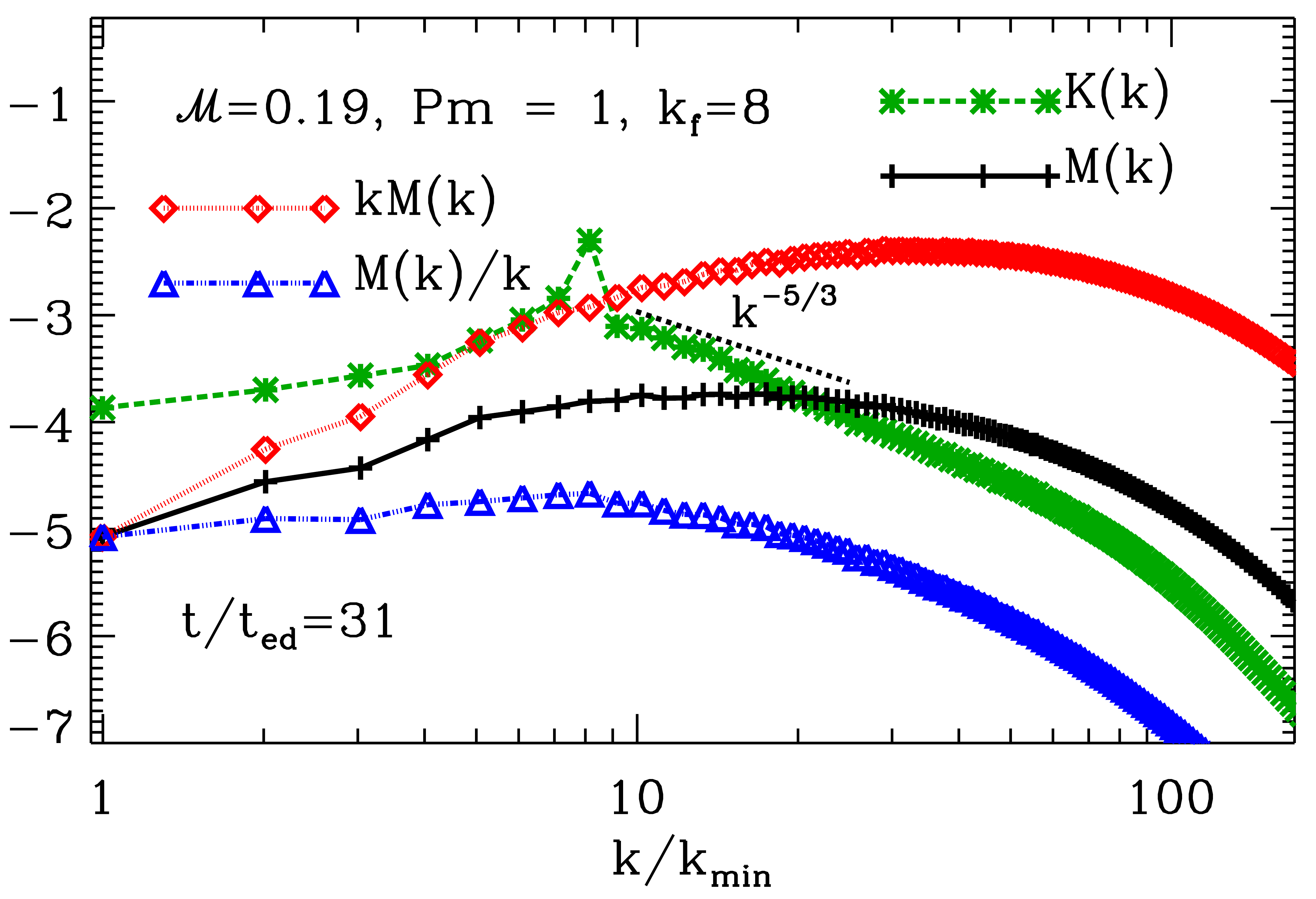}\\
\end{tabular}
\caption{Power spectra of kinetic energy, $K(k)$ (green dashed); magnetic 
energy, $M(k)$ (black, solid); $k\,M(k)$ (red, dotted); and $M(k)/k$ (blue, 
dash-dotted) for run B (left) and run C (right). These spectra are computed 
from a snapshot in the saturated state for the respective runs. The black 
dotted line shows the $k^{-5/3}$ for comparison. Here, the wave number is 
normalized in units of $k_{\rm min} = 2\pi/L$.}
\label{fig:pow_spec}
\end{figure}

In this section, we use the synthetic observations to investigate the
statistical properties of the polarized emission from the diffuse ICM. In
\citet{sur21}, it was shown that the power spectrum computed from the 2-D map
of Faraday depth (FD), the line of sight integral of $B_\|$ weighted by $n_{\rm
e}$, could be directly used to infer the magnetic integral scale, i.e., the
coherence scale of the magnetic fields, in the ICM also
see \citep{CR09}, provided FD is measured accurately. It was also found that, for
turbulence driven on 256\,kpc, broad-bandwidth spectro-polarimetric
observations $\gtrsim$3 \ghz~are well suited for detecting and inferring the
intrinsic polarization properties of the ICM, and the polarized emission could
be well recovered by applying the technique of rotation measure (RM) synthesis
\citep{brent05, heald09}. Here, we will focus on the statistical properties of
the total and polarized synchrotron emission from the ICM for turbulence driven
on different scales and the impact of smoothing the emission by a telescope
beam. It will become clear from the following sections, the different scales of
turbulent driving adopted in this work allow us to probe the effects of
magnetic fields with varying correlation scales on the statistical nature of
the polarized emission.

Before we focus our attention on the different correlation scales, we show in
Figure~\ref{fig:pow_spec}, the 1-D power spectra of  the kinetic energy $K(k)$
(green dashed with asterisks); magnetic energy, $M(k)$ (black, solid line with
plus symbols); spectra of $k\,M(k)$ (red, dotted line with diamonds)
representing the largest energy-carrying scale of the field; and that of
$M(k)/k$ (blue, dash-dotted line with triangles), which provides an estimate of
the magnetic integral scale. Here, $k$ is the wave number. These
spectra are obtained from snapshots in the non-linear saturated state of the
dynamo when turbulence is forced at $k_{\rm f} = 5$ (left-hand panel) and
$k_{\rm f} = 8$ (right-hand panel). The corresponding plot for $k_{\rm f} = 2$
can be gleaned from Figure~1 of \citet{sur21}.  The peak of $M(k)$ lies at
$\approx$1/10 of the box size for $k_{\rm f} = 5$ and at $\approx$1/20 for
$k_{\rm f} = 8$. On the other hand, the peak of $k\,M(k)$ occurs at much
smaller scales compared to the respective forcing scales in both cases. The
peak of $M(k)/k$ occurs on a scale very close to the turbulent driving scale in
both cases. These findings are qualitatively similar to the ones obtained in
\citet{sur21} when turbulence is forced at $k_{\rm f} = 2$. 

We find that, although turbulence is driven on different scales, the 
dispersion of Faraday depth ($\sigmafd$) are similar in all the cases 
having value $\sigmafd \sim100\radm$. For $k_{\rm f} = 2$, FD lies 
in the range $-438$ to $+415\radm$ with $\sigmafd \approx 118\radm$; 
for $k_{\rm f} = 5$, FD ranges between $-505$ and $+443\radm$ with 
$\sigmafd \approx 103\radm$; and for $k_{\rm f}=8$, FD lies in the 
range $-552$ to $430\radm$ having $\sigmafd \approx 93\radm$. 
There is a mild indication that $\sigmafd$ decreases with decreasing 
turbulence forcing scale $\lf$. This is because, $\sigmafd$ depends 
on the magnetic integral scale see Equation~(2) in \citep{sur21}, which, 
as we will discuss in the following sections, decreases with $\lf$.
Even though $\sigmafd$ decreases, beam and frequency-dependent 
Faraday depolarization are stronger for smaller forcing scales caused 
by the smaller scale filamentary magnetic field structures generated by 
the action of fluctuation dynamo. A quantitative investigation on the 
properties of frequency-dependent Faraday depolarization and their 
dependence on turbulence driving scale will be presented elsewhere.

\begin{figure}
	\centering
\begin{tabular}{cc}
\includegraphics[width=0.45\columnwidth]{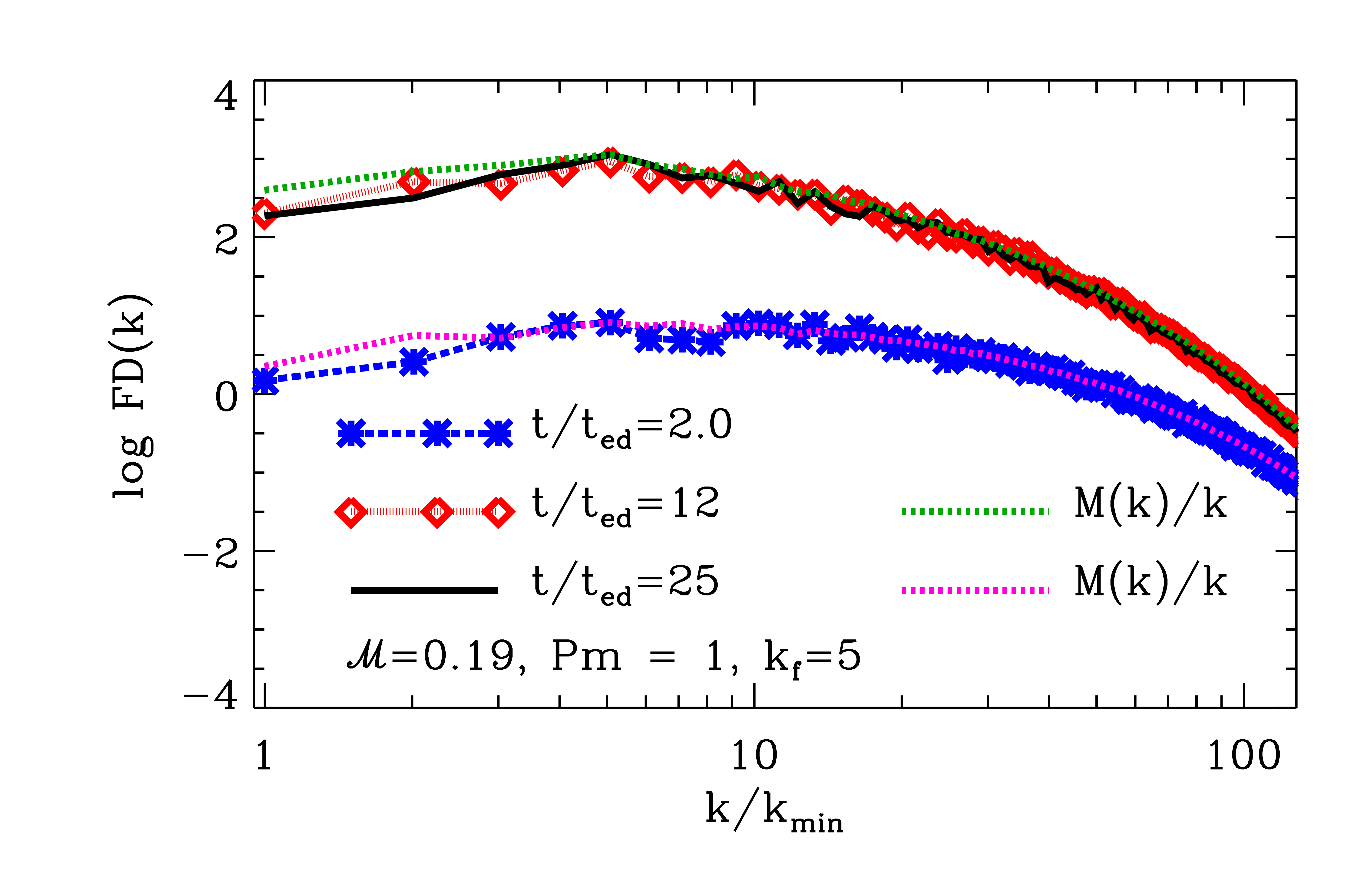} &
\includegraphics[width=0.45\columnwidth]{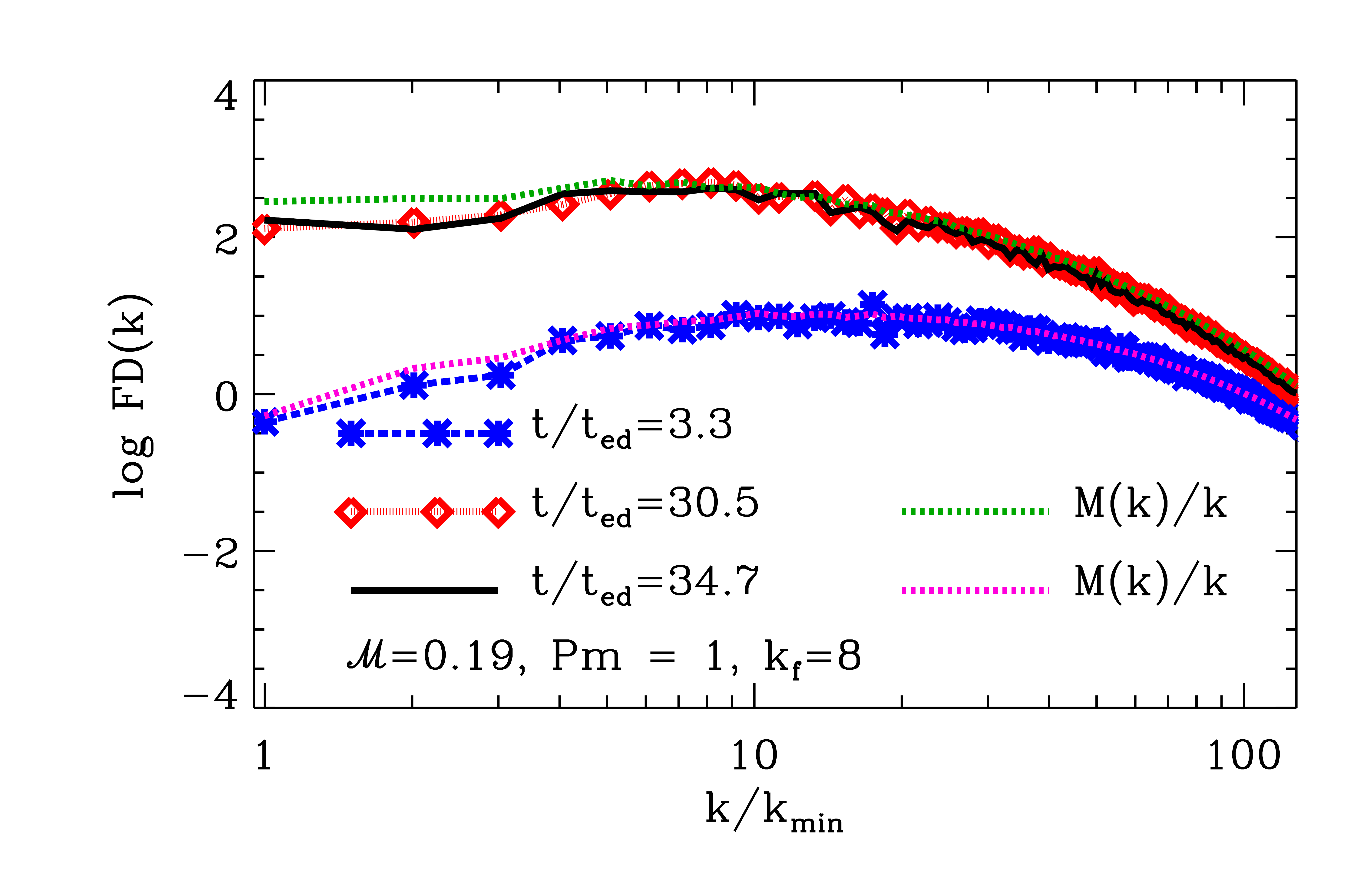}\\
\end{tabular}
\caption{Power spectra of the Faraday depth. The blue dotted curve 
corresponds to the spectra in the kinematic phase, while the red dashed
and black curves correspond the spectra in the saturated phase at 
two different times. For comparison, the magenta and green dashed 
curves show the scaled spectra of $M(k)/k$ in the kinematic and saturated 
phases, respectively. The left and right figures correspond to $k_{\rm f} = 5$ 
and $k_{\rm f} = 8$ simulations, respectively. Here, the wave number is 
normalized in units of $k_{\rm min} = 2\pi/L$.}
\label{fig:fdspec}
\end{figure}

For completeness, in Figure~\ref{fig:fdspec}, we show the power spectra of 
the map of FD in the kinematic and saturated stage of the dynamo for 
$\lf=102.4\kpc$ (left-hand panel) and $64\kpc$ (right-hand panel). For 
comparison, we also show the power spectra of $M(k)/k$ with dotted lines. 
As found for $\lf=256\kpc$ in \citet{sur21}, we find excellent match between 
the power spectra of FD and the corresponding $M(k)/k$ for all the cases, 
indicating that power spectrum of FD maps can provide valuable insight 
into the magnetic correlation scale in the ICM. However, as demonstrated 
in \citet{sur21}, estimating FD map from observations is challenging due 
to the limitations of available techniques, such as, the technique of RM 
synthesis. However, RM synthesis can recover the fractional polarization 
$p$ well, especially from broad-bandwidth spectro-polarimetric observations 
$\gtrsim$3 \ghz. Therefore, in the following, we will investigate the 
information on magnetic field properties that can be extracted from $p$.



\subsection{Correlation Scales in the ICM}\label{sec:intscales} 

The ICM can be characterized in terms of a number of distinct scales. 
Since we are primarily concerned with the effects of turbulence in this 
work, it is therefore important that the integral scales of the turbulent 
velocity and random magnetic fields be compared with the integral 
scales of the relevant observables, such as the FD, the total synchrotron 
intensity ($\Isync$) and the polarized intensity (${\rm PI}$). While 
these scales may be difficult to estimate directly from observations 
of the ICM, it can be easily derived from our simulations thereby 
providing a first hand estimate which may be confirmed with suitable 
future observations. For example, the integral scales of the turbulent 
velocity ($L_{{\rm int},V}$) and random magnetic fields 
($L_{{\rm int},M}$) can be obtained from their respective power 
spectra as, 
\begin{equation}
\label{int_scales}
L_{{\rm int},V} = \frac{2\pi\int [K(k)/k]\,dk}{\int K(k)\,dk}, \,\,\,\,\,
L_{{\rm int},M} = \frac{2\pi\int [M(k)/k]\,dk}{\int M(k)\,dk}.
\end{equation}

Estimates of the integral scales of FD, $\Isync$ and ${\rm PI}$ can 
similarly be obtained from Equation~(\ref{int_scales}) by using the 
appropriate power spectrum corresponding to the observable. 

Table~\ref{tab:integscales} lists the integral scales across different forcing
scales computed directly from the simulations and the synthetic
observations. In order to estimate the sensitivity of these scales to the
random fluctuations in the field, the values of the integral scales are shown
at two different times selected from the non-linear saturated state of the
system in each case. Irrespective of the forcing scale of turbulence, we find
that the velocity integral scale $L_{{\rm int}, V} \approx 2.2 \textrm{--}
3\,L_{{\rm int}, M}$, in agreement with the estimates obtained for subsonic
tubulence in earlier studies \citep{BS13,SBS18}. We further find that, despite
random scatter from one realization to another, the integral scales associated
with the observables, FD, $I_{\rm sync}$ and ${\rm PI}$ are all larger than
$L_{{\rm int}, M}$ by a factor $\sim$2--2.5, depending on the
forcing scale ($l_{\rm f}$). The integral scale of $I_{\rm sync}$ in each case,
appears to be somewhat larger than $L_{\rm int, FD}$ and $L_{\rm int, PI}$ at
frequencies $\gtrsim$4 \ghz.  Glimpses of the effects of Faraday depolarization
is evident from the estimates of $L_{\rm int, PI}$ at lower frequencies,
wherein, $L_{\rm int, PI}$ at $\nu < 1.4~\ghz$ are generally smaller compared to
the ones at $\nu = 5~\ghz$ due to small-scale structures introduced by Faraday
depolarization. 

\begin{table}
	\centering
\setlength{\tabcolsep}{5.0 pt}
\caption{Values of the integral scales (in kpc) of the velocity ($L_{{\rm int},V}$), 
magnetic fields ($L_{{\rm int},M}$), Faraday depth ($L_{\rm int, FD}$), the total 
synchrotron intensity ($L_{{\rm int},I}$) at $1~\ghz$, and, the polarized intensity 
($L_{{\rm int},{PI}}$) at $0.6$, $1.2$ and $5~\ghz$ at two different times in the steady 
state obtained from three simulations where turbulence is driven on scales 
of $l_{\rm f} = 256, 102.4$ and $64~\kpc.$}
\scalebox{0.995}{\begin{tabular}{*{10}{c}}
    \toprule
\boldmath{$k_{\rm f}$} & \boldmath{$t/t_{\rm ed}$}  & \boldmath{$L_{{\rm int},V}$} &  \boldmath{$L_{{\rm int},M}$} & \boldmath{$L_{\rm int, FD}$} &
\boldmath{$L_{{\rm int},I}$} & \multicolumn{3}{c}{\boldmath{$L_{\rm int, PI} ~(\kpc)$}}  \\
\textbf{(Forcing Scale)} &      & \boldmath{$(\kpc)$} & \boldmath{$(\kpc)$}  & \boldmath{$(\kpc)$} &\boldmath{ $(\kpc)$} & \boldmath{$0.6\ghz$} & \boldmath{$1.2\ghz$} & \boldmath{$5\ghz$}  \\ \midrule
2 & 16.6 & 320  &  106  &  212.5  &  224 &  125.5 & 157.5 &  194  \\
(256\,kpc) & 23 & 340 & 112.4 & 216 & 227.6 & 140 & 159 & 188 \\ 
 5 & 20.2 & 98.7  &  40.7  &  78.8  &  130 & 57.3 & 55.0 & 88.6  \\
(102.4\,kpc) & 24.5 & 99.1 & 40.8 & 87.5 & 120.2 & 57.7 & 60.3 & 87.0 \\
 8 & 30.6 & 62.4  & 27.6  &  58.0  &  86.0 & 38.3 & 37.1 & 70.2  \\
(64\,kpc) & 34.7 & 62.4 & 28.0 & 61.6 & 89.5 & 35.0 & 34.0 & 74.2 \\
\bottomrule
\end{tabular}}
\label{tab:integscales}
\end{table}

\subsection{Smoothing of Total Intensity}
\label{sec:totI}


\begin{figure}
\centering
\begin{tabular}{ccc}
{\large $l_{\rm f}=256\kpc$} & {\large $l_{\rm f}=102.4\kpc$} & {\large $l_{\rm f}=64\kpc$} \\
& & \\
\includegraphics[width=5.5cm, trim=10mm 0mm 0mm 10mm, clip]{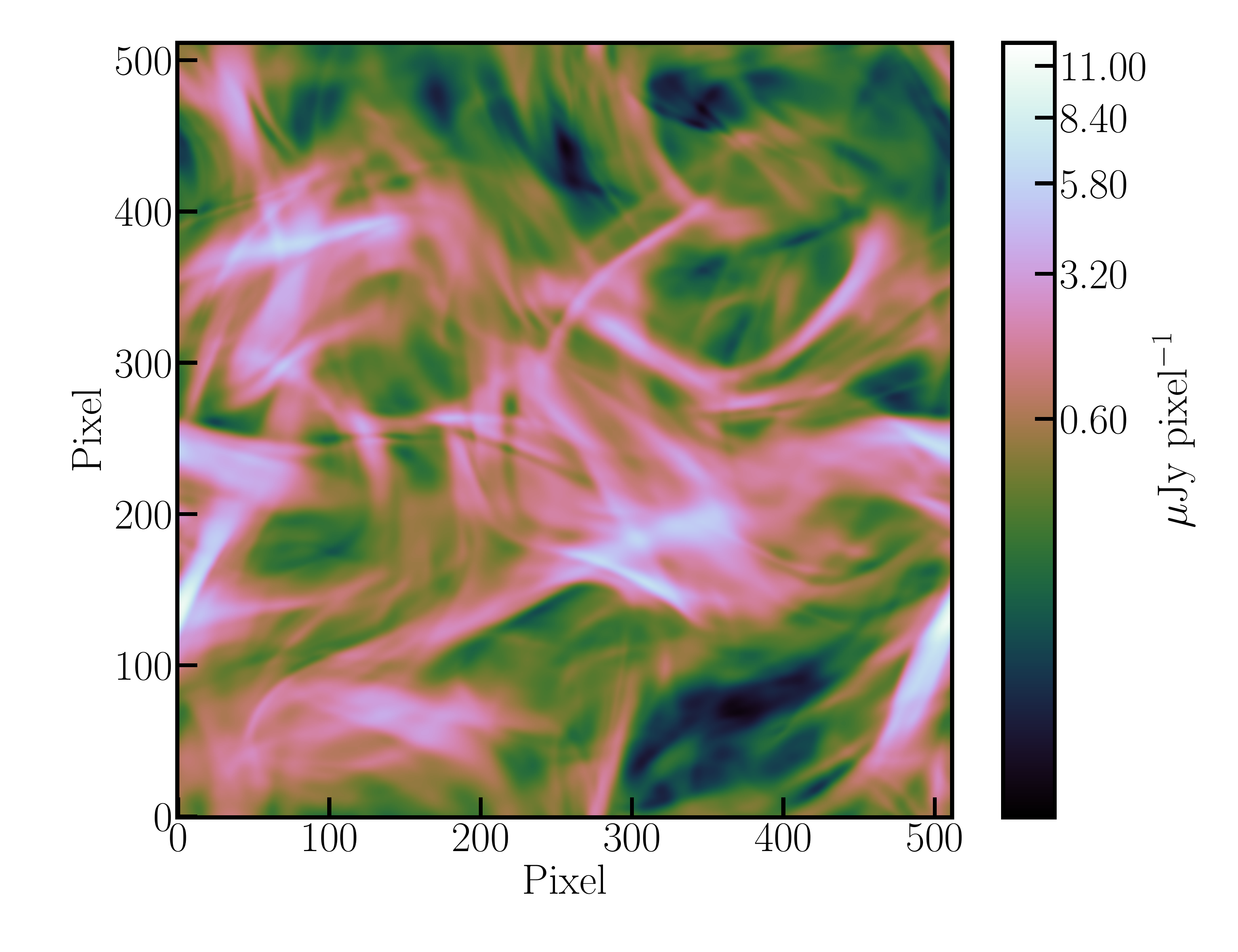} &
\includegraphics[width=5.5cm, trim=10mm 0mm 0mm 10mm, clip]{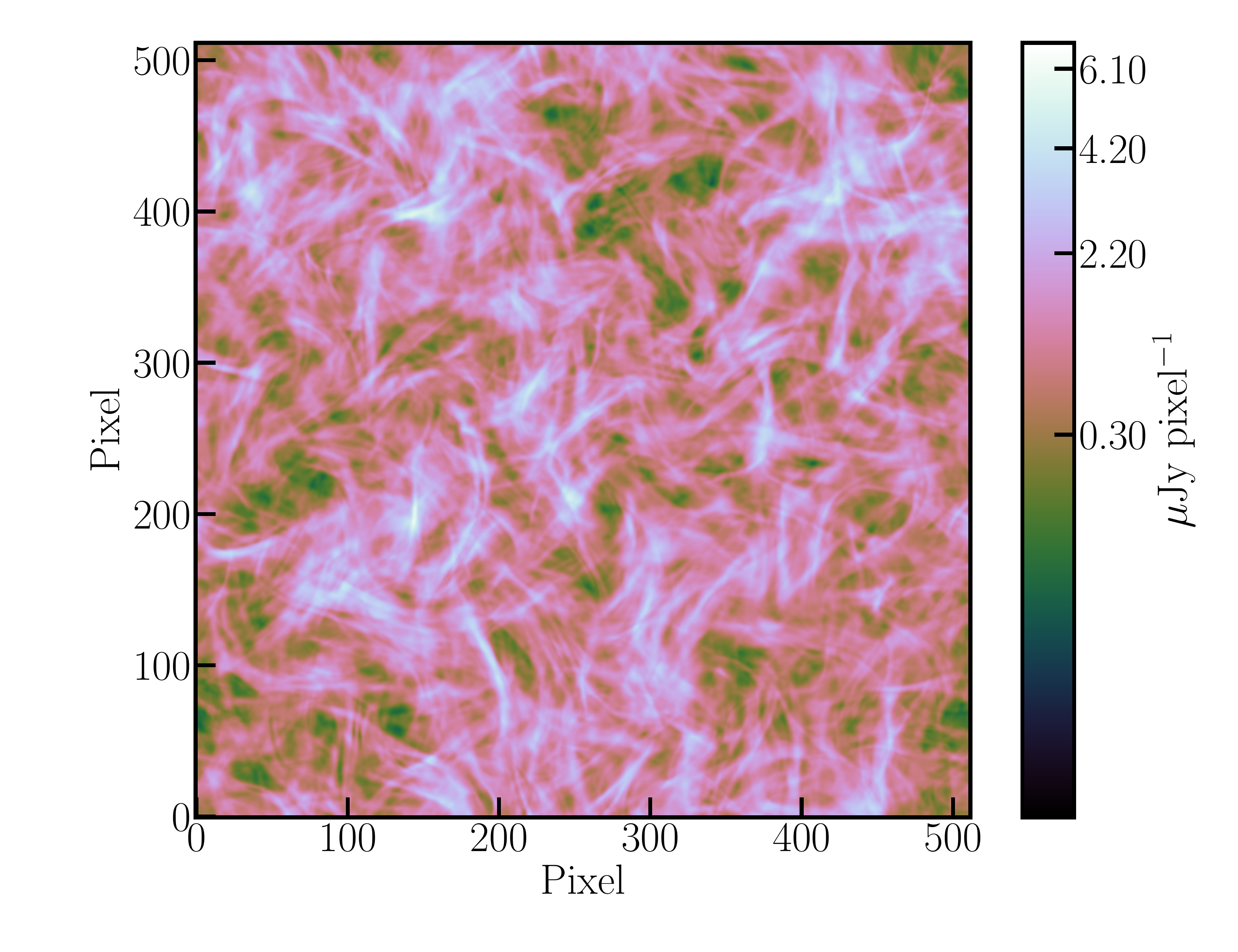} &
\includegraphics[width=5.5cm, trim=10mm 0mm 0mm 10mm, clip]{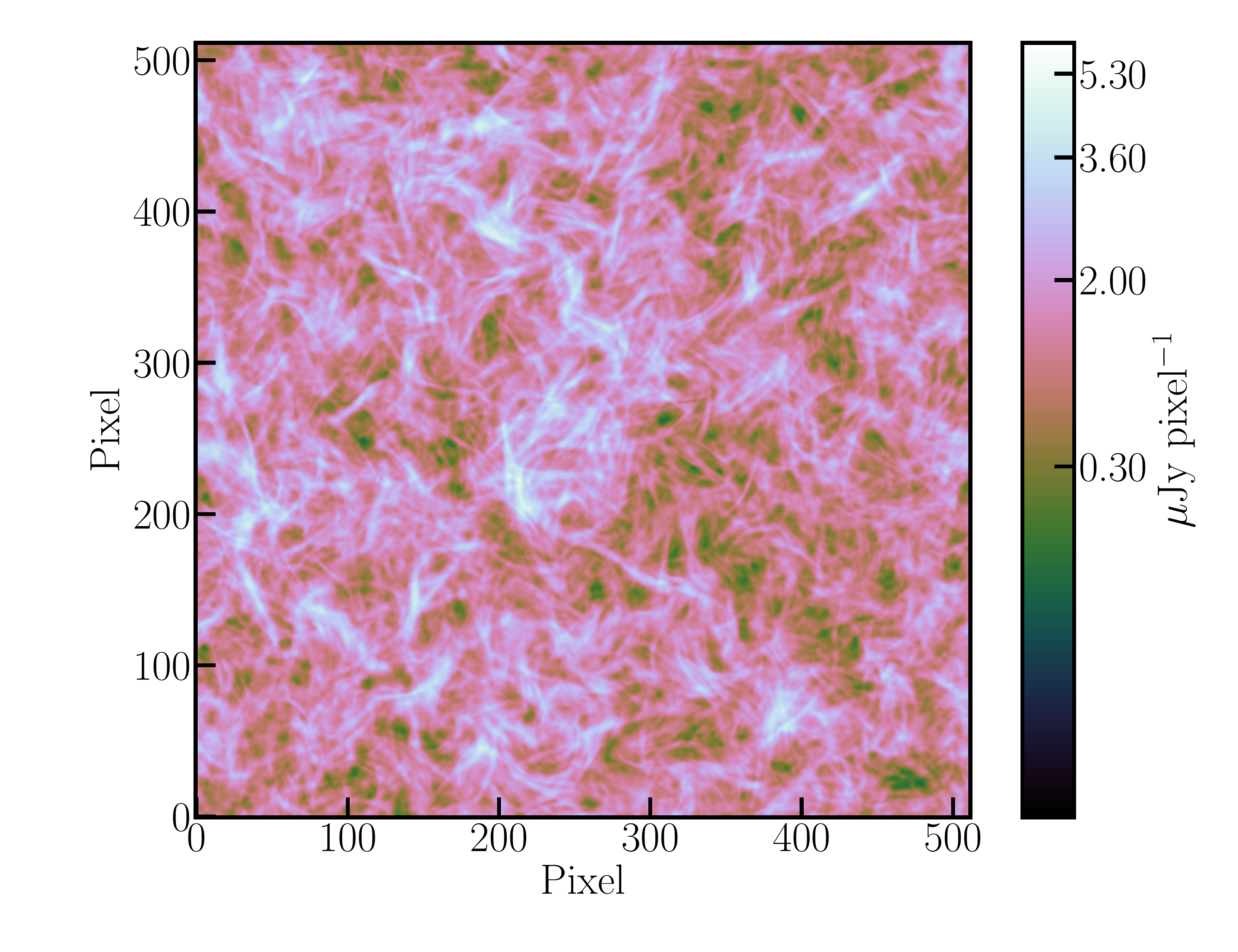} \\
\includegraphics[width=5.5cm, trim=10mm 0mm 0mm 10mm, clip]{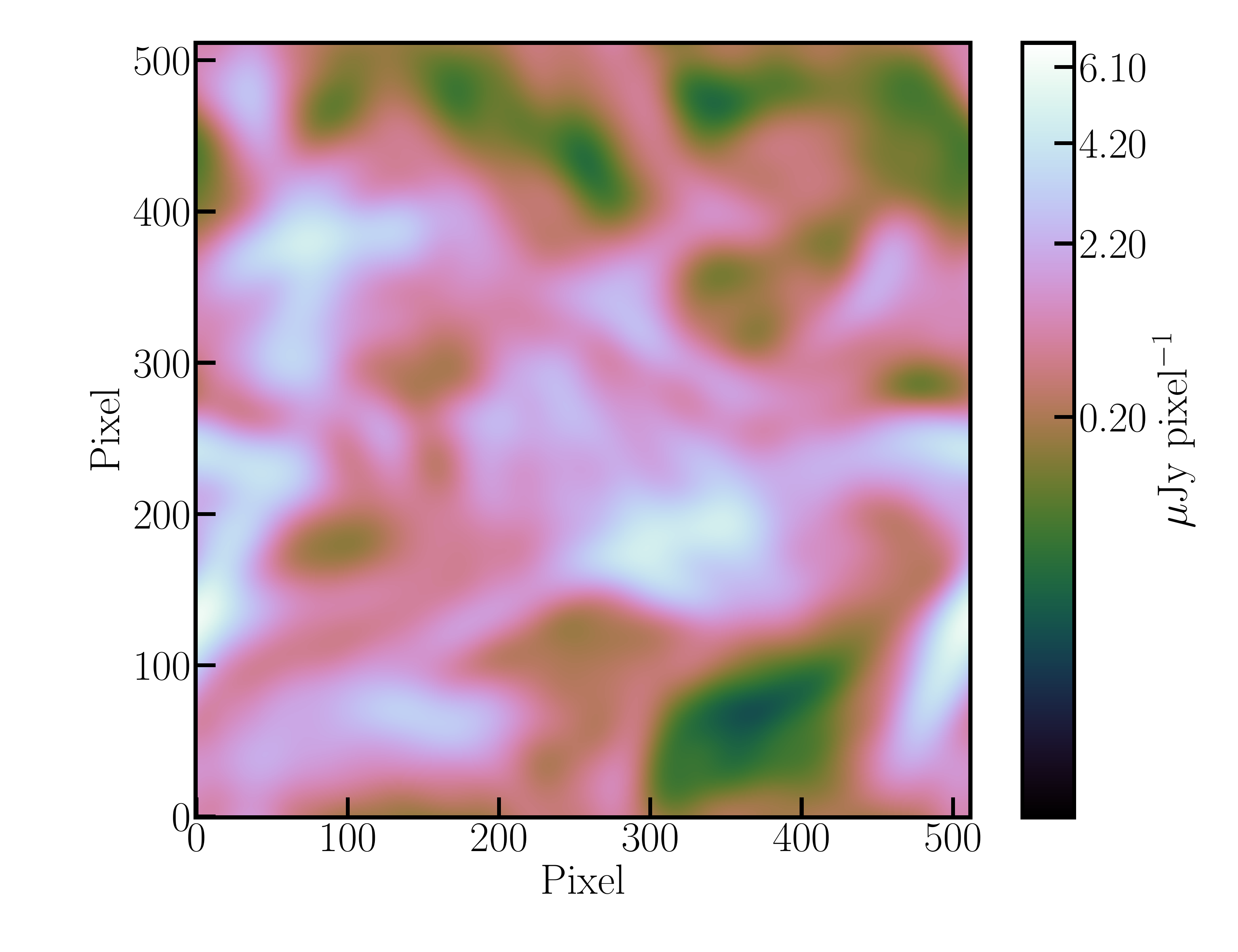} &
\includegraphics[width=5.5cm, trim=10mm 0mm 0mm 10mm, clip]{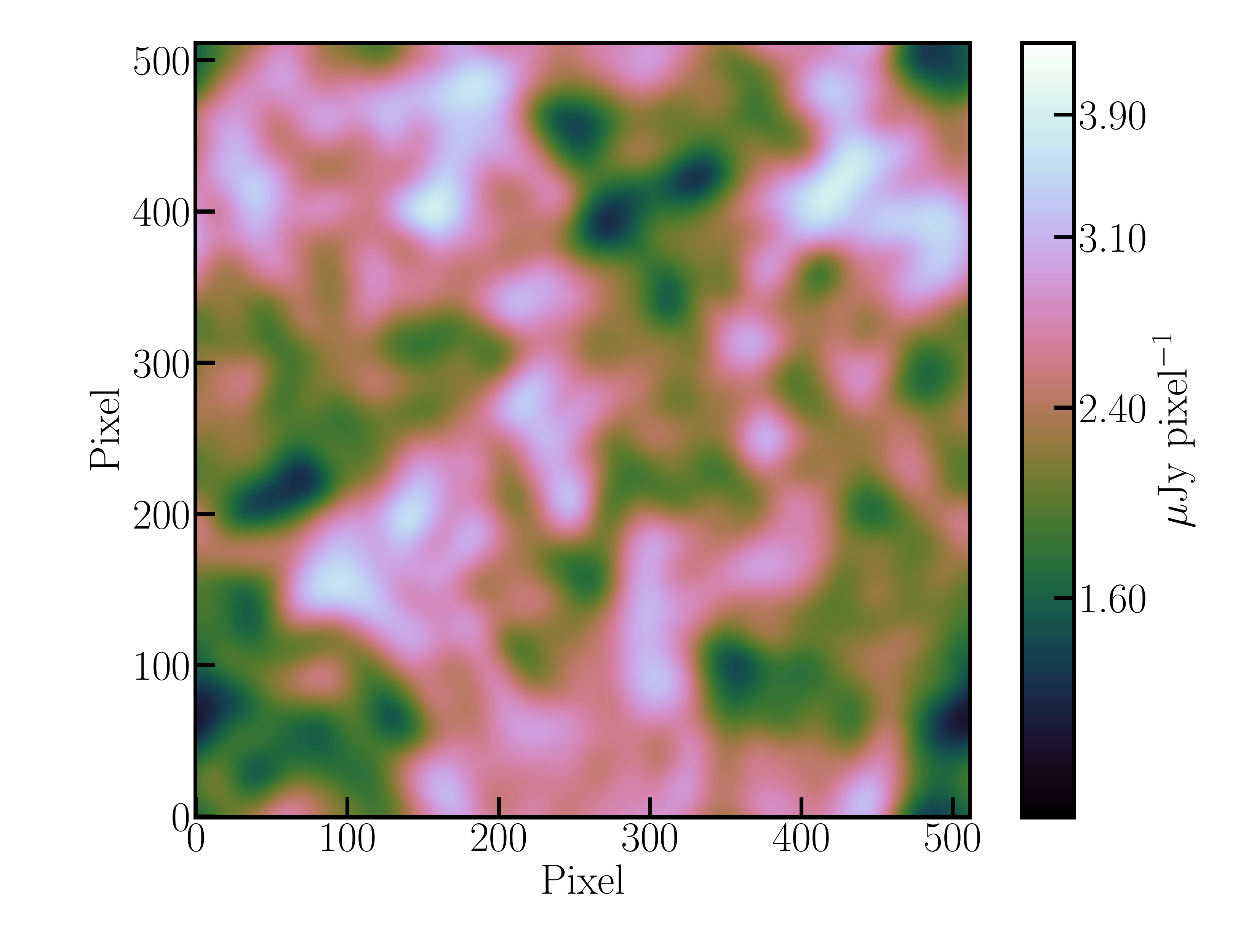} &
\includegraphics[width=5.5cm, trim=10mm 0mm 0mm 10mm, clip]{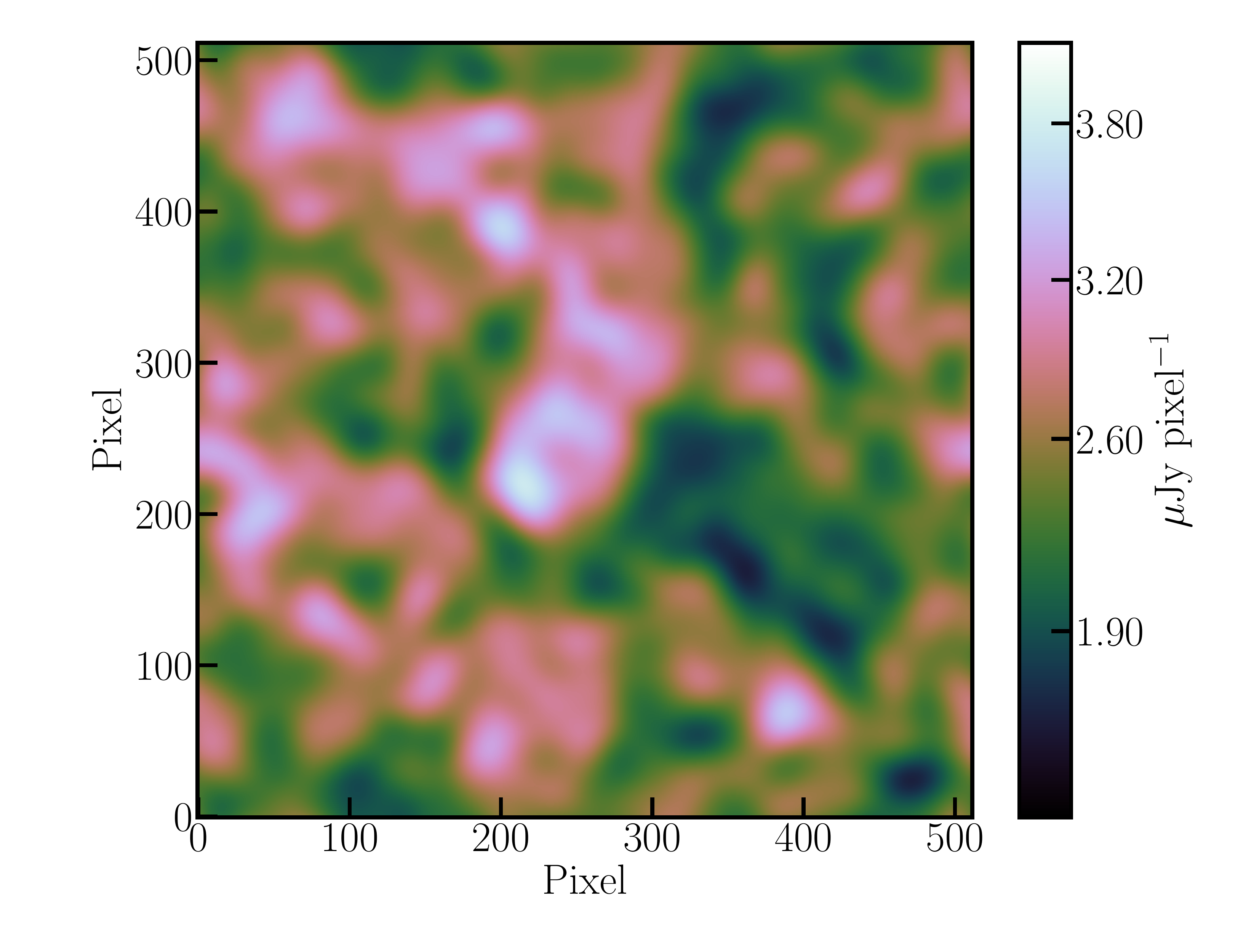} \\
\includegraphics[width=5.5cm]{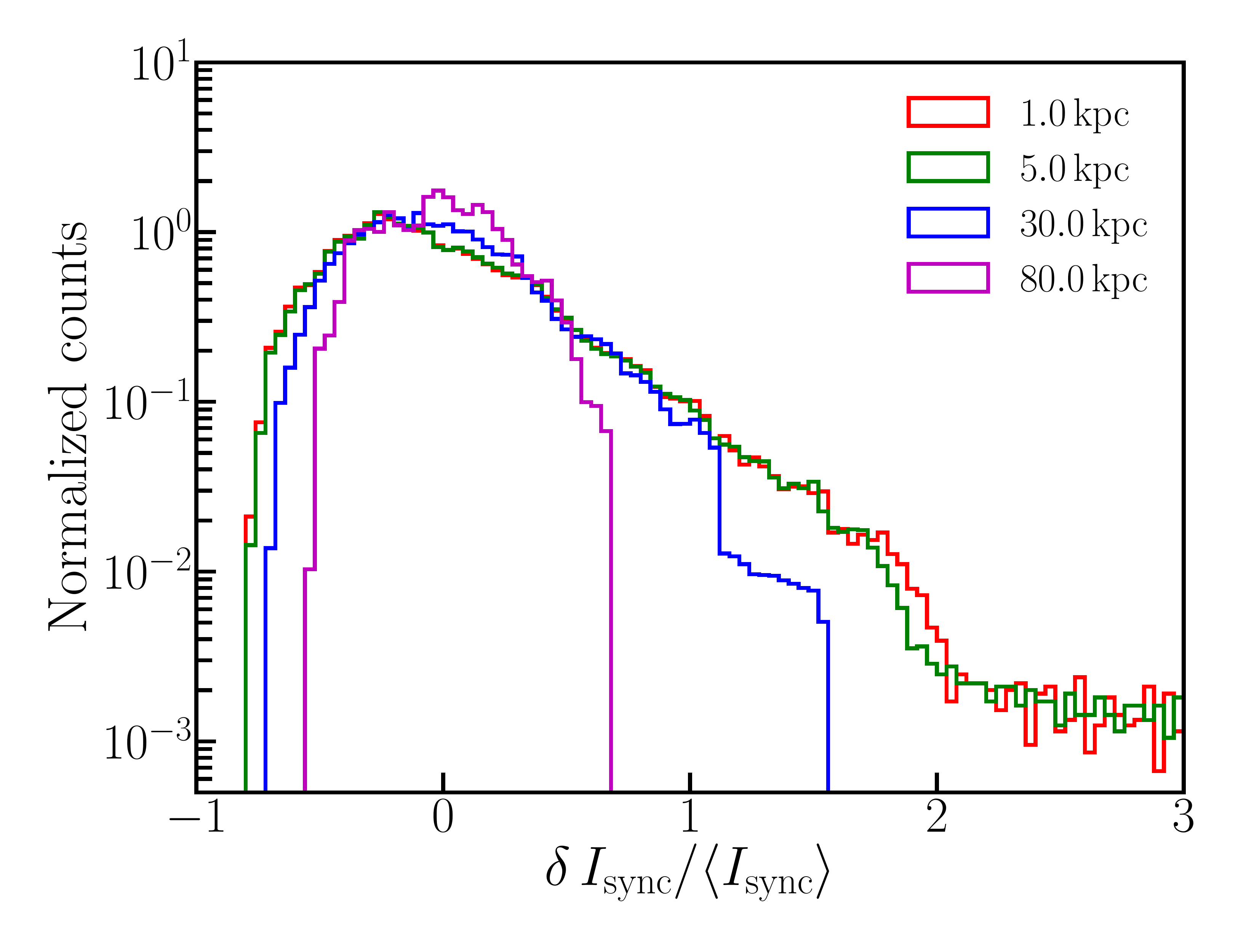} &
\includegraphics[width=5.5cm]{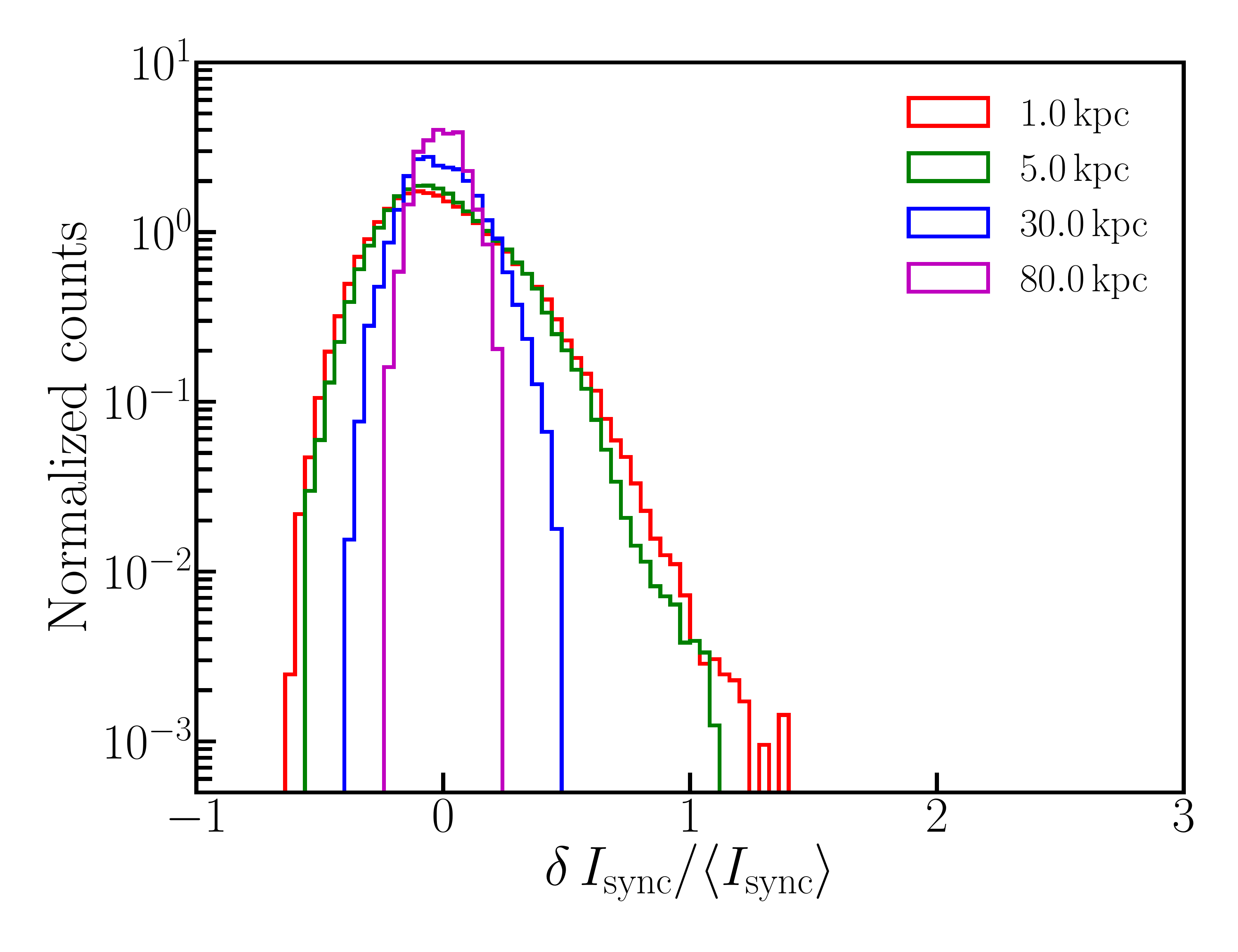} &
\includegraphics[width=5.5cm]{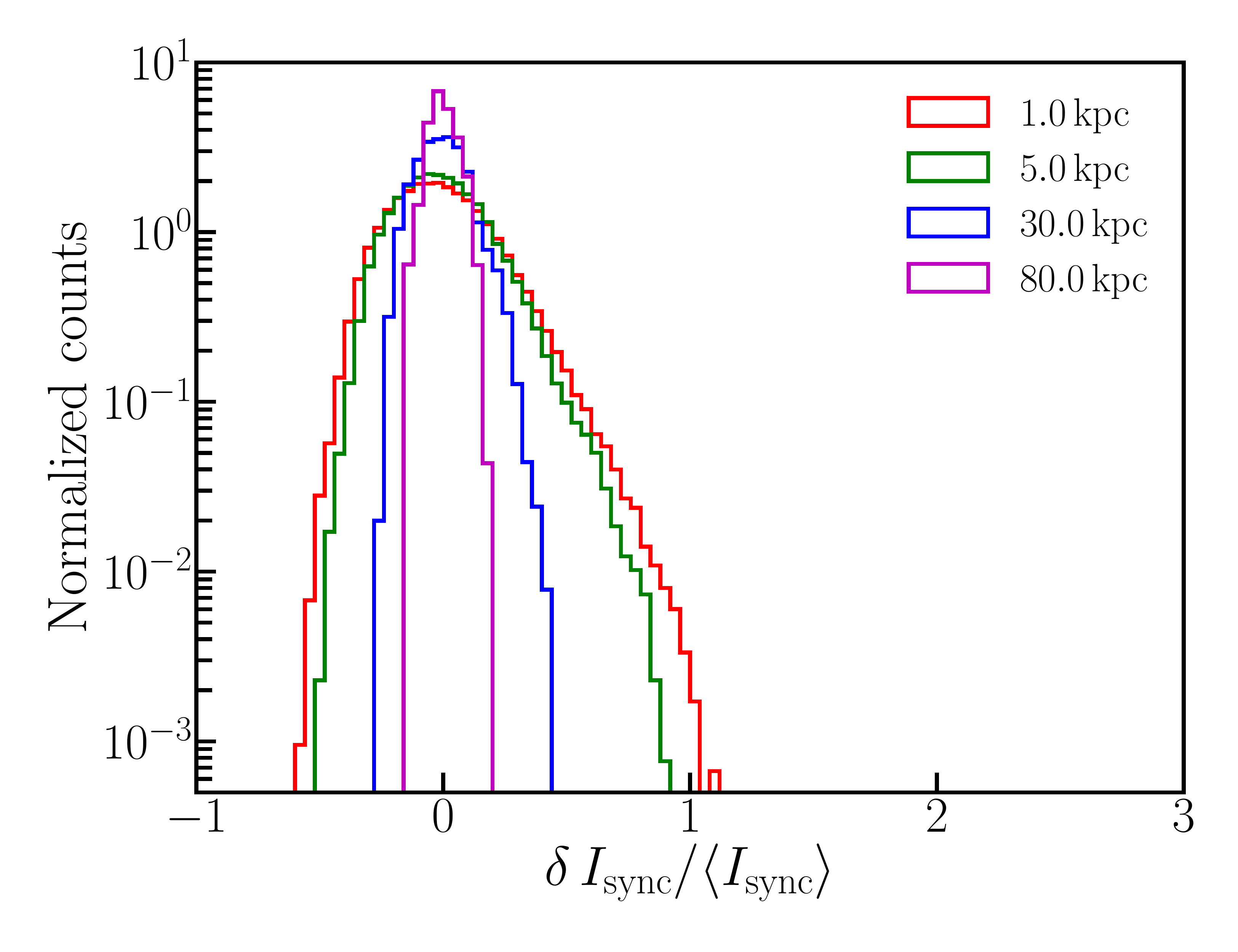} \\
\end{tabular}
\caption{\textbf{Top row:} The 2-D maps of surface brightness of the 
total synchrotron emission ($I_{\rm sync}$) at the native 1\,kpc resolution of the 
simulations. \textbf{Middle row:} Synchrotron total intensity maps 
in the top row smoothed by a Gaussian kernel with $\rm FWHM = 30\kpc$.
\textbf{Bottom row:} Pixel-wise distribution of $\delta\,\Isync/\langle \Isync\rangle$ 
for smoothing the total synchrotron intensity on different scales shown 
with different colours. The \textbf{Left}, \textbf{middle} and 
\textbf{right} columns are for synthetic observations of $l_{\rm f}=256\kpc$ 
at $t/t_{\rm ed} = 23$, $l_{\rm f}=102.4\kpc$ at $t/t_{\rm ed} = 20.2$, and $l_{\rm f}=64\kpc$ 
at $t/t_{\rm ed} = 30.6$, respectively.}
\label{fig:totIoverdens}
\end{figure}


For our assumed constant spatial distribution of $n_{\rm CRE}$ in each mesh
point, the synchrotron intensity ($I_{\rm sync}$) in the plane of the sky is
directly dependent on $B_\perp$ as $I_{\rm sync} \propto
\int\,B_\perp^{1-\alpha}\,dl$.  Hence, the filamentary magnetic field
structures generated due to the action of fluctuation dynamo are also observed
in the synthetic synchrotron intensity map. These structures, along with
non-linear dependence of the $I_{\rm sync}$ on $B_\perp$, gives rise to a
log-normal distribution of the $I_{\rm sync}$ with long tails when observed
with resolution comparable to that of the simulations also
see \citep{sur21}. In the top row of Figure~\ref{fig:totIoverdens} we show the maps
of $\Isync$, where, the left, the middle and the right columns are for forcing
at 256, 102.4 and 64\,kpc. It is clearly seen that when turbulence is forced on
smaller scales, the filamentary structures in $\Isync$ become more volume
filling. These filaments have smaller extent for smaller $\lf$, which
is also evident from the integral scale of the synchrotron emission ($L_{{\rm
int},I}$) given in Table~\ref{tab:integscales}, and seen in
Figure~\ref{fig:depol_forcing} (right). In the absence of shocks (due to
subsonic turbulence) these filamentary structures are a direct signature of
fluctuation dynamos in the ICM. We find that both $L_{{\rm int},I}$ and
$L_{{\rm int},M}$ increases linearly with $l_{\rm f}$, except that $L_{{\rm
int},I}$  is higher than $L_{{\rm int},M}$ by roughly a factor of 2.
Furthermore, the intensity contrast decreases significantly when turbulence is
forced on smaller scales. The middle row of Figure~\ref{fig:totIoverdens} shows
the total intensity maps in top row smoothed on 30\,kpc (FWHM).  It is clear
that observations with a telescope smears the filamentary features and
significantly reduces the intensity contrast.

Interestingly, radio continuum observations reveal that the emission from halos
of galaxy clusters are spatially smooth \citep{cuciti21}, devoid of the
filamentary structures seen at the native resolution of the simulations.
Hence, it is important to assess the extent to which telescope beams smear out
such structures, and whether they can be discerned with respect to the diffuse
background emission. In the bottom row of Figure~\ref{fig:totIoverdens}, we show
the pixel-wise distribution of overdensity of $I_{\rm sync}$, i.e.,
$\delta\,I_{\rm sync}/\langle I_{\rm sync} \rangle$ at the native 1\,kpc
resolution of the simulations and for smoothing over different spatial scales
mimicking astronomical observations performed with different telescopes
resolutions. Here, $\bra{I_{\rm sync}}$ is the mean $I_{\rm sync}$ over the
map, and $\delta\,\Isync = \Isync - \bra{\Isync}$. The bright filamentary
structures with $\delta\,\Isync/\bra{\Isync} > 1$ are clearly seen as long
tails of the distribution at 1\,kpc resolution and for smoothing on 5\,kpc
scales when turbulence is forced on 256\,kpc scale. However, for forcing on
102.4 and 64\,kpc scales, the maximum overdensity decreases significantly to
$\sim$1. In fact, the Fisher-Pearson coefficient of skewness ($g_1$) of the
distributions decreases from 1.4 for forcing on 256\,kpc scale to 0.62 for
forcing on 102.4 and 64\,kpc scales, indicating that the distribution of
surface brightness of the synchrotron emission for turbulence driving on scales
$\lesssim$100~\kpc~to be relatively symmetrical. This is due to the fact that,
the filamentary structures are more volume filling when turbulence is driven on
smaller scales. These results suggest that, in the presence of realistic
telescope noise, coupled with generically low surface brightness, the
filamentary structures in total intensity emission from the radio halos of
galaxy clusters would be difficult to discern even when observed with
relatively high spatial resolutions of $\sim$10--20 \kpc.

Smoothing the synthetic $\Isync$ maps on larger spatial scales of 30 and
80\,kpc, which are the typical spatial resolutions achieved with existing
observations of radio halos, we find that $\delta\,\Isync/\bra{\Isync}$
decreases significantly below 1 in almost every pixel, making the distributions
symmetric with $g_1 < 0.2$ for all the three scales of turbulence driving.  The
situation is expected to be further aggravated in the presence of telescope
noise and demonstrates why such filamentary structures seen in the simulations
have not been observed. In order to detect the fluctuations in $\Isync$,
sensitive, high resolution observations ($<$5 \kpc) are of paramount importance.
It is noteworthy, in contrast to the nearly uniform $n_{\rm e}$ (fluctuations
in overdensity are at $3\%$ level for $\mathcal{M} \approx$ 0.18--0.19) and
uniform distribution of $n_{\rm CRE}$ assumed in our models for the core region
in ICM, the magnetic field strength, $n_{\rm e}$ and $n_{\rm CRE}$ are likely
to be stratified in galaxy clusters, decreasing away from the cluster core on
scales of hundreds of kpc over the entire volume. Hence, stratification of
these physical quantities could give rise to skewness in the distribution of
the synchrotron surface brightness even if observed using significantly
smoothed beam.

\subsection{Intrinsic Polarization and Forcing Scale} \label{sec:intpol}

Although, results from numerical simulations of fluctuation dynamos in the ICM
suggest that cluster halos are expected to be intrinsically polarized at
10--30\% level \citep{sur21}, unambiguous detection of polarized emission
remains elusive. Work by~\mbox{\citet{GF04}} found cluster radio halos to be polarized
with $p < 2\textrm{--}10\%$ at $1.4\ghz$.  \citet{TKW03} failed to detect any
significant diffuse polarized emission in the Coma cluster at 2.67 and
$4.85~\ghz$ with linear resolutions of $110$ and $60\kpc$, respectively. This
could be due to a combination of Faraday and beam depolarization, and
insufficient sensitivity.  In light of these results, we present here the
statistical properties of the polarized synchrotron emission when smoothed on
different scales by a telescope beam. Before delving into them in detail, we
first discuss the intrinsic properties of the fractional polarization ($p_{\rm
int}$) at the native 1\,kpc resolution of the simulations, and in the absence
of frequency-dependent Faraday depolarization (equivalent to $p$ at wavelength
$\lambda=0$).

In the presence of random magnetic fields, polarized emission originating at
different depths in the ICM is expected to undergo random walk as the LOS
passes through a number of magnetic cells, within which the field is ordered
but, randomly oriented from cell to cell. This leads to rotation of the plane
of polarization by random angles resulting in a random degree of polarization.
Therefore, the volume-averaged intrinsic fractional polarization ($\pmean_{\rm
int}$) is expected to depend as $\pmean_{\rm int} \approx p_{\rm
max}/\sqrt{N_{\rm mag}}$ \citep{sokol98}.  Here, $p_{\rm max} = (1-\alpha)/(5/3
- \alpha) = 0.75$ (for $\alpha = -1$) is the maximum fractional polarization
and, $N_{\rm mag} = L/L_{{\rm int}, M}$ is the number of magnetic cells along
the LOS. The above relation hints to a decrease of $\pmean_{\rm int}$ with
decreasing $l_{\rm f}$. This can be intuitively understood from the fact that a
smaller $l_{\rm f}$ implies a larger value of $N_{\rm mag}$ as $L_{{\rm int},M}
< l_{\rm f} < L$.\footnote{Note that, the evolution of $L_{{\rm int}, M}$ is
controlled by the Lorentz force which tends to order the field as the system
approaches saturation \citep{BS13,SBS18}.  Since magnetic fields generated by
fluctuation dynamos can be ordered at the most on the scale of turbulent
driving, the integral scale of the field is expected to be smaller than $l_{\rm
f}$.} In fact, Table~\ref{tab:integscales} shows that $L_{{\rm int},
M}$ is smaller by a factor of $\approx$4 in each dimension for $l_{\rm f} =
64\kpc$, compared to $l_{\rm f} = 256\kpc$. This translates to $N_{\rm mag}$
being larger by the same factor for $l_{\rm f} = 64\kpc$, which would result in
a larger cancellation of the polarized emission along the LOS, and hence, lower
$\pmean_{\rm int}$.

\begin{figure*}[t]
	\centering
\begin{tabular}{cc}
\includegraphics[width=0.45\columnwidth]{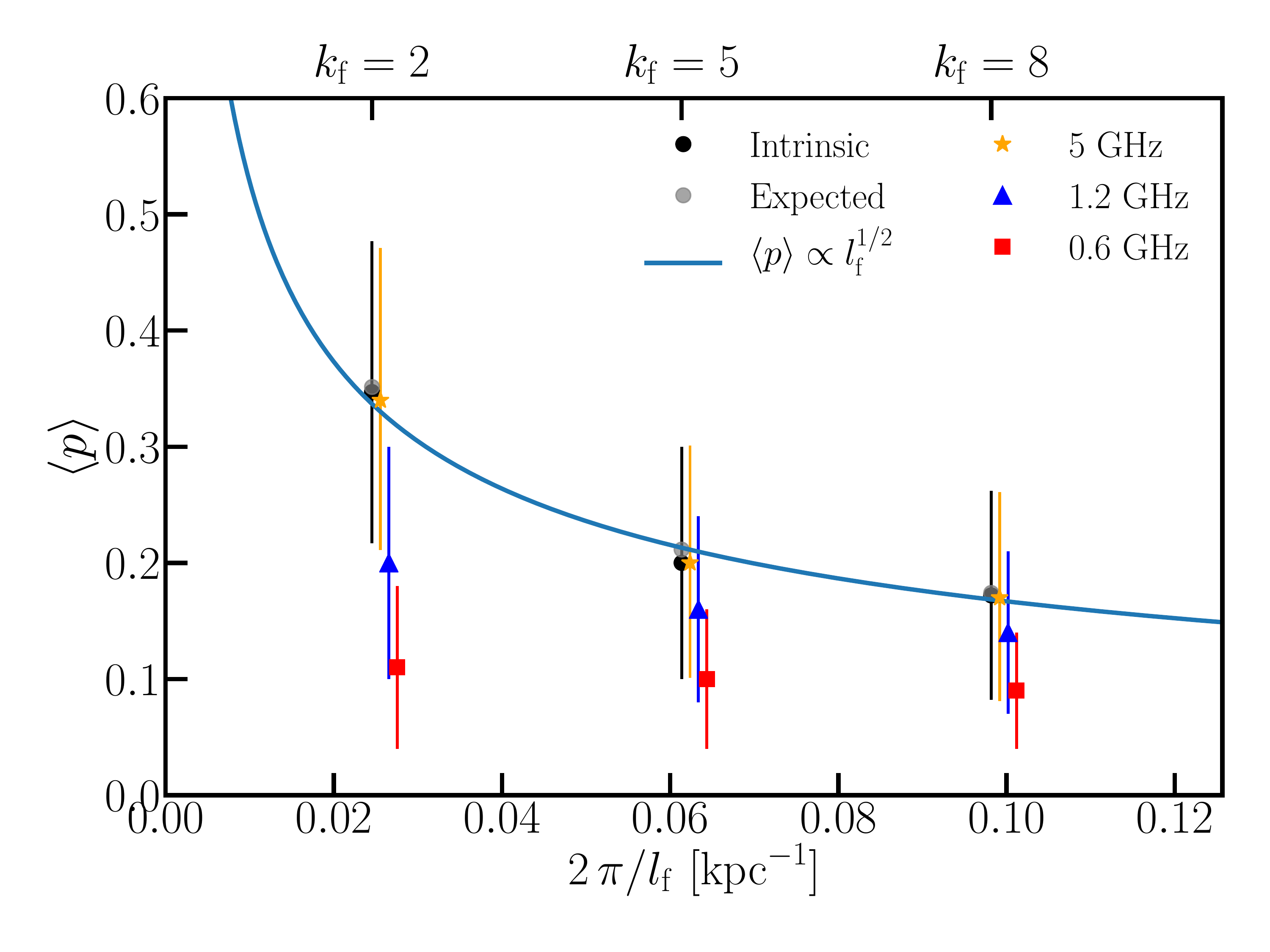} &
\includegraphics[width=0.45\columnwidth]{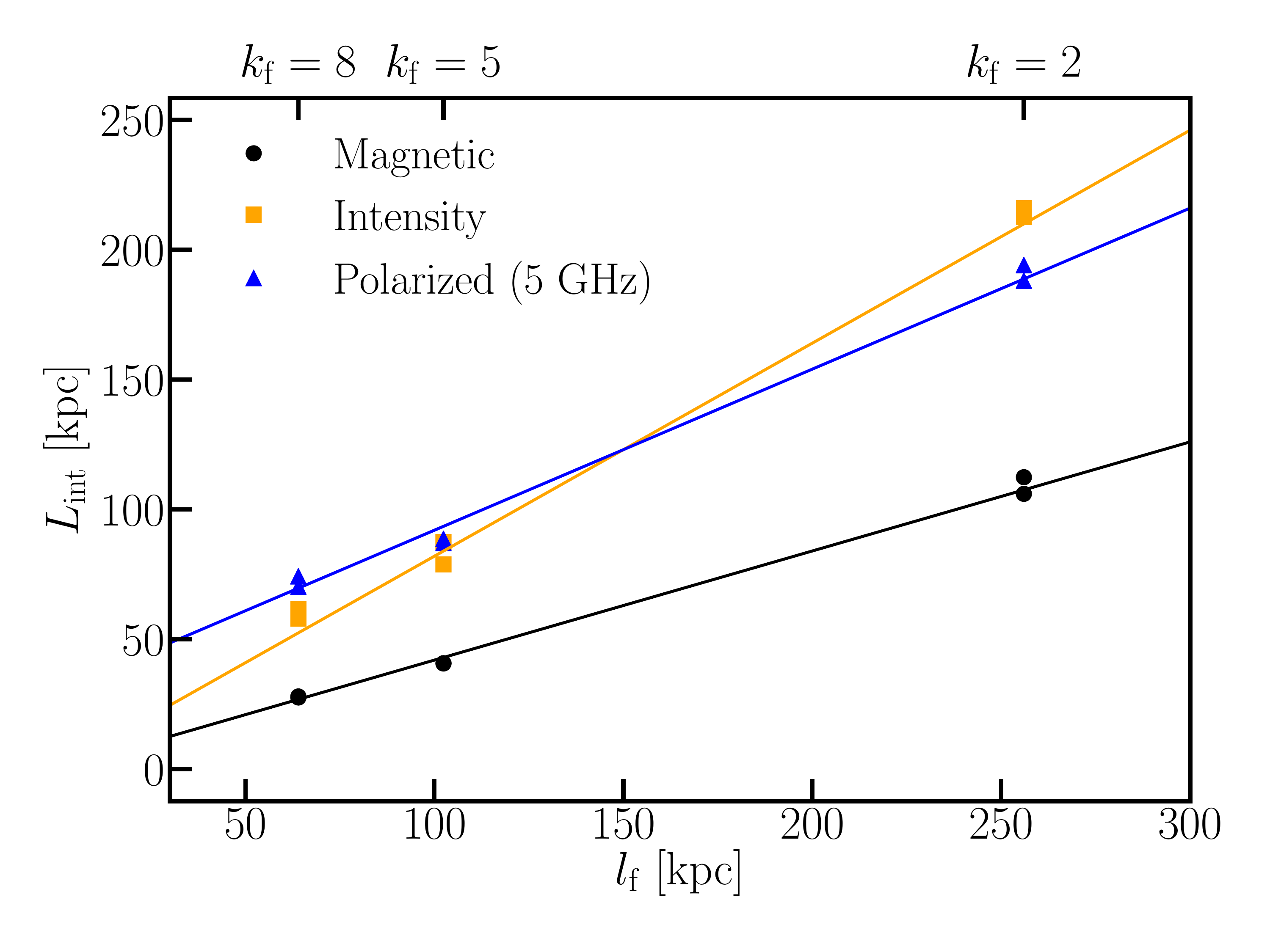}\\
\end{tabular}
\caption{\textbf{Left:} Variation of mean fractional polarization 
$\langle p \rangle$ with the turbulence driving scale $l_{\rm f}$ in the 
saturated stage at the native resolution of the simulations.
The black points show the intrinsic $\pmean$ from the simulated 
volumes computed in the absence of Faraday rotation. The grey 
points show the expected $\pmean$ computed from the magnetic 
integral scale. The blue curve shows $\langle p \rangle \propto l_{\rm f}^{1/2}$ 
variation. The red squares, blue triangles and orange stars correspond 
to 0.6, 1.2 and 5\,GHz in the presence of Faraday rotation, and they 
are plotted with slight offset to avoid overlap. All errors show the 
standard deviation of pixels in the fractional polarization maps.
\textbf{Right:} Variation of the integral scales with 
$l_{\rm f}$. The black, orange and blue points are for the magnetic 
field, total intensity and polarized intensity at 5\,GHz, respectively.
The lines shows $L_{\rm int} \propto l_{\rm f}$.}
\label{fig:depol_forcing}
\end{figure*}

In order to understand how $\pmean_{\rm int}$ changes with the driving scale
$l_{\rm f}$ due to different number of magnetic correlation scales within the
volume, we computed the fractional polarization by turning off the effects of
Faraday rotation when generating the synthetic observations.  Note that, in the
absence of Faraday rotation, $p$ is independent of $\nu$.  In
Table~\ref{tab:depolscales}, we list the frequency-independent $\pmean_{\rm
int}$ and its dispersion computed from the maps of $p_{\rm int}$ for different
$\lf$. The values of $\pmean_{\rm int} = \langle {\rm PI}/I\rangle$ computed
from the maps of polarized intensity and the total synchrotron intensity are
also depicted by black points in the left-hand panel of
Figure~\ref{fig:depol_forcing}. In the same figure, we also show, for comparison,
the expected $\pmean_{\rm int}$, computed as $\pmean = p_{\rm max}/\sqrt{N_{\rm
mag}}$ by using $L_{{\rm int},M}$ from Table~\ref{tab:integscales}, as the grey
points.  The variation of $\pmean_{\rm int}$ with $l_{\rm f}$, computed
directly from the synthetic maps and from expectation, remarkably follows the
$\pmean \propto l_{\rm f}^{1/2}$ relation, shown as the blue curve in
\mbox{Figure~\ref{fig:depol_forcing} (left)}. Such a relation is expected for Gaussian
random magnetic fields, and finding this relation for intermittent magnetic
fields generated by the fluctuation dynamo indicates that the polarization
vector undergo random walk similar to Gaussian random fields when the LOS is
integrated over several magnetic integral scales.  To our knowledge, this is
the first time we confirm such a relation for fluctuation dynamo-generated
intermittent magnetic fields.

\begin{table}
	\centering
\setlength{\tabcolsep}{5.0 pt}
\caption{Values of $\pmean$ and $\sigma_p$ obtained from the synthetic maps 
at 5, 1.2 and 0.6\,GHz for the three forcing scales $\lf=256, 102.4$ and $64\kpc$. 
The `intrinsic' values refers to the frequency-independent values, equivalent at 
$\lambda=0$, obtained by turning off the effects of Faraday rotation. $p_0$ and 
$l_{1/2}$ are the best-fit values obtained by fitting the variation of $\pmean$ 
with smoothing scale modelled by Equation~(\ref{eq:depol}).} 
\tabcolsep=0.475cm
\begin{tabular}{ *{5}{c}}
    \toprule
\textbf{Forcing Scale} & \textbf{Frequency} & \boldmath{$\bra{p}, \sigma_p$} & \boldmath{$p_0$} & \boldmath{$l_{1/2}$}  \\
\textbf{(kpc)}    & \textbf{(GHz)} &    &  & \textbf{(kpc)} \\ \midrule
256  & Intrinsic & $0.35,0.13$ & $0.356\pm0.004$ & $197.0\pm14.9$ \\
& 5.0 & $0.34,0.13$ & $0.352\pm0.003$ &   $155.1\pm7.8$     \\
& 1.2 & $0.20,0.10$ & $0.20\pm0.01$ &   $9.3\pm1.0$     \\
& 0.6 & $0.11, 0.07$ & $0.14\pm0.04$ & $2.1\pm0.6$ \\
&&&&\\
102.4 & Intrinsic & $0.20,0.10$ & $0.22\pm0.01$ & $36.2\pm4.0$ \\
& 5.0 & $0.20,0.10$ & $0.22\pm0.01$ &   $33.3\pm2.9$     \\
& 1.2 & $0.16,0.08$ & $0.15\pm0.03$ &   $4.7\pm1.2$     \\
& 0.6 & $0.10, 0.06$ & $0.20\pm0.11$ & $0.8\pm0.5$ \\
&&&&\\
64 & Intrinsic & $0.17, 0.09$ & $0.20\pm0.01$  &  $22.8\pm2.4$ \\
& 5.0 & $0.17,0.09$ & $0.19\pm0.01$ &   $21.5\pm1.9$     \\
& 1.2 & $0.14,0.07$ & $0.17\pm0.04$ &   $2.5\pm0.6$     \\
& 0.6 & $0.09, 0.05$ & $0.80\pm0.80$ & $0.12\pm0.12$ \\
\bottomrule
\end{tabular}
\label{tab:depolscales}
\end{table}

The fact that  $\pmean_{\rm int}$ varies as $l_{\rm f}^{1/2}$, indicates a 
possible linear relationship between $l_{\rm f}$ and $L_{{\rm int}, M}$. 
This is indeed seen in our simulations and is shown in the right-hand panel 
of Figure~\ref{fig:depol_forcing}. Here, we plot the variation of the integral 
scales of magnetic field (in black) and synchrotron intensity (in orange), 
$L_{{\rm int},M}$ and $L_{{\rm int}, I}$, respectively, as a function of 
$l_{\rm f}$. This linear dependence of $L_{{\rm int},M}$ on $l_{\rm f}$ 
suggests that $\pmean_{\rm int}$ of the ICM is a direct indicator of the 
forcing and/or magnetic integral scale. 

In the following, we investigate how frequency-dependent Faraday depolarization
affect the variation of $\pmean$ with $\lf$. Details of numerical
computations of the frequency-dependent Faraday depolarization is presented in
\citet{basu19b} and \citet{sur21}. In the left-hand panel of
Figure~\ref{fig:depol_forcing}, we also show variation of $\pmean$ with $\lf$
determined from the synthetic maps in the presence of Faraday rotation at 5,
1.2 and 0.6\,GHz with orange stars, blue triangles and red squares,
respectively. The values of $\pmean$ and its pixel-wise dispersion $\sigma_p$
obtained at the native resolution of 1\,kpc at these frequencies are listed in
Table~\ref{tab:depolscales}. It is immediately evident that, due to relatively
low Faraday depolarization at 5\,GHz, the variation of $\pmean_{\rm 5\,GHz}$
with $\lf$ matches excellently with that of $\pmean_{\rm int}$ and dropping by
a factor of two as $l_{\rm f}$ decreases from $256$ to $64\kpc$. This
emphasizes the fact that measurement of polarized emission from the ICM at
frequencies $\gtrsim$4~\ghz, where Faraday rotation is low, can be directly used
to gain insights into the nature of turbulence driving in the ICM of galaxy
clusters.  At lower frequencies, $\pmean_{\rm 1.2\ghz}$ and $\pmean_{\rm
0.6\ghz}$ change mildly with $\lf$ (see Table~\ref{tab:depolscales}), deviating
significantly from the $\lf^{1/2}$ dependence. Furthermore, as
mentioned earlier, the dispersion of Faraday depth, $\sigma_{\rm FD}$, for all
the three forcing scales are similar with $\sigma_{\rm FD} \approx 100\radm$.
That means, small-scale structures introduced due to strong Faraday
depolarization at frequencies $\lesssim$3~\ghz~see \citep{sur21} makes
$\pmean$ insensitive to $L_{{\rm int},M}$. Therefore, even if
polarized emission from the diffuse ICM at frequencies $\lesssim$3 \ghz~are
detected, they will be unsuitable to glean any meaningful insight into the
magnetic field properties.

\subsection{Smoothing of Polarization Parameters} \label{sec:smoothPol}

\begin{figure*}[ht]
	\centering
\begin{tabular}{ccc}
{\large 5\,GHz} & {\large 1.2\,GHz} & {\large 0.6\,GHz} \\
{\mbox{\includegraphics[width=5cm]{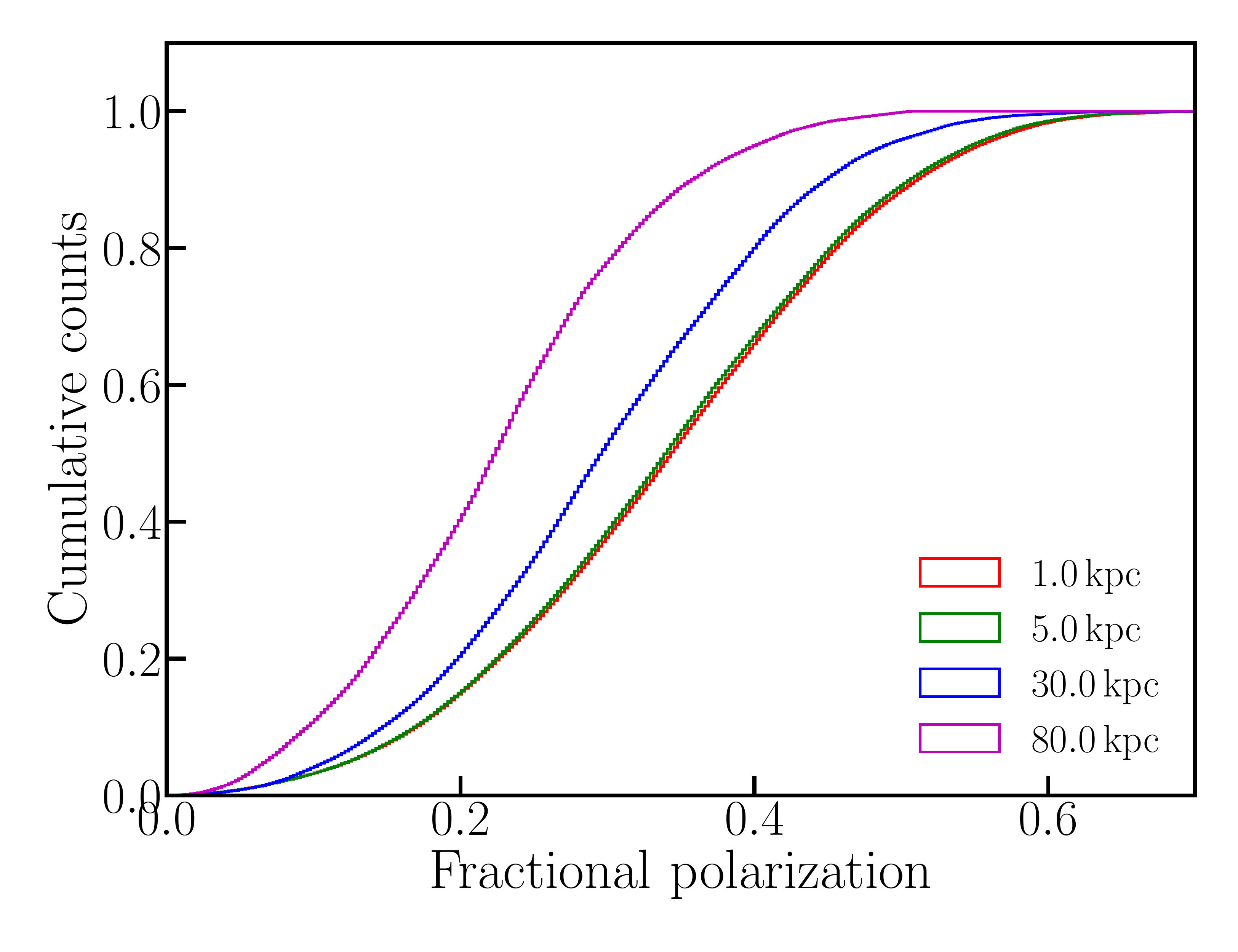}}}&
{\mbox{\includegraphics[width=5cm]{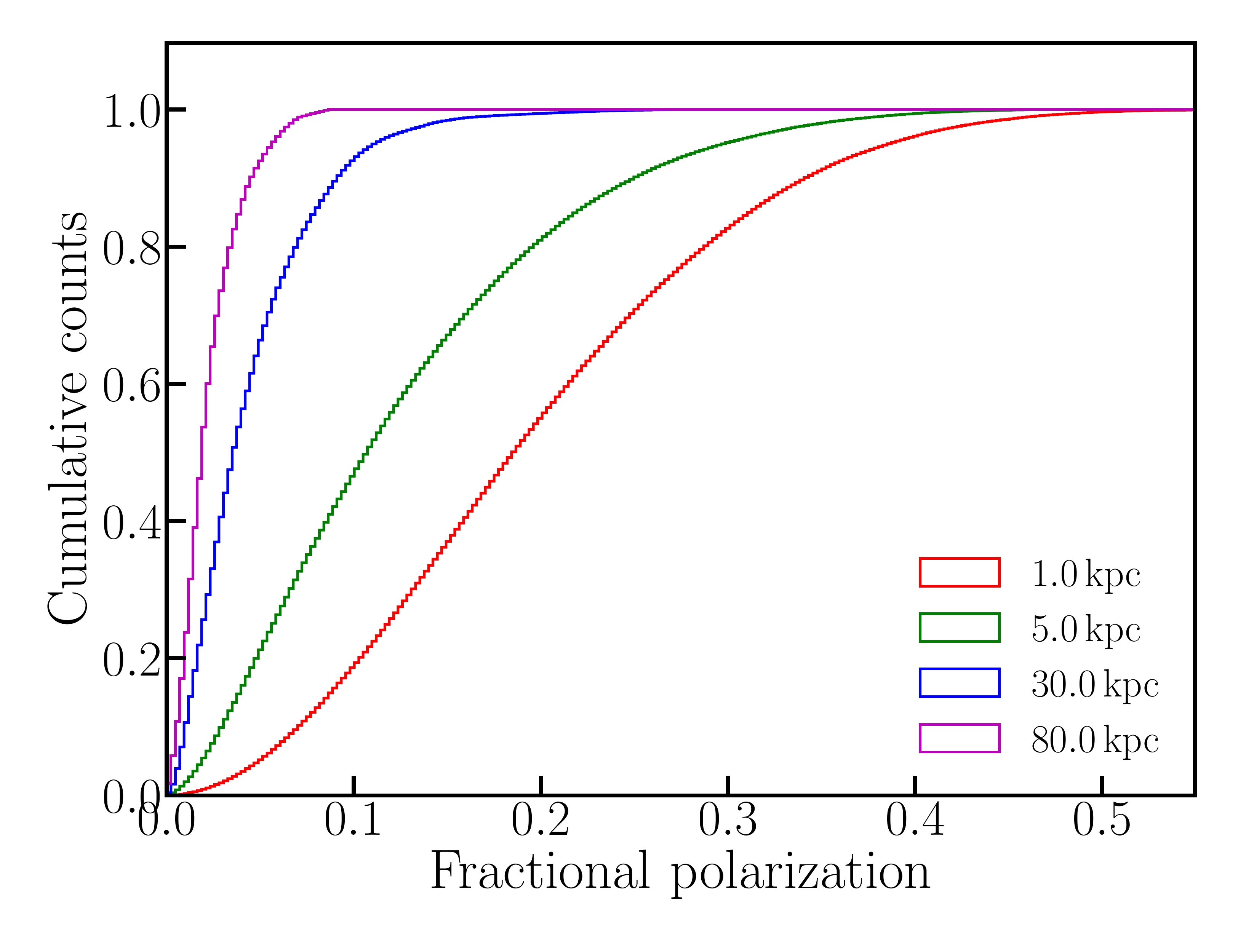}}}&
{\mbox{\includegraphics[width=5cm]{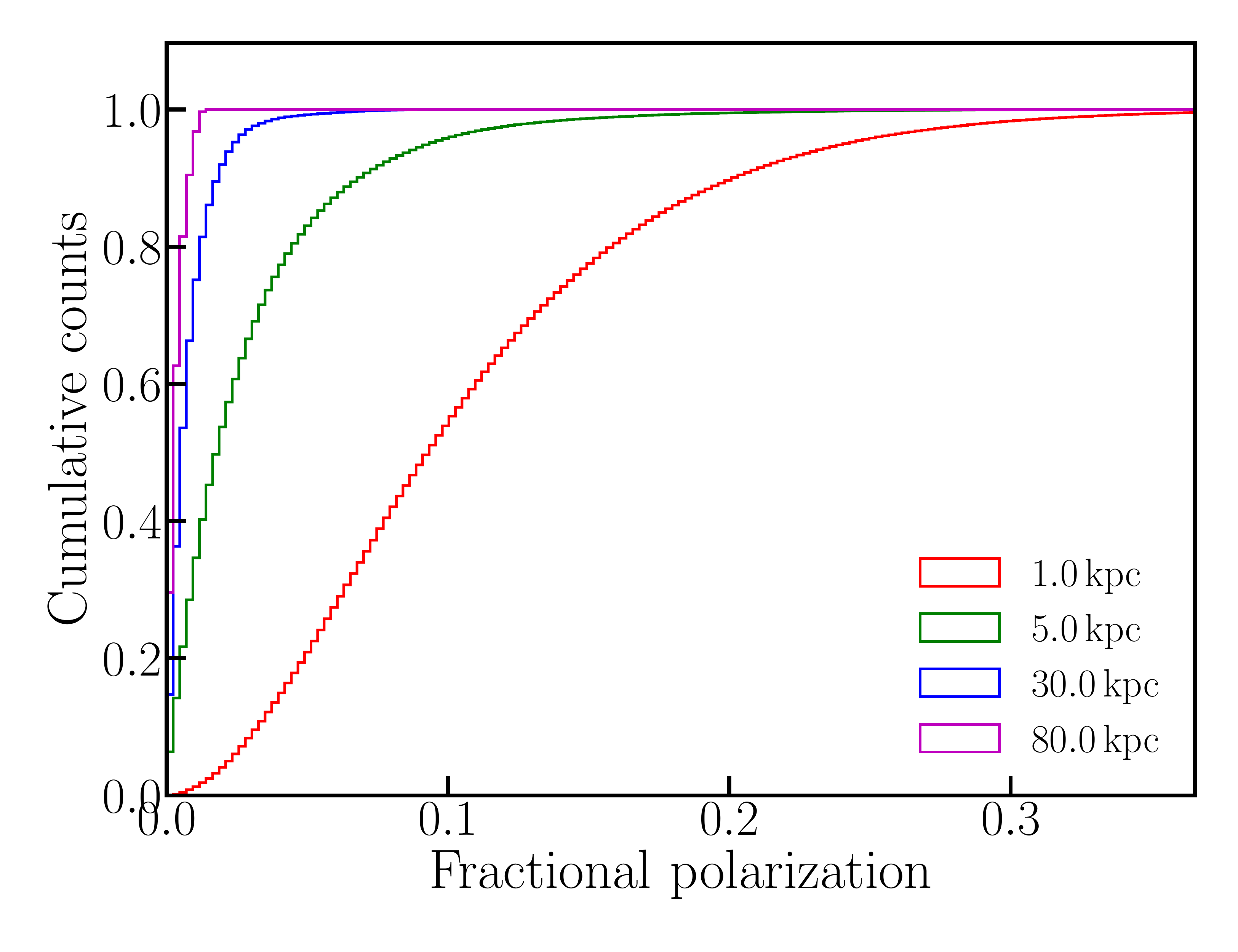}}}\\
{\mbox{\includegraphics[width=5.3cm]{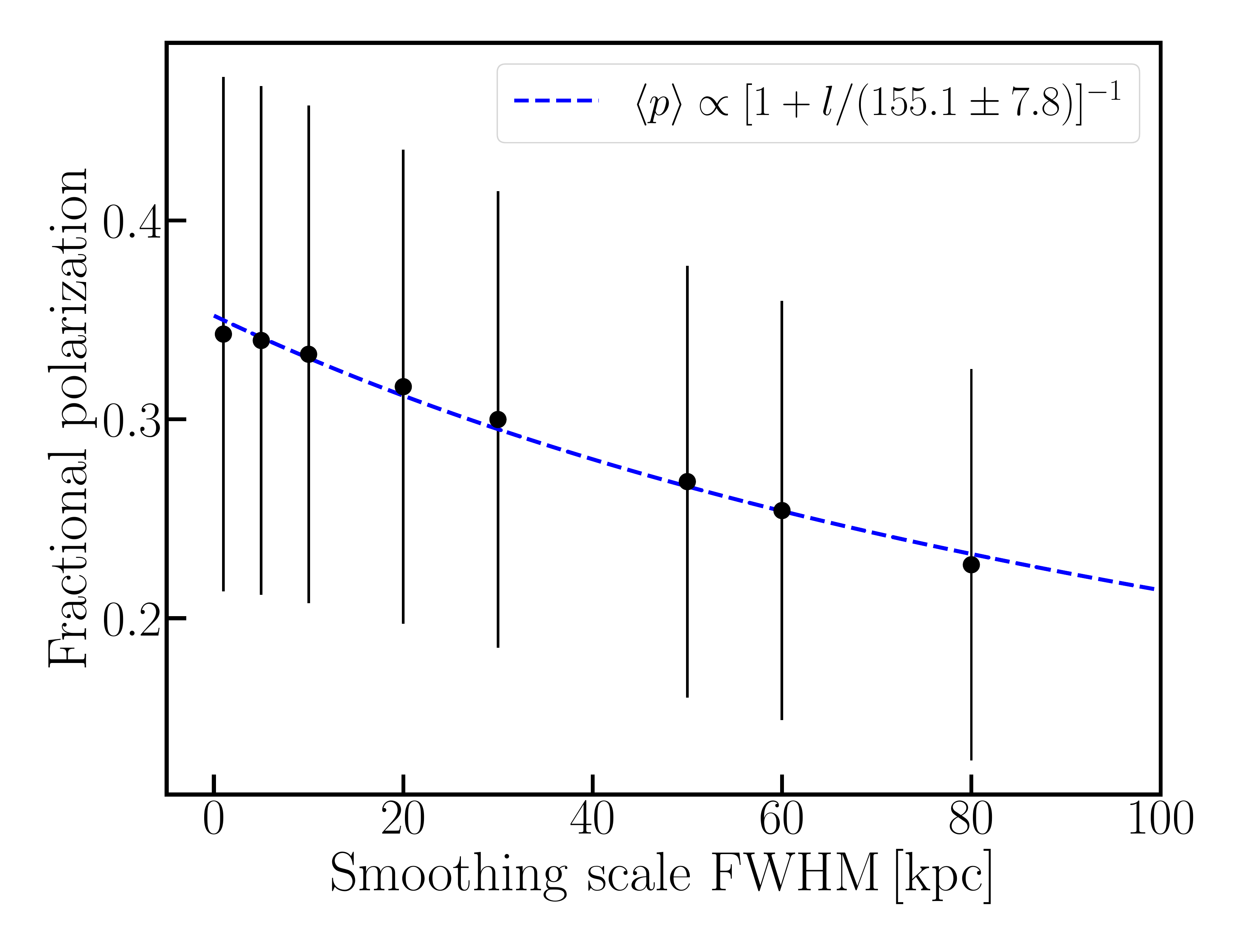}}}&
{\mbox{\includegraphics[width=5.3cm]{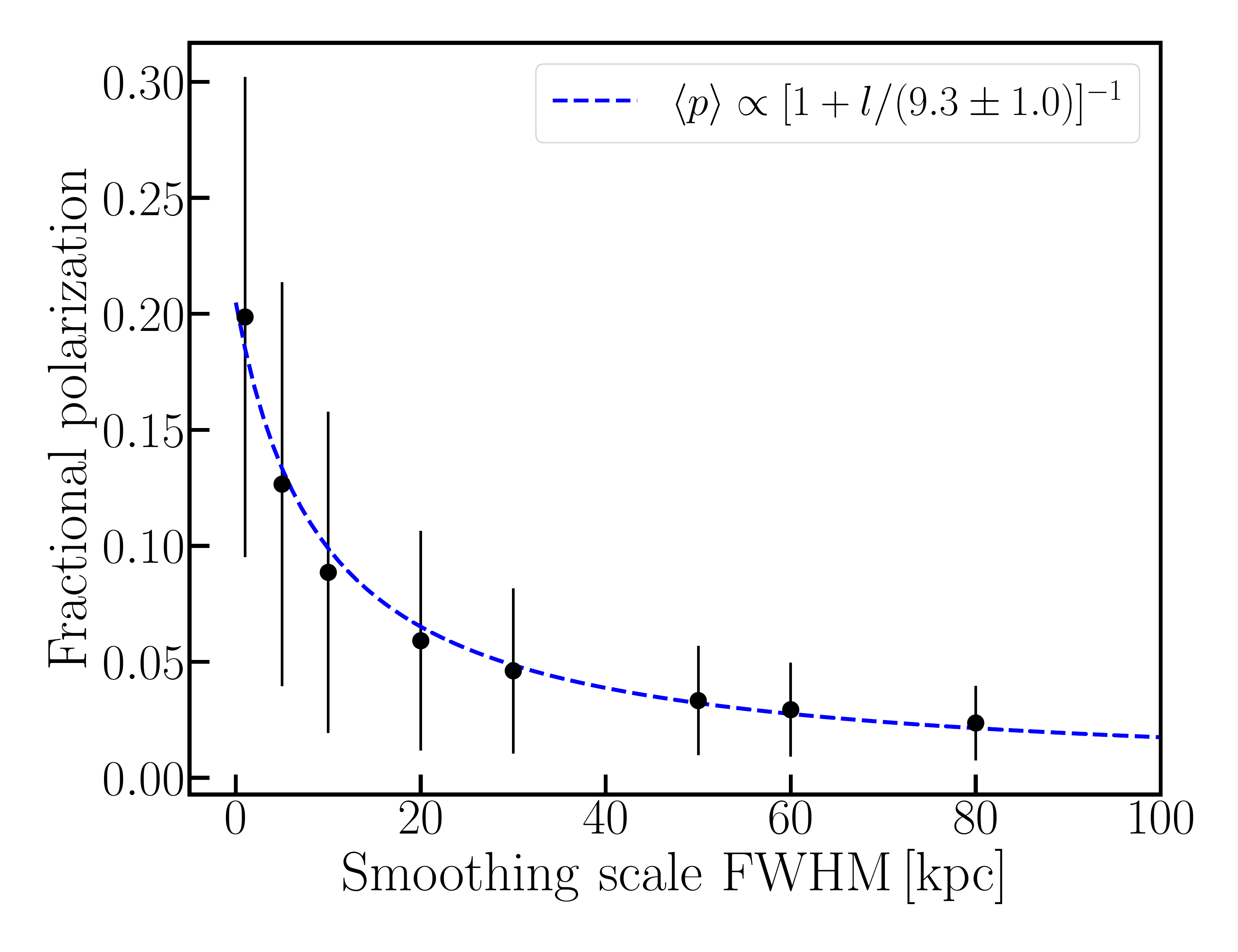}}}&
{\mbox{\includegraphics[width=5.3cm]{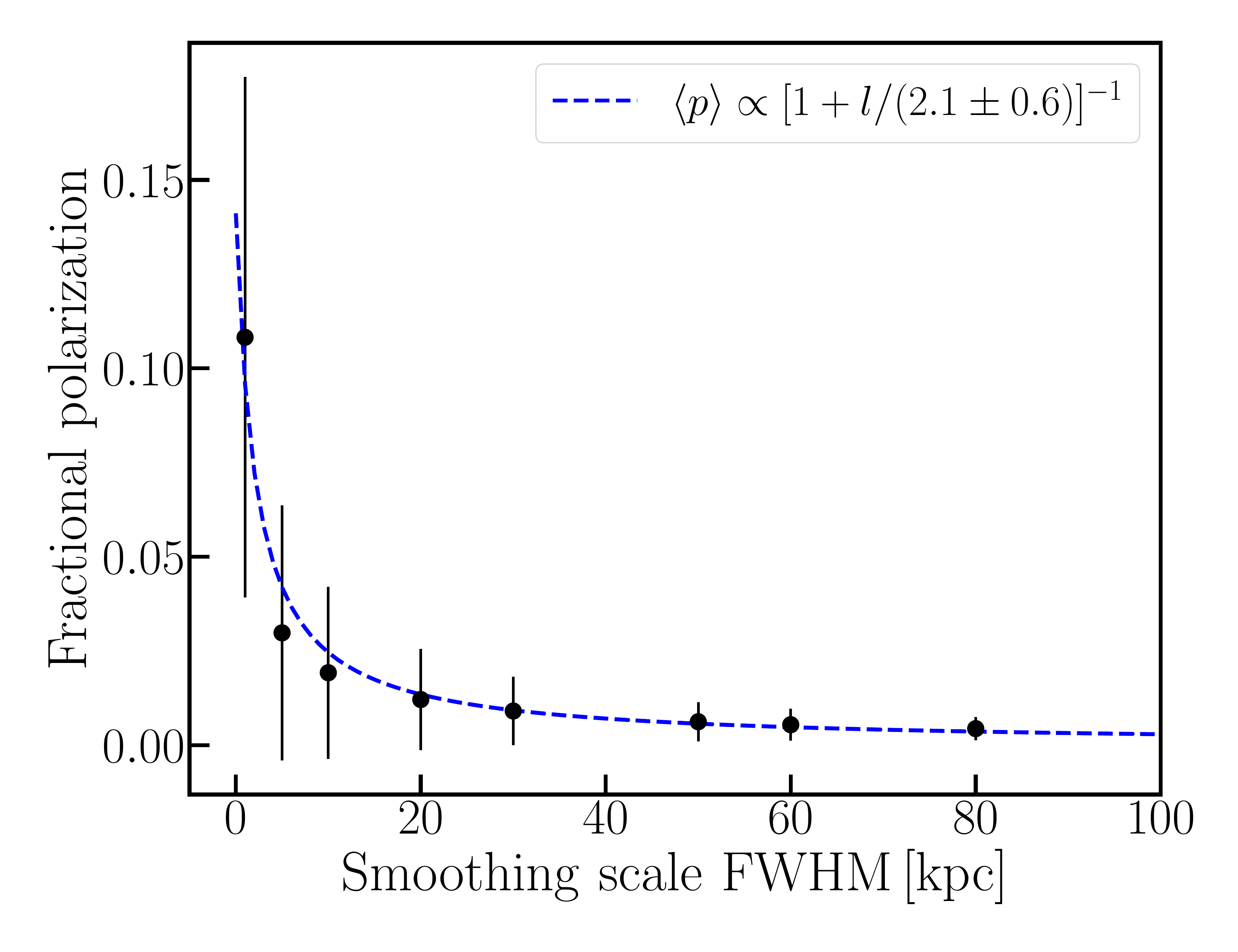}}}\\
\end{tabular}
\caption{Effect of smoothing by different scales on the fractional polarization 
($p$) obtained from $\lf = 256$\,kpc at $t/t_{\rm ed}=23$. 
\textbf{Top panel:} Cumulative distribution of $p$ for smoothing by 
different scales. \textbf{Bottom panel:} Variation of the mean fractional 
polarization $\langle p\rangle$ as a function of the smoothing scale ($l$). 
The dashed blue curve show the best-fit to the points using Equation~(\ref{eq:depol}). 
\textbf{Left, middle and right rows} are for synthetic observations at 
5, 1.2 and 0.6~GHz, respectively.}
\label{fig:convlk2}
\end{figure*}

In the previous subsection, we demonstrated that the observed $\pmean$ of the
ICM measured at frequencies $\gtrsim$4~\ghz~directly provides an estimate of
the turbulent forcing scale when observations are performed with spatial
resolution comparable to or higher than the 1\,kpc resolution of the
simulations. However, achieving such resolutions is challenging with currently
available radio telescopes. Here, we consider the statistical properties of
$\pmean$ measured at different frequencies smoothed on various scales in the
presence of frequency-dependent Faraday depolarization.  In the top-panel of
Figure~\ref{fig:convlk2}, we show the pixel-wise empirical cumulative
distribution function (CDF) of $p$ smoothed on various scales at 5, 1.2 and
0.6\,GHz for $\lf=256\kpc$. The CDFs for $\lf=102.4$ and $64\kpc$ are shown in
\mbox{Figures~\ref{fig:convlk5} and \ref{fig:convlk8}}, respectively.  From the top-row
of Figure~\ref{fig:convlk2} it is evident that the CDFs shift towards the left
with increasing smoothing scale, indicating $\pmean_\nu$ decreases with
smoothing on larger scales due to a combination of beam and Faraday
depolarization. At 5\,GHz, $\pmean_{5\ghz}$ decreases by about 30\% from 0.35
at the native resolution to 0.25 when smoothed on 80\,kpc scales for
$\lf=256\kpc$. However, as a consequence of stronger Faraday depolarization at
lower frequencies, the decrease in $\pmean_\nu$ with smoothing scale is
stronger, wherein $\pmean_{1.2\ghz}$ decreases from 0.2 at native resolution to
significantly below 0.05 for smoothing on 80\,kpc, and at 0.6\,GHz,
$\pmean_{0.6\ghz}$ decreases from 0.11 to about 0.01.  For lower $\lf$ of 102.4
and 64\,kpc, in general we find that, due to larger $N_{\rm mag}$ (see
Secion \ref{sec:intpol}), the rate of decrease of $\pmean_\nu$ with increasing
smoothing scale is significantly larger compared to $\lf=256\kpc$ at all the
three frequencies.

We model the depolarization as a function of smoothing scale in the presence of
Faraday rotation seen in the top panel of Figure~\ref{fig:convlk2} using the
form,
\begin{equation}
\bra{p(l)} = {p_0}\,{\left(\frac{1}{1 + l/l_{1/2}} \right)}.
\label{eq:depol}
\end{equation}
Here, $l$ is the spatial smoothing scale, $p_0$ provides estimate of $\bra{p}$
at infinitesimal resolution, and $l_{1/2}$ is the smoothing scale at which the
mean fractional polarization decreases by 50\%, i.e., $\bra{p(l_{1/2})} =
p_0/2$.  In the bottom panel of Figure~\ref{fig:convlk2}, we show the variation
of $\pmean_\nu$ for $\lf=256\kpc$ as the data points, and the best-fit model
using Equation~(\ref{eq:depol}). The corresponding plots for $\lf=102.4$ and
$64\kpc$ are shown in the bottom panels of Figures~\ref{fig:convlk5} and
\ref{fig:convlk8}, respectively. The best-fit parameters, $p_0$ and $l_{1/2}$,
for all the three $\lf$ at the three representative frequencies and for
$\pmean_{\rm int}$ are presented in Table~\ref{tab:depolscales}. Firstly we
note that, for the frequency-independent $\pmean_{\rm int}$, $l_{1/2}$ reduces
drastically from $\approx$200~\kpc~for $\lf=256\kpc$ to $l_{1/2} \approx 20\kpc$
for $\lf = 64\kpc$.  This decrease in $l_{1/2}$ with $\lf$ is significantly
non-linear, in contrast to what was found in the right-hand panel of
Figure~\ref{fig:depol_forcing} for the different integral scales, and could be
qualitatively understood as a consequence of increasing variance of the random
magnetic fields within the beam for lower $\lf$ see, e.g., \citep{sokol98}.
Therefore, we believe that it is difficult to use $l_{1/2}$ as an indicator of
$\lf$ in a straightforward way. Secondly, in the presence of
frequency-dependent Faraday depolarization, for a given $\lf$, $l_{1/2}$
becomes drastically smaller at lower frequencies. For example, when turbulence
is driven on $256\kpc$, $l_{1/2}$ decreases from $155.1 \pm 7.8\kpc$ at $5\ghz$
to only $2.1 \pm 0.6\kpc$, comparable to the resolution of the simulations, at
$0.6\ghz$. Thirdly, the decrease in $l_{1/2}$ with decreasing frequency is
significantly stronger for smaller $\lf$, so that $l_{1/2}$ becomes comparable
to, or smaller than, the resolution of the simulations for $\lf \lesssim
100\kpc$ at frequencies $\lesssim$1~\ghz~(see Table~\ref{tab:depolscales}).  Our
findings imply that due to the small-scale structures introduced by the effect
of Faraday depolarization, beam depolarization completely wipes out the
intrinsic polarized structures. This reduces the level of polarization
from the ICM to insignificant levels at frequencies below $1\ghz$, even if
observations are performed with high spatial resolutions of $\sim$1--5~\kpc.  
This re-emphasizes the finding in \citet{sur21} that, high
frequency observations $\gtrsim$3~\ghz~with high spatial resolutions
$\lesssim$20~\kpc~are of paramount importance in order to detect diffuse
polarized emission from the radio halo of galaxy clusters.

\section{\label{sec:discuss}Discussion and Conclusions}

In this paper, we presented detailed investigation of the expected statistical
properties of polarized emission from the ICM of galaxy clusters for different
scales of turbulent driving of fluctuation dynamo, and the effect of beam
smoothing when observations are performed using a finite telescope resolution.
To obtain synthetic observations, covering the frequency range 0.6 to 5\,GHz,
we made use of three non-ideal MHD simulations of fluctuation dynamos with
turbulence driven on 256, 102.4 and 64\,kpc scales. The resultant rms Mach
numbers in these simulations are in the range $\mathcal{M} = 0.18 \textrm{--}
0.19$.  Thus, in the absence of any noticeable density fluctuations, these
simulations allowed us to probe the effects of magnetic fields on the
properties of the polarized emission. The simulations were performed over
$512\kpc^3$ volume with a resolution of $1\kpc$ along each axes. The 2-D maps
of various observables have spatial size of $512\times512\kpc^2$ in the plane
of the sky with each pixel separated by 1\,kpc. For the purpose of studying the
impact of a telescope beam on the level of polarization in the ICM, we have
smoothed the synthetic maps on various scales ranging between 5 and 80\,kpc. 

Fluctuation dynamo generates highly non-Gaussian, spatially intermittent
distribution of the magnetic field components giving rise to filamentary
structures in the synchrotron total intensity maps that are extended roughly on
scales of turbulence driving (see \mbox{Table~\ref{tab:integscales}}). These filaments
give rise to long tails in the surface brightness distribution of the
synchrotron emission as presented in Section~\ref{sec:totI}.  However, these
structures are significantly smeared-out when smoothed on scales
$\gtrsim$30~\kpc, especially for the smaller turbulence driving scales $\lf
\lesssim 100\kpc$ where the filamentary structures are more volume filling and
thereby results in lower intensity contrast. Our work shows that, in the presence
of realistic noise and due to the faint surface brightness of the diffuse
synchrotron emission from the ICM, such filamentary structures are difficult to
discern in current observations.

\begin{figure}[H]
	\centering
\begin{tabular}{c}
\hspace{-6pt}\includegraphics[width=0.7\columnwidth]{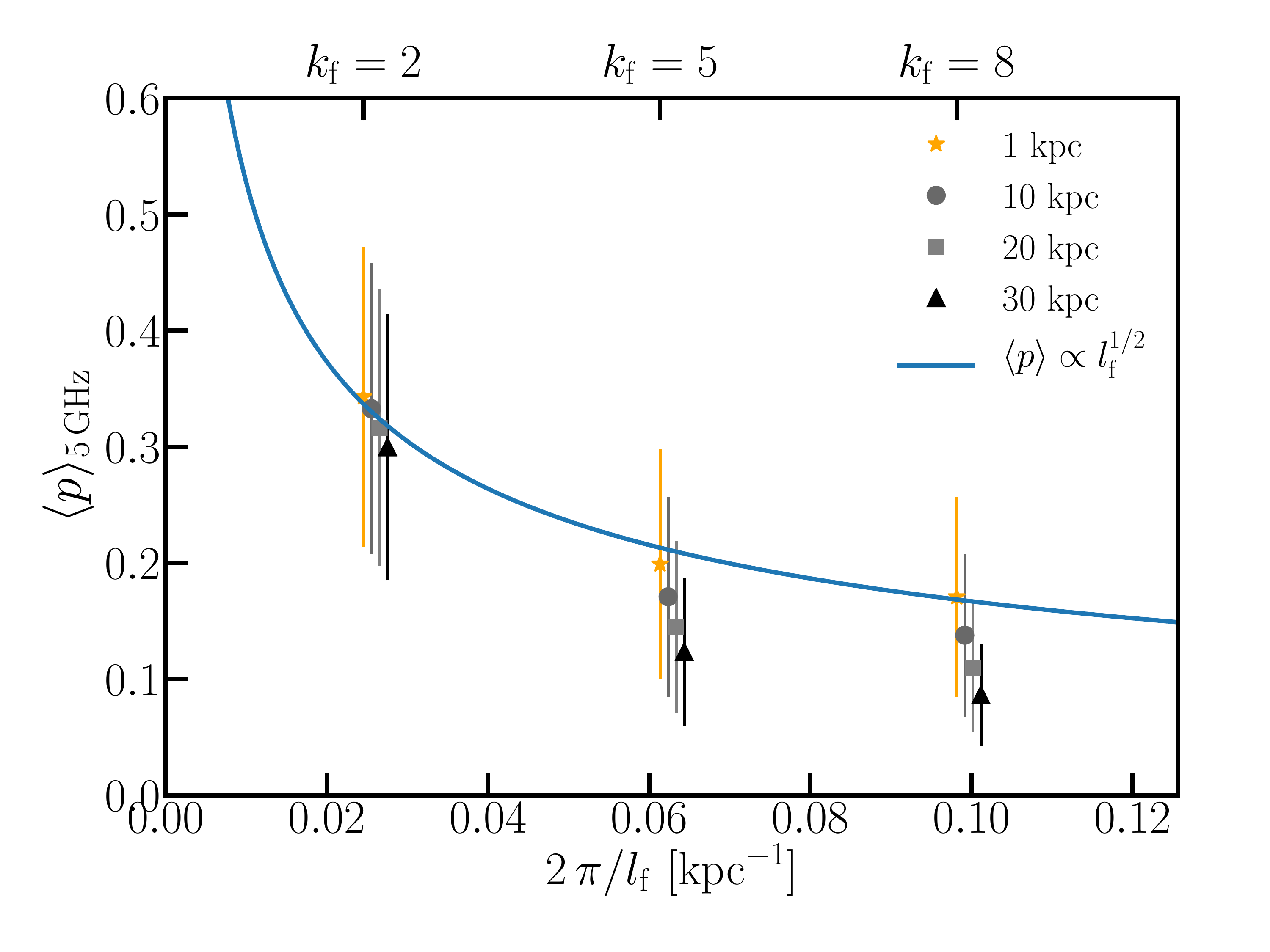}
\end{tabular}
\caption{Variation of the mean fractional polarization at 5\,GHz, 
$\pmean_{5\ghz}$, with the turbulence driving scale $l_{\rm f}$ in the 
saturated stage. The different coloured symbols are $\pmean$ obtained 
by smoothing on different scales. The blue curve shows the 
$\pmean \propto l_{\rm f}^{1/2}$ variation, same as in Figure~\ref{fig:depol_forcing} 
(left-hand panel).}
\label{fig:beamdepol_forcing}
\end{figure}

In Section~\ref{sec:intpol}, we show for the first time that in the presence of
spatially intermittent magnetic fields the mean intrinsic fractional
polarization $\pmean_{\rm int}$ at 1\,kpc resolution of the simulations varies
with the turbulence driving scale $\lf$ as $\pmean_{\rm int} \propto
\lf^{1/2}$.  This implies that $\pmean_{\rm int}$ when estimated at $\lambda=0$
using the technique of Stokes $Q,U$ fitting \citep{sulli12, sulli15} applied to
broad-bandwidth spectro-polarimetric observations of radio halos could be used
to directly infer the correlation scale of the magnetic fields in the ICM.
However, due to the extremely faint surface brightness of the polarized
emission which is expected to be of the order of a fraction of
$\mu$Jy\,arcsec$^{-2}$, applying Stokes $Q,U$ fitting would be a challenging
proposition. This limitation can be easily circumvented by performing
polarization observation at frequencies $\gtrsim$4~\ghz.  As shown in
Figure~\ref{fig:depol_forcing}, due to significantly low Faraday depolarization,
$\pmean$ can be confidently used to infer the magnetic correlation scales in
the ICM, provided observations are performed with sufficiently high spatial
resolution. This is evident from Figure~\ref{fig:beamdepol_forcing} where we
compare the variation of the mean fractional polarization at 5\,GHz,
$\pmean_{5\ghz}$, versus $l_{\rm f}$ at the native resolution of the
simulations, with those obtained by smoothing on different scales.  It is clear
from the plot that even at high frequencies ($\nu > 4\ghz$), $\pmean_{5\ghz}$
deviates significantly from $\propto l_{\rm f}^{1/2}$ relation when the maps
are smoothed to a resolution of $30\kpc$, especially for $\lf<100\kpc$.
On the other hand, at frequencies $\lesssim$2~\ghz, Faraday
depolarization gives rise to small-scale structures in the observed Stokes\,$Q$
and $U$ maps. Therefore, even if polarized emission from radio halos are
detected at these frequencies, they would be of limited use for inferring about
the intrinsic properties of the magnetic fields in the ICM.

The expected $\pmean$ in the ICM is significantly reduced due to a combination
of frequency-dependent Faraday depolarization and when observations are
smoothed by a telescope beam (see, e.g., Figure~\ref{fig:beamdepol_forcing}).  In
general, reduction of $\pmean$ due to beam smoothing is stronger for turbulence
driven on smaller scales. In Section~\ref{sec:smoothPol}, we find that at
frequencies $\lesssim$3~\ghz, $\pmean$ reduces by more than 50\% when smoothed
on about $10\kpc$ scales for turbulent driving on any of the three scales
investigated in this study. In fact, at low frequencies, smoothing the emission
from the ICM on about 30\,kpc scales reduces $\pmean$ by more than a factor of
five, making the detection of diffuse polarized emission challenging. However,
pockets of clumpy emission, extended on scales of a few beam, could be
substantially polarized up to about 10--20\% level. Detection of such features
at low frequencies would contain limited information on the nature of the
magnetic fields also see \citep{sur21}. At frequencies $\gtrsim$4~\ghz, the
diffuse polarized emission from the ICM is largely unaffected by Faraday
depolarization, and $\pmean$ is only mildly reduced for turbulence driven on
larger scales $\gtrsim$100~\kpc.  For turbulence driven on scales below
$\sim$100~\kpc, $\pmean$ of the diffuse ICM reduces to half when the emission is
smoothed on scales $>$20~\kpc. That means, the level of polarized emission from
the ICM when observed with resolutions better than 6\,arcsec at frequencies
$>$4~\ghz~for galaxy clusters located up to redshift $z=0.2$ contains valuable
information on the turbulence driving scale.

Our results on the effect of beam smoothing on the expected mean fractional
polarization $\pmean$ for different driving scale of turbulence, in combination
with the $\pmean \propto \lf^{1/2}$ relation, brings to light an important fact
that $\pmean$ measured at frequencies above $4\ghz$ can be used to glean
information on $\lf$. It is clear from Figure~\ref{fig:beamdepol_forcing} that
$\pmean_{5\ghz}$ is relatively less affected by smoothing for large $\lf$. This
implies, detection of diffuse polarized emission near $5\ghz$ at $\gtrsim$20\%
level for spatial resolutions up to $30\kpc$ would indicate $\lf \gtrsim
100\kpc$ by directly using the blue curve shown in
Figure~\ref{fig:beamdepol_forcing}, which represents the intrinsic $\pmean
\propto \lf^{1/2}$ relation.  We note that, the value of $\pmean$ in the
$\pmean \propto \lf^{1/2}$ relation also depends on the path-length, $L$ (which
is the same, 512\,kpc, in all our simulations), as $\pmean \propto
(\lf/L)^{1/2}$. This implies that, increasing the simulation domain by a factor
of 2 would also reduce the values of $\pmean$ by $\sqrt{2}$. The
simulations in this work represents a small volume covering the core region of
ICM. Since we are considering polarized emission from the ICM where synchrotron
emission and Faraday rotation are mixed, the emission along the LOS is likely
to be dominated by the core regions as compared to that in the outer parts for
$L \gg 512\kpc$.  This is because, due to the radial stratification of the ICM,
the gas densities and the magnetic field strengths decreases away from the
cluster core, and therefore, a larger ICM volume is unlikely to change our
results substantially. In addition, note that external FD contribution along the LOS from
large-scale cosmic structures in the foreground, e.g., cosmic filaments, is
comparatively smaller than the ICM \citep{akaho16,vazza14, sulli19}.  The FD in
these cosmic structures fluctuate on $\gtrsim$300~kpc~scales \citep{CR09},
comparable to size of the our simulation domain, and have $\sigma_{\rm FD}$
only $\mathcal{O}(1\radm)$ which is significantly smaller than $\sigmafd \sim
100\radm$ in the ICM of galaxy clusters.  FD fluctuations in the Galactic
foreground is also low for the typical angular extent of galaxy clusters
$<$1~degree. Therefore, external FD fluctuations in the cosmic filaments and
in the Milky Way will not affect Faraday and beam polarization presented for
ICM in our work.

As discussed above, it is important to compare $\pmean$ with $\lf$
normalized to $L$, i.e., with $k_{\rm f}$ (as shown in the top \emph{x}-axis of
Figures~\ref{fig:depol_forcing} and \ref{fig:beamdepol_forcing}), or normalized
to an equivalent length-scale. The largest scale of turbulent driving $l_{\rm
f} = 256\kpc$ in our simulations roughly correspond to the core radius ($r_{\rm
c}$) of galaxy clusters. For example, for the Coma cluster $r_{\rm c} \approx
300\kpc$ \citep{BHB92, bonaf10}. 
On the other hand, $l_{\rm f} = 64\kpc$ is of the
order of the scale height of the cluster core. In the following, we discuss
about inferring $\lf$ normalized to $r_{\rm c}$.  From
Figure~\ref{fig:beamdepol_forcing}, $\pmean_{5\ghz} \gtrsim 0.2$ directly implies
$\lf/2\,r_{\rm c} \approx 1/4\textrm{--}1/2$. On these scales, turbulence is
likely to be driven by the cascade of vortical motions generated in oblique
accretion shocks and instabilities during cluster formation on Mpc scales
\citep{NB99, SSH06, RKCD08, Xu+12, Miniati15}.  On the other hand, detection
of, or constrain on (in the case of non-detection), the polarized emission
within $r_{\rm c}$ at $\pmean_{5\ghz} \lesssim 0.05$ using a spatial resolution
up to $30\kpc$ implies $\lf \lesssim 10\kpc$, i.e., $\lf/2\,r_{\rm c} \lesssim
1/60$. On these scales, turbulence could be driven by energy input on galactic
scales, perhaps driven by gas accretion and/or star formation driven feedback
from galaxies~\mbox{\citep{donnert2009, dubois2012, pakmor2016, wiene17}}.  When the
diffuse emission in the ICM is polarized in the intermediate range with
$\pmean_{5\ghz} \sim 0.05 \textrm{--} 0.2$ for resolutions below $\sim$30~\kpc,
$\lf/2\,r_{\rm c}$ is expected to lie in the range $1/30$ to $1/6$, roughly
corresponding to $\lf$ between 20--100\,kpc.  Turbulent energy input within
such range of scales is expected to be driven by feedback from active galactic
nuclei (AGN) \citep{fabian2012, bourne2017,ehlert2021}. Therefore, $\pmean$ of
the diffuse polarized emission from the ICM above $4\ghz$ contains valuable
information on the turbulence driving scale in galaxy clusters.  We emphasize
that, the diffuse polarized emission at frequencies $\lesssim$3~GHz is expected
to be severely depolarized within the beam, but it is possible for some regions
to be locally polarized at up to $\sim$20\% level for smoothing on up to
$\sim$30~kpc scales. Detection of such clumpy polarized regions would contain
limited information on the structure of magnetic fields and on the scale of
turbulence driving in the ICM.

In light of the above discussions, we qualitatively explore the
prospect of detecting polarized emission from the radio halo of galaxy clusters
using the SKA.  The Band\,5a covering the frequency range 4.6 to 8.5\,GHz is
expected to achieve a rms noise of $1.3\,\upmu$Jy\,beam$^{-1}$ for angular
resolution in the range 0.13 to 17\,arcsec in one hour \citep{braun2019}. As
discussed above, a spatial resolution of 20--30\,kpc, and sensitivity to
emission polarized down to 0.05 level is required for broadly distinguishing
the driving scale of turbulence in the ICM by using $\pmean$ above 4\,GHz. The
median redshift of clusters detected at radio frequencies is $\sim$0.21, 
e.g., Refs. \citep{yuan2015, weere19}, and therefore angular resolution between 6
to 10\,arcsec is sufficient. Excluding radio relics and cluster minihalos,
radio halos have a median flux density of $\sim$25~mJy at 1.4\,GHz, and have
median angular extent of $\sim$6~arcmin estimated from table~1
of \citep{yuan2015}. This corresponds to surface brightness of $\sim$6 and
$2\,\upmu$Jy\,beam$^{-1}$ for a resolution of 10 and 6\,arcsec, respectively, at
the reference frequency of 6.7\,GHz in Band\,5a (assuming $\alpha = -1$). That
means, the surface brightness of the emission with fractional polarization
$>$0.05, is expected to be $\gtrsim$0.3 $\upmu$Jy\,beam$^{-1}$ which can be
achieved with $\sim$20 h of observation time with the SKA in Band\,5.
However, about 20\% of the known radio halos have flux density $\gtrsim$60 mJy
at 1.4 GHz. The polarized emission from the halos of these galaxy clusters can
be comfortably detected in Band\,5a of the SKA. We are currently investigating
in detail the prospect of detecting polarized emission from the diffuse ICM by
normalizing our MHD simulations tuned to the properties of known galaxy
clusters, and will be presented elsewhere.

Although substantial diffuse polarization at about 5\% level above 4\,GHz is
expected for smoothing the ICM emission on scales up to 30\,kpc and roughly
distinguish between the scales of turbulent energy input, $\pmean$ alone is
insufficient to distinguish the driving mechanisms. The different drivers,
i.e., galactic and AGN feedback or cluster mergers and accretion from filaments
are expected to have varying volume filling factors and possibly generate
different structural properties of the magnetic field. These differences are
expected to be imprinted on the frequency-dependent Faraday depolarization of
the polarized emission and on the properties of the Faraday depth spectrum see, e.g., 
\citep{basu19b}.  Interestingly, the role of spatially intermittent
magnetic field structures on $\pmean$ at different frequencies can already be
gleaned from the variation of $\pmean$ with $\lf$ . For a synchrotron emitting
media which is also Faraday rotating in the presence of Gaussian random fields,
$\bra{p(\lambda)}$ varies as $\bra{p(\lambda)} = \pmean_{\rm int}\,[1 -
\exp(-2\,\sigmafd^2\,\lambda^4)]/2\,\sigmafd^2\,\lambda^4$ \citep{sokol98}.
Since for all $\lf$, $\sigmafd \approx \mathcal{O}(100\radm)$, Faraday
depolarization due to Gaussian random fields should have resulted in $\pmean
\ll 0.01$ for $\nu \lesssim 3\ghz$. In contrast, due to the intermittent
magnetic field structures, substantially polarized emission are expected, as
indicated by our study. A detailed investigation of the properties of
frequency-dependent depolarization, and the nature of Faraday depth spectrum
based on the magnetic field structures generated by the action of fluctuation
dynamo driven on different scales for different $r_{\rm c}$ will form the topic
of our future work.

\vspace{6pt}
\authorcontributions{Conceptualization and methodology; software and analysis; 
writing---review and editing, A.B. and S.S. All authors have read and agreed to the published version of the manuscript.}

\funding{S.S. thanks the Science and Engineering Research Board (SERB) of the 
Department of Science \& Technology (DST), Government of India, for support through 
research grant ECR/2017/001535.}

\dataavailability{The simulation data, synthetic observations, and, the
\texttt{COSMIC} package will be made publicly available, until which they can
be shared with reasonable request to the authors.}

\acknowledgments{We thank Kandaswamy Subramanian for very insightful
discussions and encouraging us to pursue this study. We also thank
 Matthias Hoeft for helpful discussions on cluster observations and
critical comments which improved the presentation of the paper. We
thank the two anonymous referees for constructive comments. S.S. acknowledges
computing time awarded at CDAC National Param supercomputing facility, India,
under the grant `Hydromagnetic-Turbulence-PR' and the use of the High
Performance Computing (HPC) resources made available by the Computer Center of
the Indian Institute of Astrophysics. The software used in this work was in
part developed by the DOE NNSA-ASC OASCR Flash Center at the University of
Chicago. This research also made use of
Astropy,\footnote{http://www.astropy.org} a community-developed core Python
package for Astronomy \citep{astropy:2013, astropy:2018}, NumPy
\citep{numpy11}, Matplotlib \citep{matplotlib07} and Joblib.}

\conflictsofinterest{The authors declare no conflict of interest. The funders
had no role in the design of the study; in the collection, analyses, or
interpretation of data; in the writing of the manuscript, or in the decision to
publish the results.}

\appendixtitles{yes} 
\appendix

\section{Distribution of \textit{p}} \label{sec:smoothing_app}

\begin{figure}[H]
\appendix
\begin{tabular}{ccc}
{\large 5\,GHz} & {\large 1.2\,GHz} & {\large 0.6\,GHz} \\
{\mbox{\includegraphics[width=5cm]{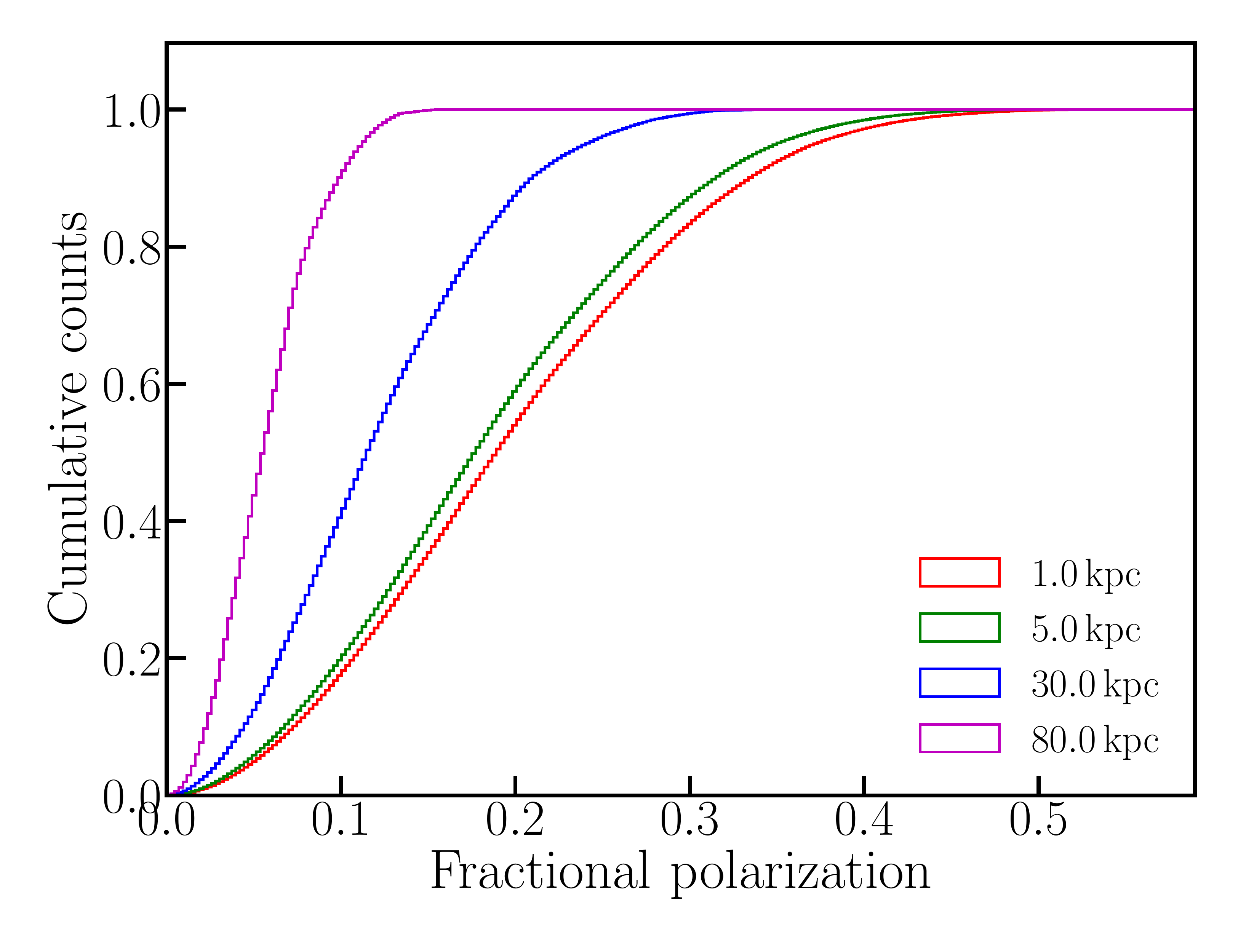}}}&
{\mbox{\includegraphics[width=5cm]{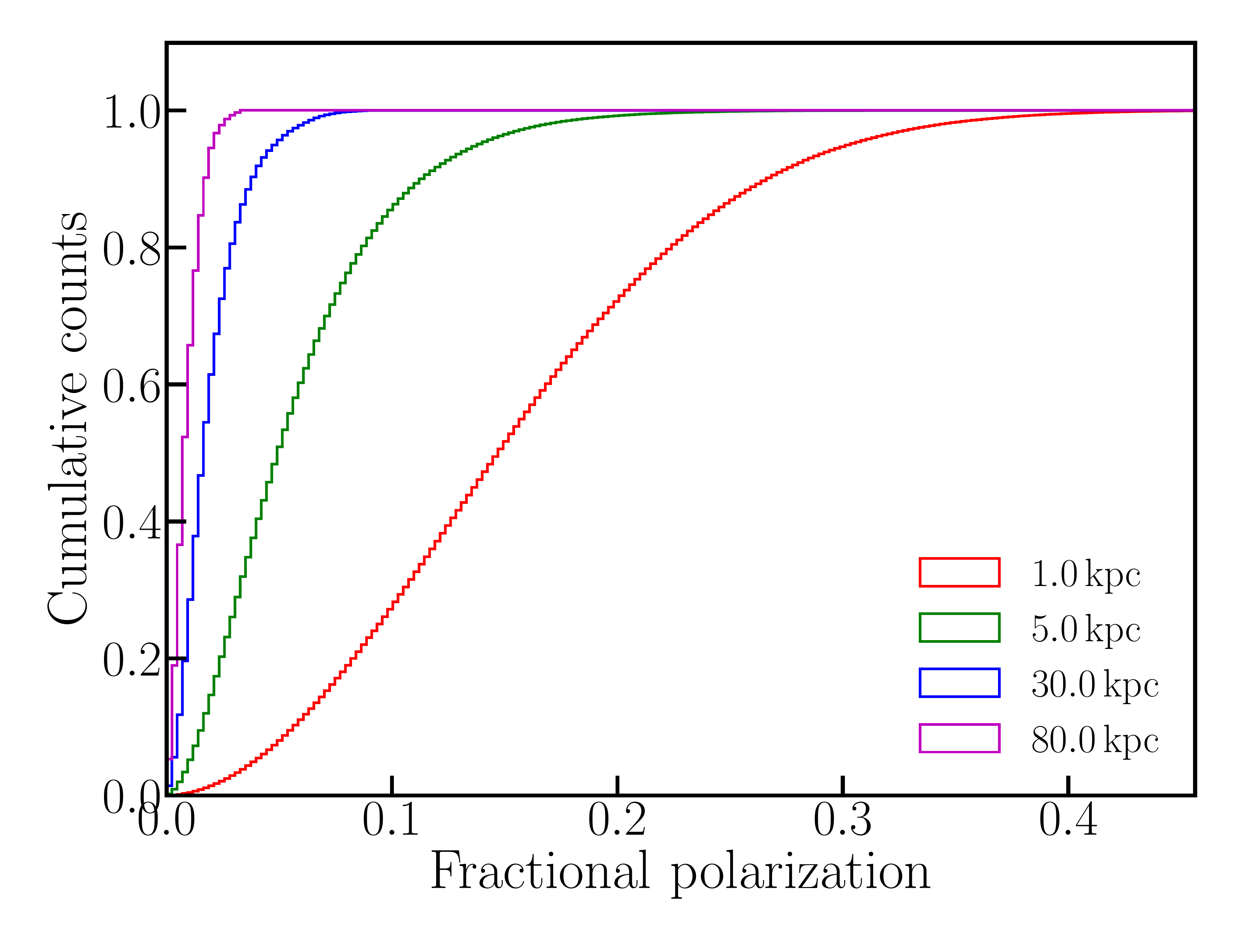}}}&
{\mbox{\includegraphics[width=5cm]{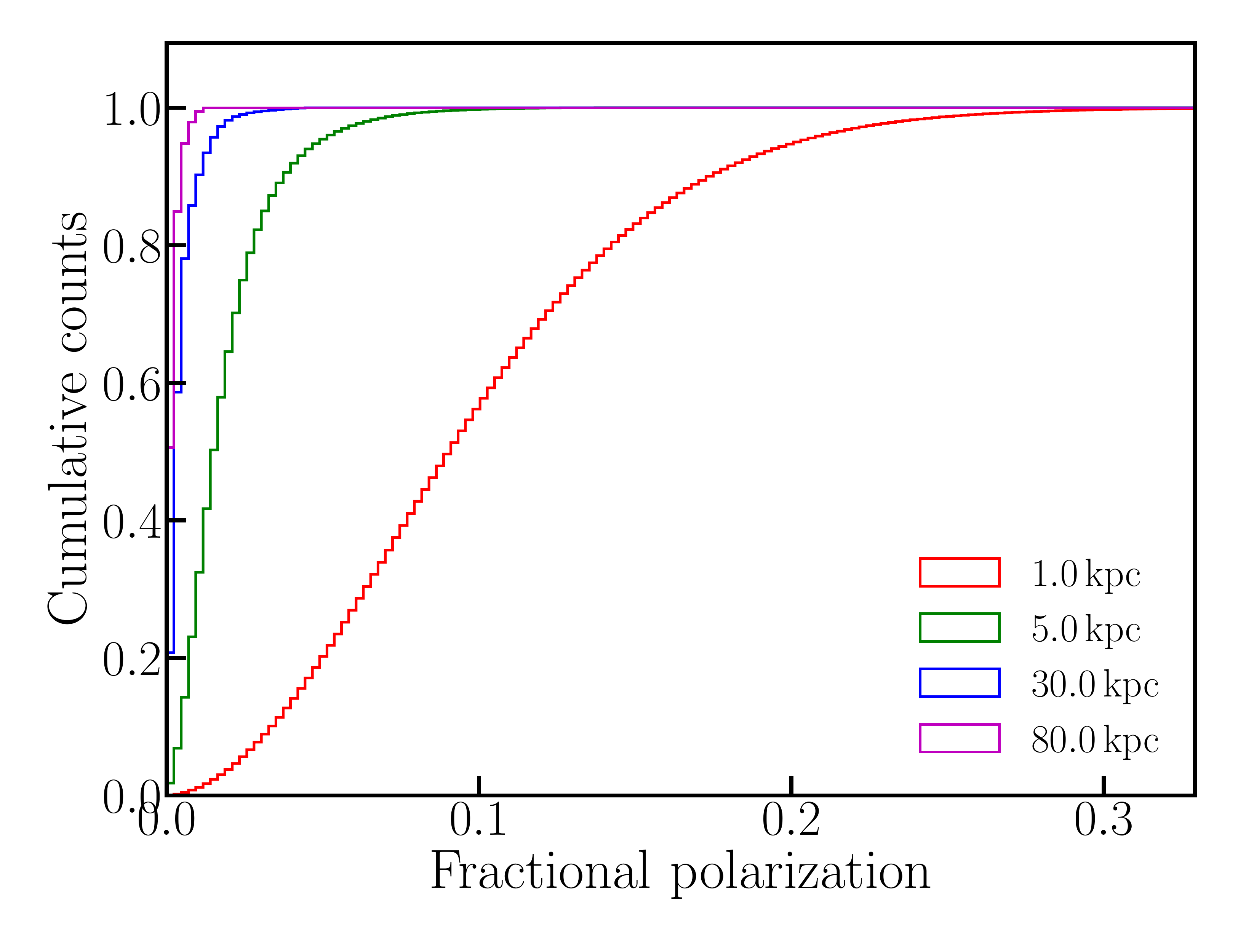}}}\\
{\mbox{\includegraphics[width=5.3cm]{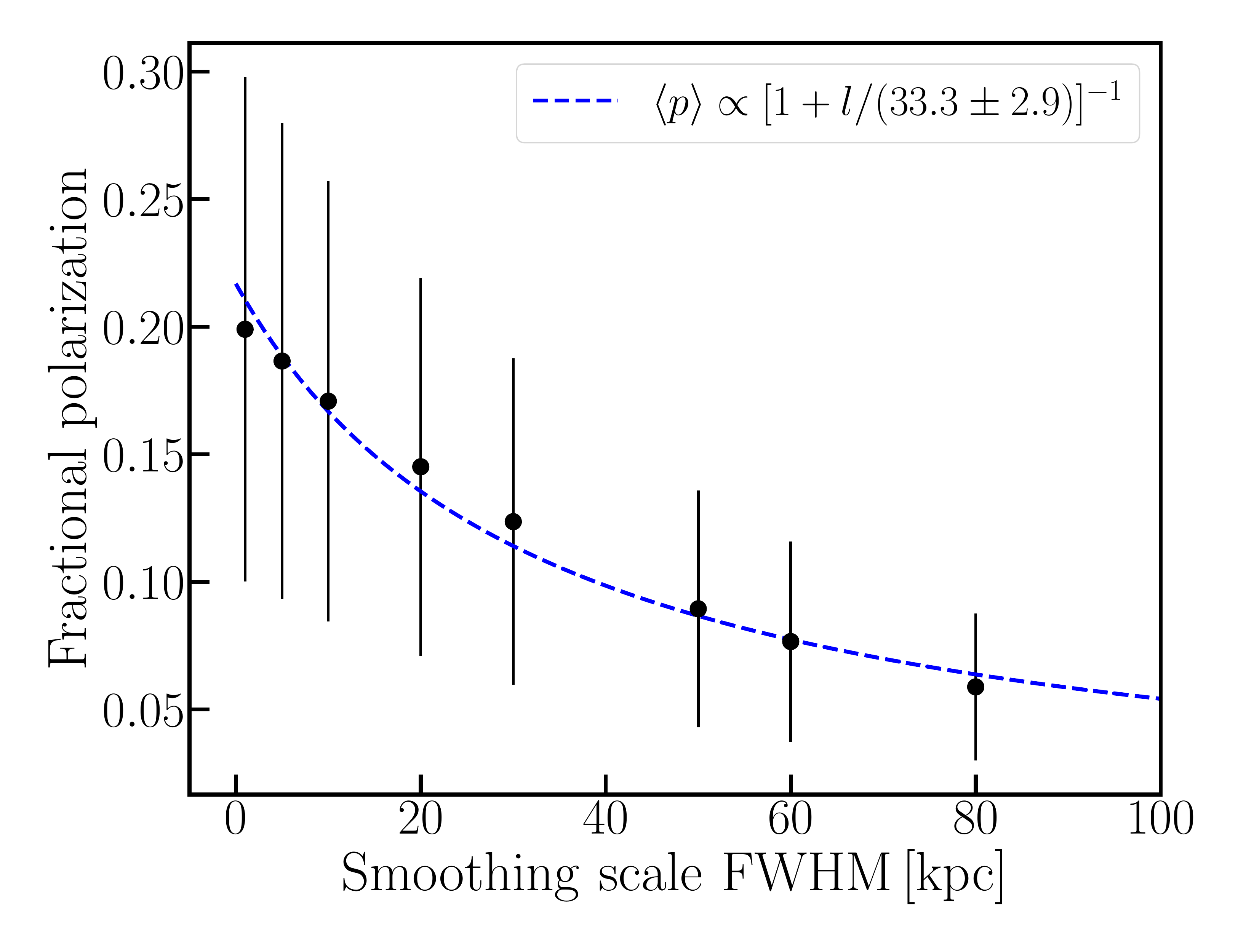}}}&
{\mbox{\includegraphics[width=5.3cm]{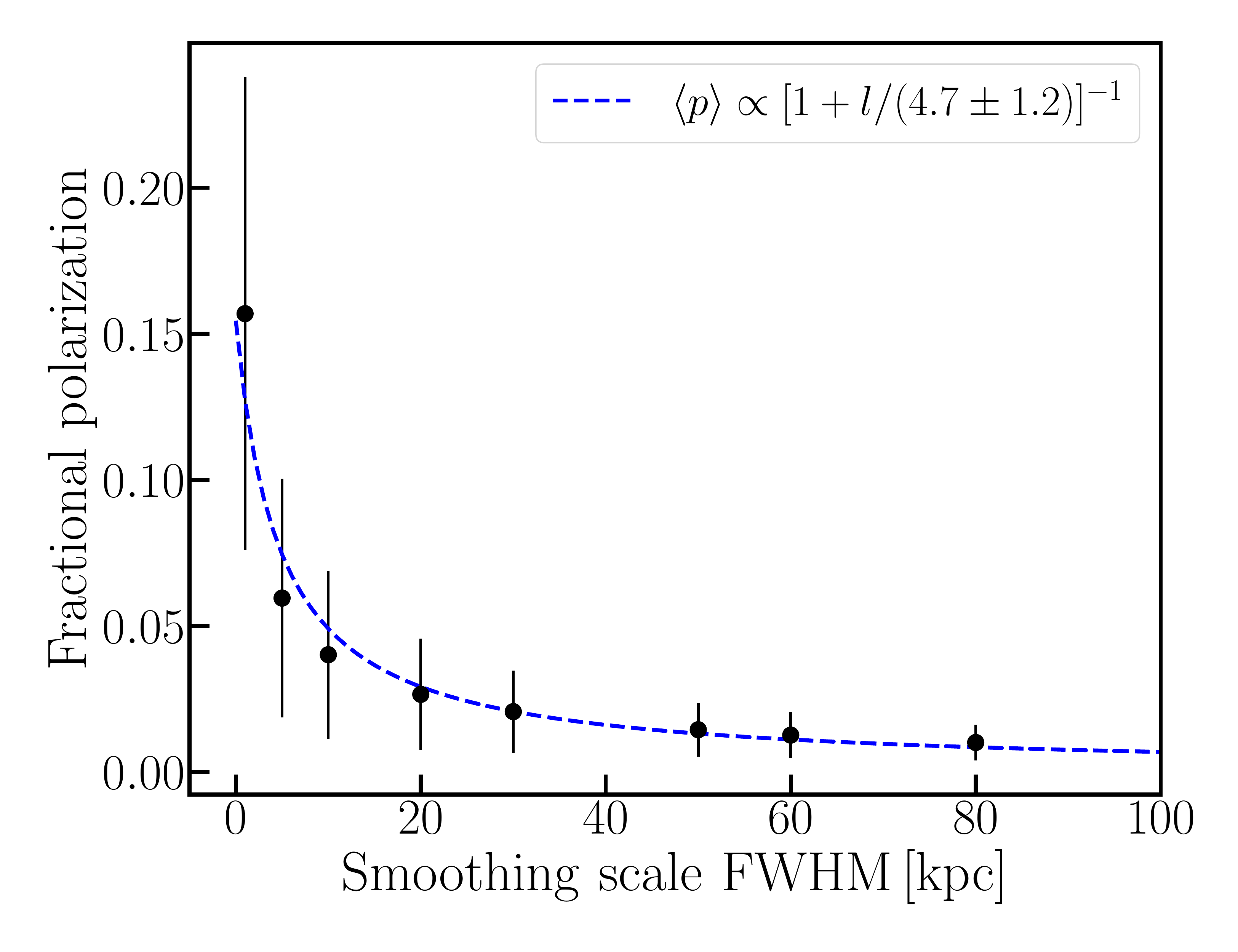}}}&
{\mbox{\includegraphics[width=5.3cm]{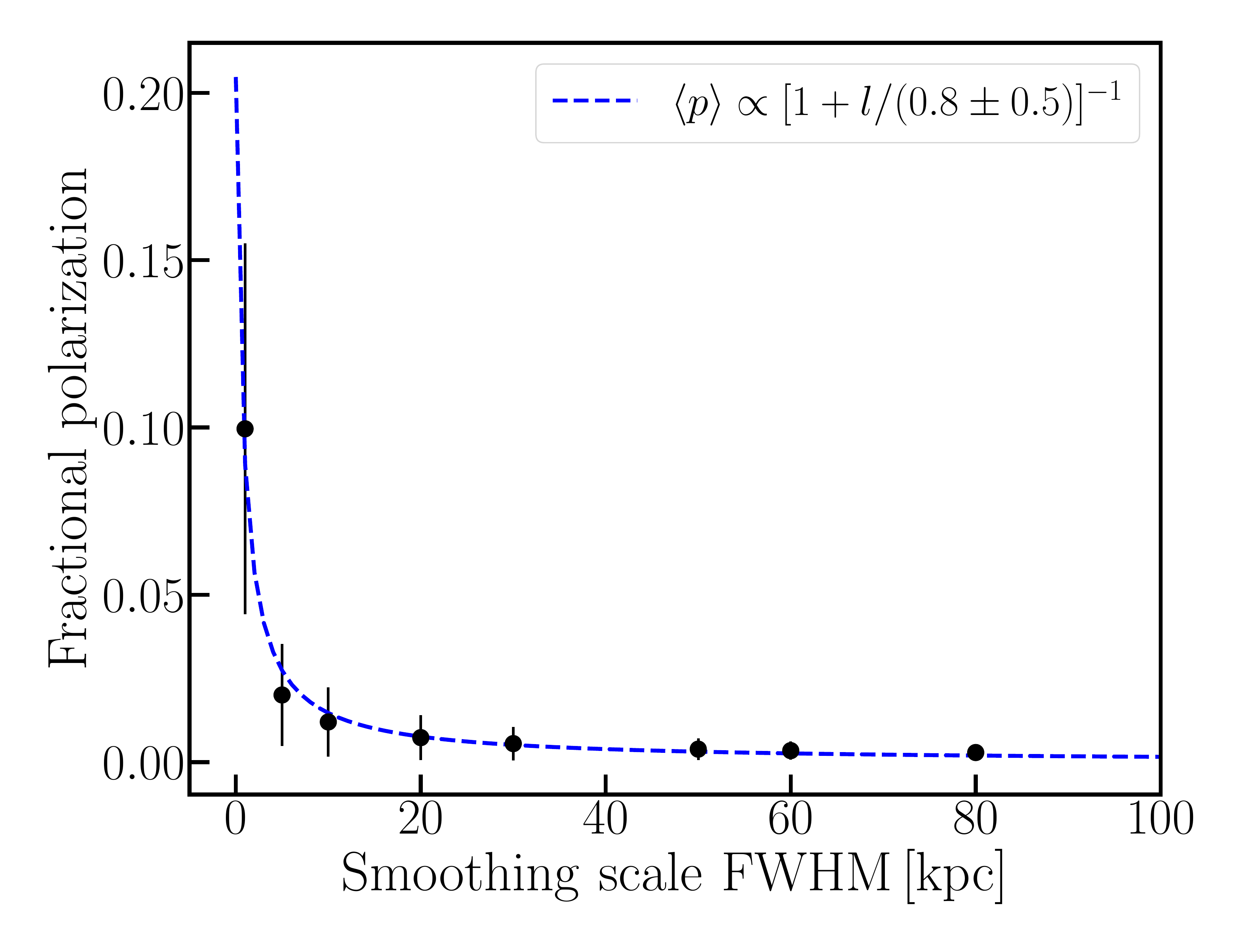}}}\\
\end{tabular}
\caption{Same as Figure~\ref{fig:convlk2} for 
$\lf = 102.4$\,kpc at $t/t_{\rm ed}= 20.2$.}
\label{fig:convlk5}
\end{figure}

\begin{figure}[H]
\appendix
\begin{tabular}{ccc}
{\large 5\,GHz} & {\large 1.2\,GHz} & {\large 0.6\,GHz} \\
{\mbox{\includegraphics[width=5cm]{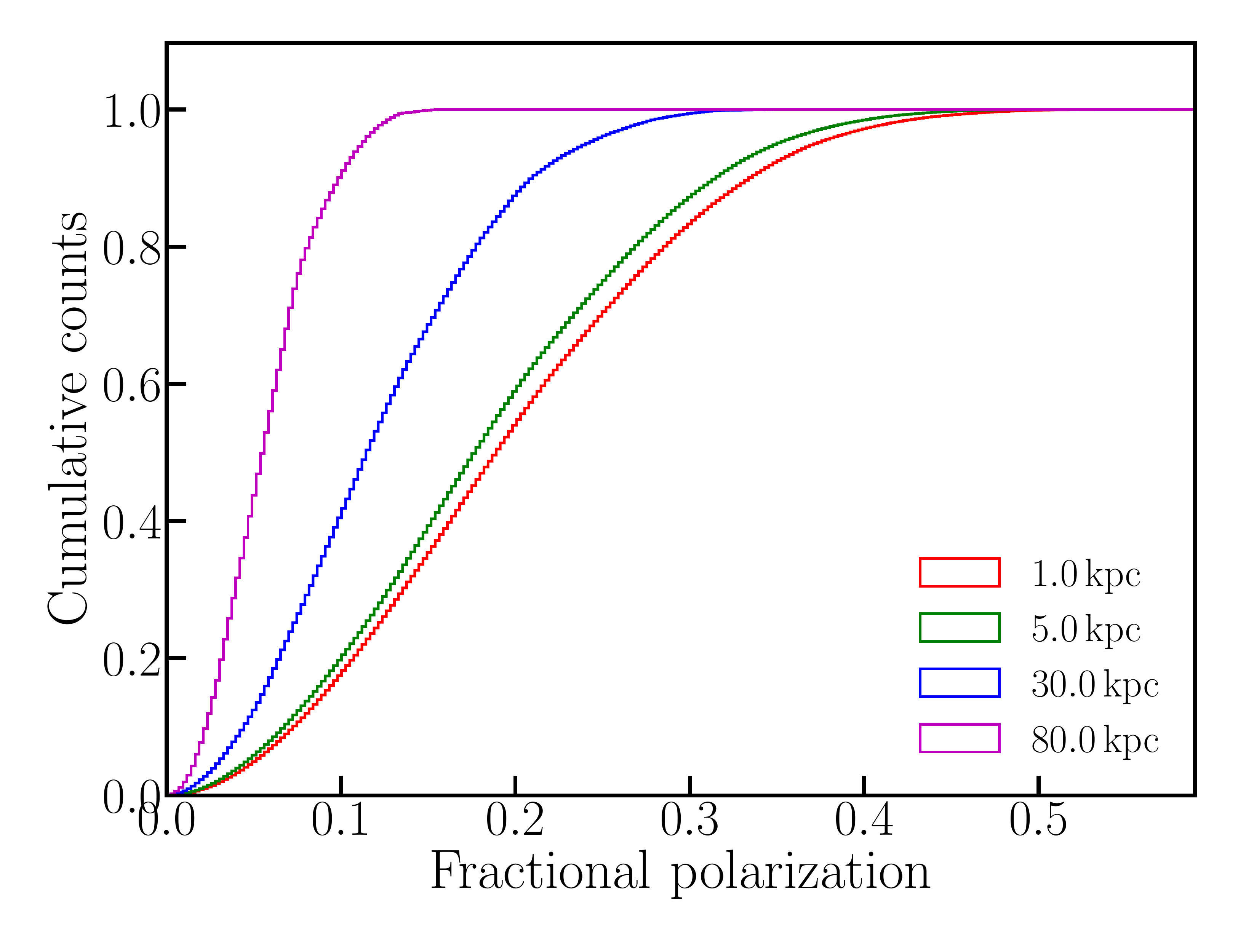}}}&
{\mbox{\includegraphics[width=5cm]{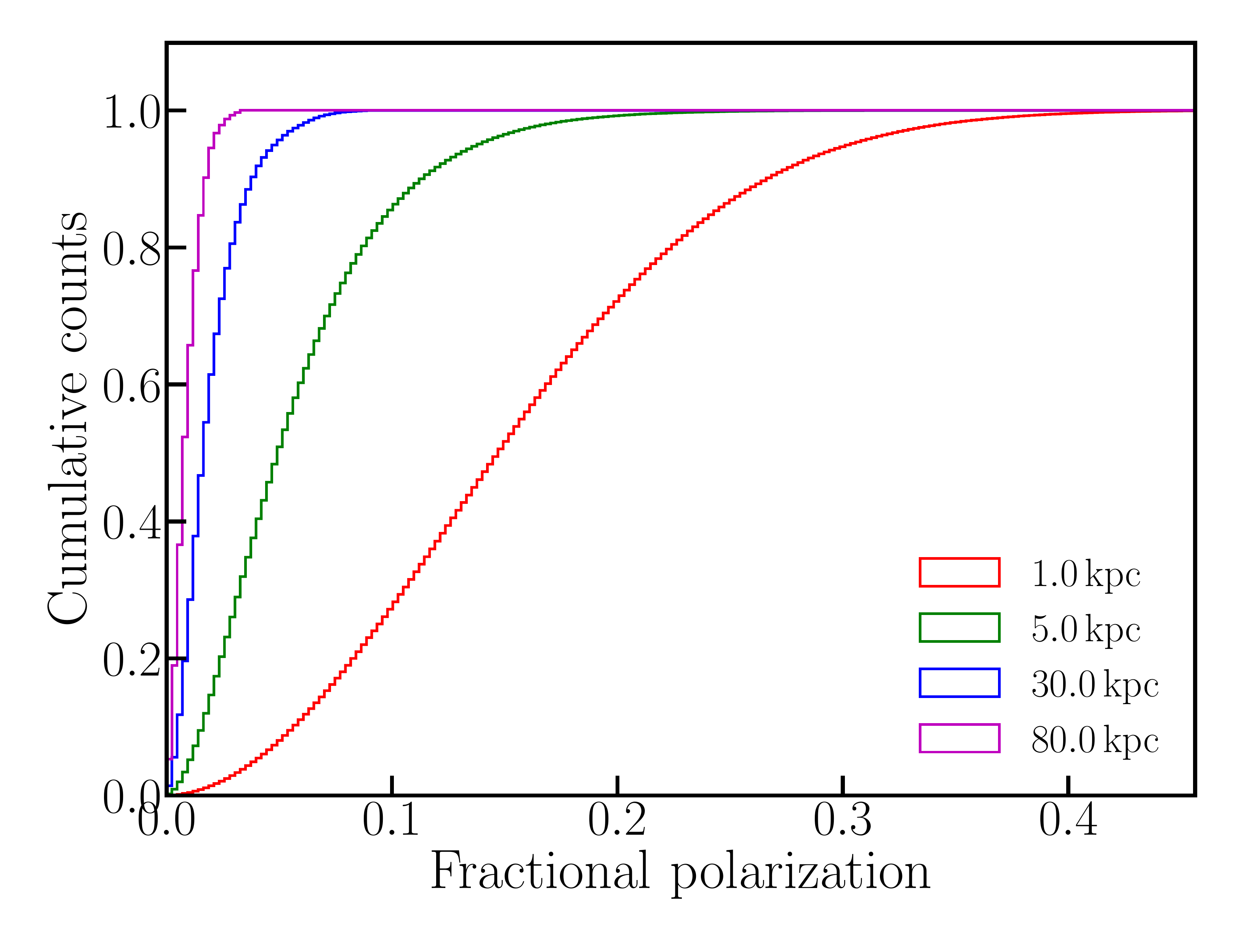}}}&
{\mbox{\includegraphics[width=5cm]{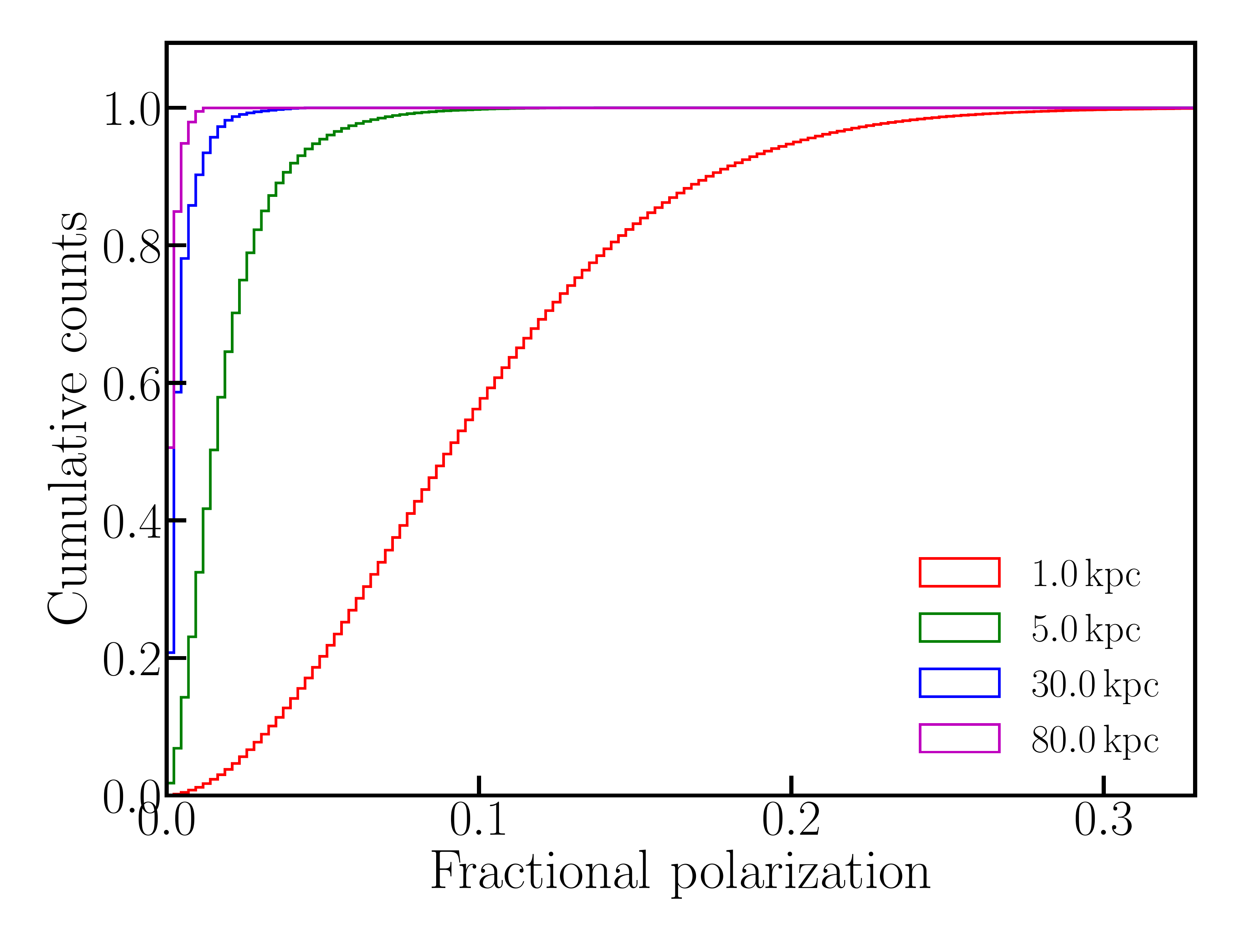}}}\\
{\mbox{\includegraphics[width=5.3cm]{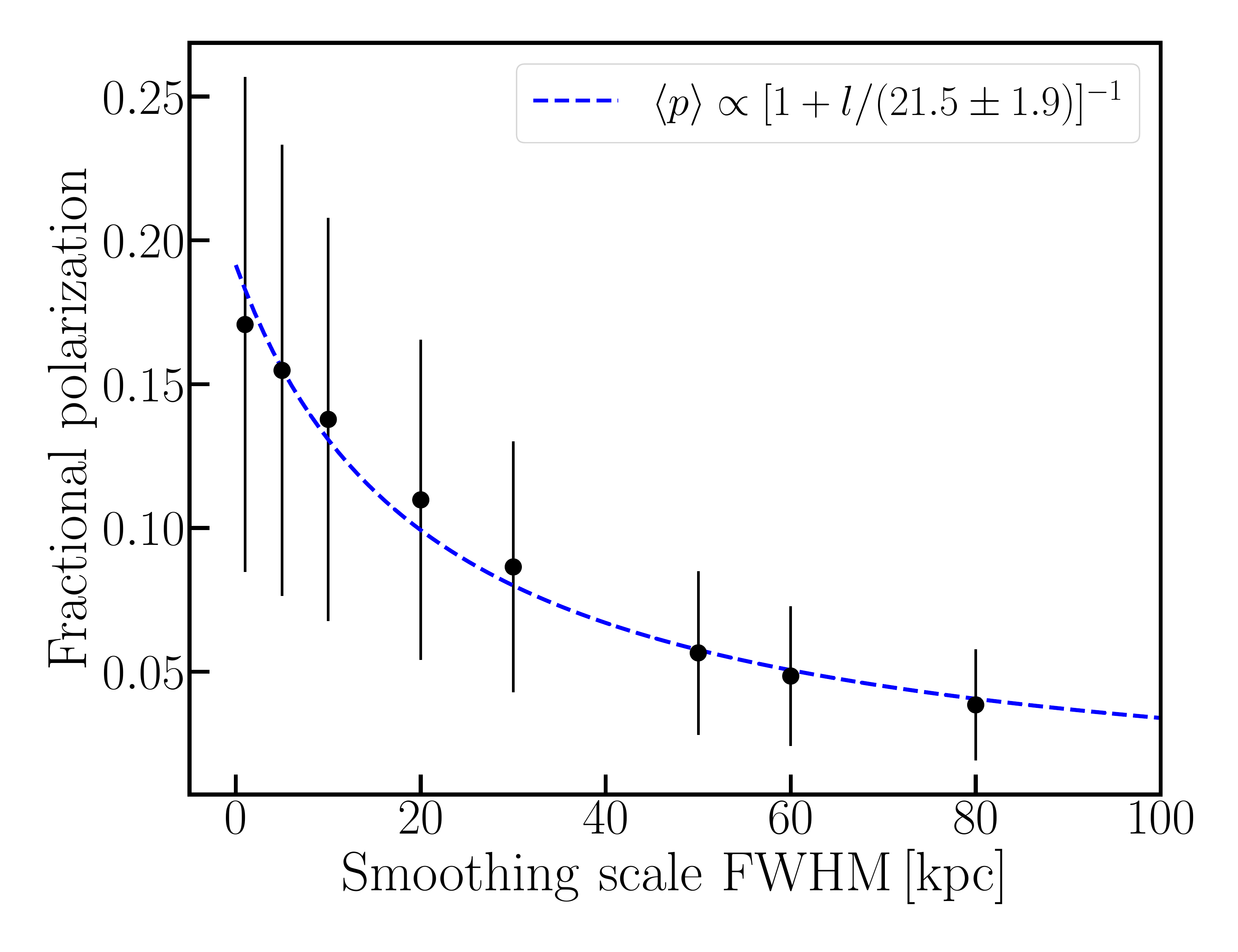}}}&
{\mbox{\includegraphics[width=5.3cm]{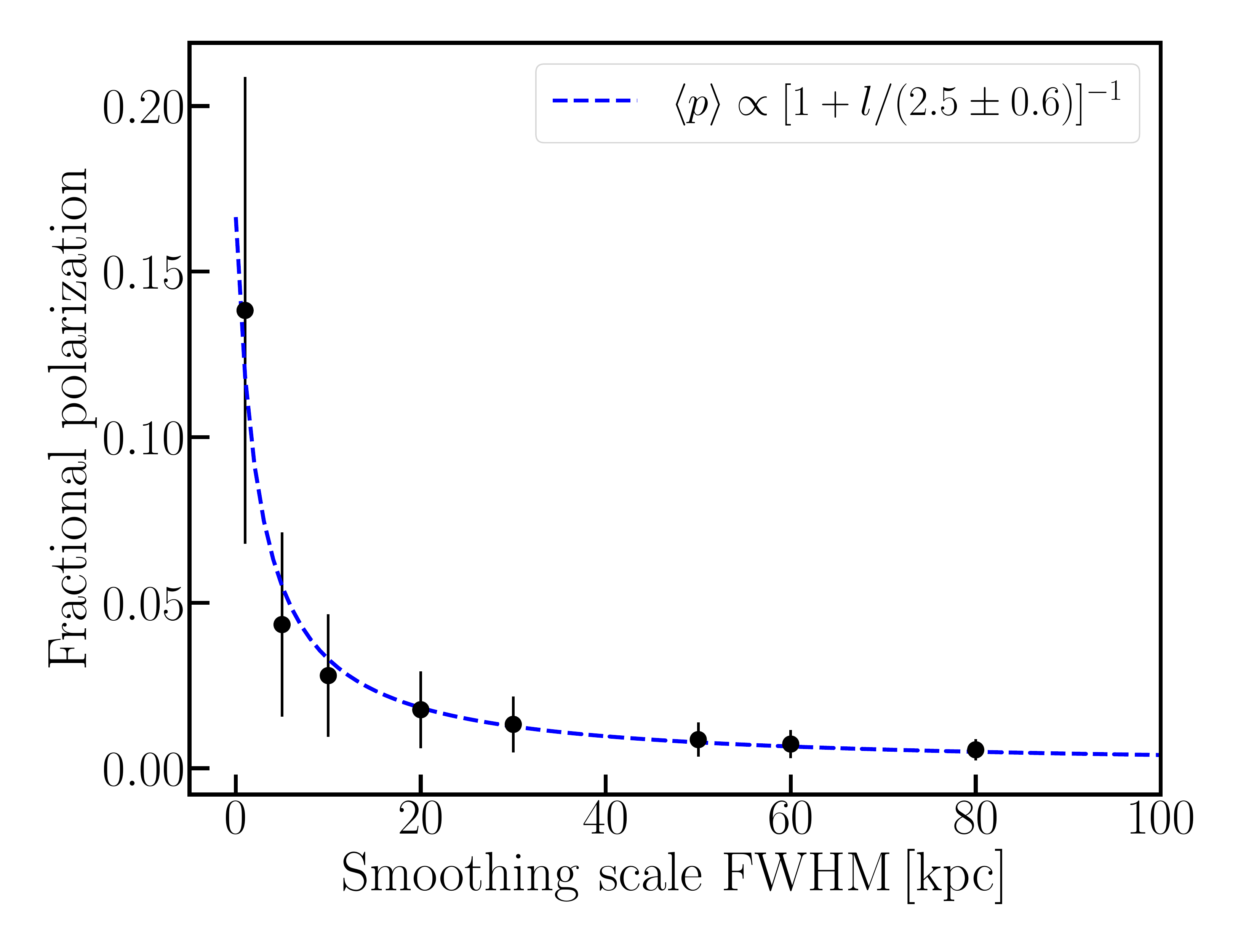}}}&
{\mbox{\includegraphics[width=5.3cm]{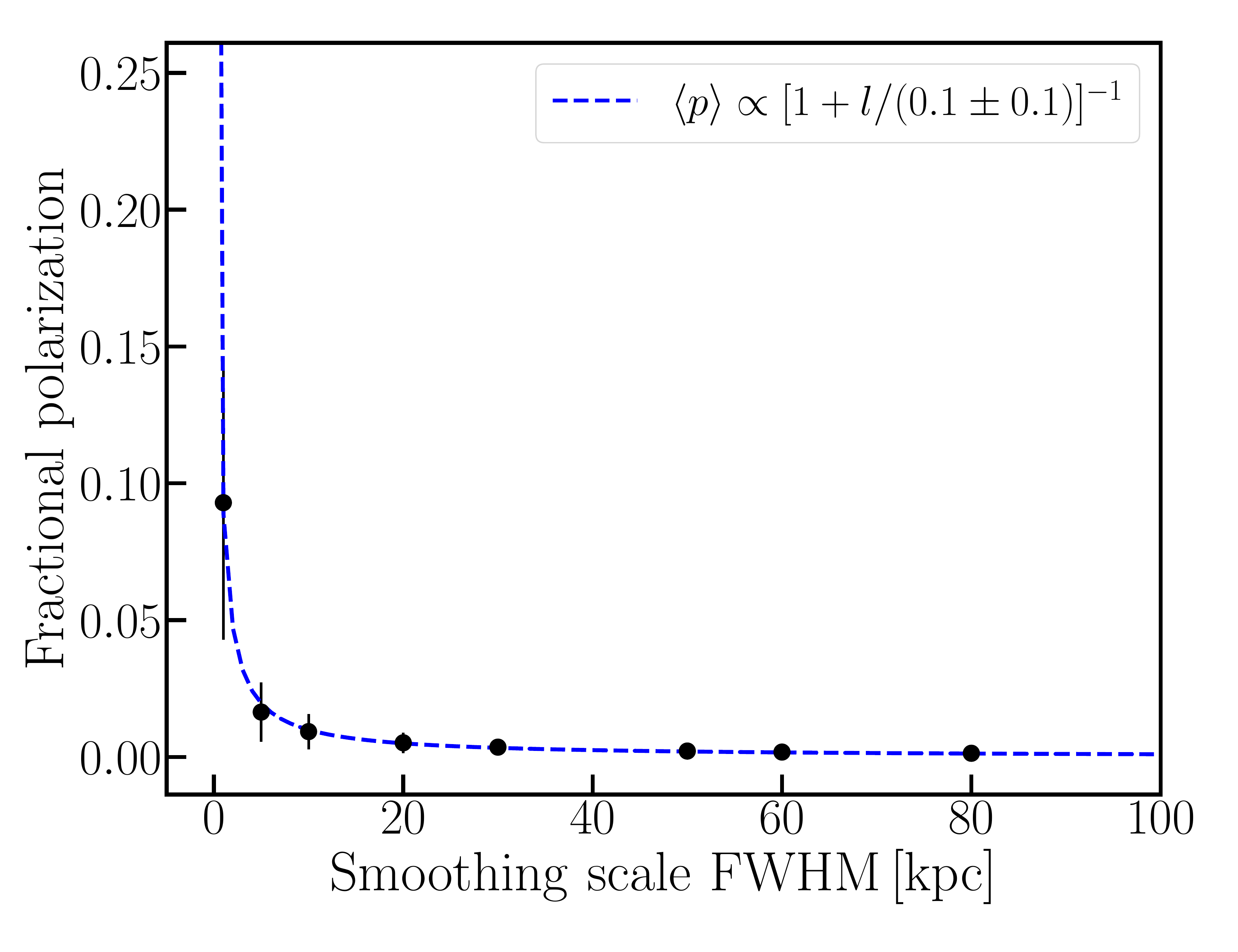}}}\\
\end{tabular}
\caption{Same as Figure~\ref{fig:convlk2} for 
$\lf = 64$\,kpc at $t/t_{\rm ed}=30.6$.}
\label{fig:convlk8}
\end{figure}

Here, we present the distributions of $p$ and the variation of $\pmean$ with
smoothing scale for turbulent forcing scales $\lf = 102.4\kpc$ in
Figure~\ref{fig:convlk5}, and for $\lf=64\kpc$ in Figure~\ref{fig:convlk8},
respectively.
\vspace{-6pt}



\vspace{-10pt}
\begin{adjustwidth}{-4.6cm}{0cm}


\end{adjustwidth}
\reftitle{References}


\begin{thebibliography}{-------}
\providecommand{\natexlab}[1]{#1}

\bibitem[{Feretti} \em{et~al.}(2012){Feretti}, {Giovannini}, {Govoni}, and
  {Murgia}]{Fer+12}
{Feretti}, L.; {Giovannini}, G.; {Govoni}, F.; {Murgia}, M.
\newblock {Clusters of galaxies: observational properties of the diffuse radio
  emission}.
\newblock {\em \aapr} {\bf 2012}, {\em 20},~54,
  \href{http://xxx.lanl.gov/abs/1205.1919}{{\normalfont
  [arXiv:astro-ph.CO/1205.1919]}}.
\newblock
  doi:{\changeurlcolor{black}\href{https://doi.org/10.1007/s00159-012-0054-z}{\detokenize{10.1007/s00159-012-0054-z}}}.

\bibitem[{van Weeren} \em{et~al.}(2019){van Weeren}, {de Gasperin}, {Akamatsu},
  {Br{\"u}ggen}, {Feretti}, {Kang}, {Stroe}, and {Zandanel}]{weere19}
{van Weeren}, R.J.; {de Gasperin}, F.; {Akamatsu}, H.; {Br{\"u}ggen}, M.;
  {Feretti}, L.; {Kang}, H.; {Stroe}, A.; {Zandanel}, F.
\newblock {Diffuse Radio Emission from Galaxy Clusters}.
\newblock {\em \ssr} {\bf 2019}, {\em 215},~16,
  \href{http://xxx.lanl.gov/abs/1901.04496}{{\normalfont
  [arXiv:astro-ph.HE/1901.04496]}}.
\newblock
  doi:{\changeurlcolor{black}\href{https://doi.org/10.1007/s11214-019-0584-z}{\detokenize{10.1007/s11214-019-0584-z}}}.

\bibitem[{Vazza} \em{et~al.}(2018){Vazza}, {Brunetti}, {Br{\"u}ggen}, and
  {Bonafede}]{Vazza+18}
{Vazza}, F.; {Brunetti}, G.; {Br{\"u}ggen}, M.; {Bonafede}, A.
\newblock {Resolved magnetic dynamo action in the simulated intracluster
  medium}.
\newblock {\em \mnras} {\bf 2018}, {\em 474},~1672--1687,
  \href{http://xxx.lanl.gov/abs/1711.02673}{{\normalfont
  [arXiv:astro-ph.CO/1711.02673]}}.
\newblock
  doi:{\changeurlcolor{black}\href{https://doi.org/10.1093/mnras/stx2830}{\detokenize{10.1093/mnras/stx2830}}}.

\bibitem[{Dom{\'\i}nguez-Fern{\'a}ndez}
  \em{et~al.}(2019){Dom{\'\i}nguez-Fern{\'a}ndez}, {Vazza}, {Br{\"u}ggen}, and
  {Brunetti}]{Domin+19}
{Dom{\'\i}nguez-Fern{\'a}ndez}, P.; {Vazza}, F.; {Br{\"u}ggen}, M.; {Brunetti},
  G.
\newblock {Dynamical evolution of magnetic fields in the intracluster medium}.
\newblock {\em \mnras} {\bf 2019}, {\em 486},~623--638,
  \href{http://xxx.lanl.gov/abs/1903.11052}{{\normalfont
  [arXiv:astro-ph.CO/1903.11052]}}.
\newblock
  doi:{\changeurlcolor{black}\href{https://doi.org/10.1093/mnras/stz877}{\detokenize{10.1093/mnras/stz877}}}.

\bibitem[{Sur} \em{et~al.}(2021){Sur}, {Basu}, and {Subramanian}]{sur21}
{Sur}, S.; {Basu}, A.; {Subramanian}, K.
\newblock {Properties of polarized synchrotron emission from fluctuation-dynamo
  action - I. Application to galaxy clusters}.
\newblock {\em \mnras} {\bf 2021}, {\em 501},~3332--3349,
  \href{http://xxx.lanl.gov/abs/2011.08499}{{\normalfont
  [arXiv:astro-ph.CO/2011.08499]}}.
\newblock
  doi:{\changeurlcolor{black}\href{https://doi.org/10.1093/mnras/staa3767}{\detokenize{10.1093/mnras/staa3767}}}.

\bibitem[{Clarke} \em{et~al.}(2001){Clarke}, {Kronberg}, and
  {B{\"o}hringer}]{CKB01}
{Clarke}, T.E.; {Kronberg}, P.P.; {B{\"o}hringer}, H.
\newblock {A New Radio-X-Ray Probe of Galaxy Cluster Magnetic Fields}.
\newblock {\em \apj} {\bf 2001}, {\em 547},~L111--L114,
  \href{http://xxx.lanl.gov/abs/astro-ph/0011281}{{\normalfont
  [arXiv:astro-ph/astro-ph/0011281]}}.
\newblock
  doi:{\changeurlcolor{black}\href{https://doi.org/10.1086/318896}{\detokenize{10.1086/318896}}}.

\bibitem[{Bonafede} \em{et~al.}(2009){Bonafede}, {Feretti}, {Giovannini},
  {Govoni}, {Murgia}, {Taylor}, {Ebeling}, {Allen}, {Gentile}, and
  {Pihlstr{\"o}m}]{Bonafede+09}
{Bonafede}, A.; {Feretti}, L.; {Giovannini}, G.; {Govoni}, F.; {Murgia}, M.;
  {Taylor}, G.B.; {Ebeling}, H.; {Allen}, S.; {Gentile}, G.; {Pihlstr{\"o}m},
  Y.
\newblock {Revealing the magnetic field in a distant galaxy cluster: discovery
  of the complex radio emission from MACS J0717.5 +3745}.
\newblock {\em \aap} {\bf 2009}, {\em 503},~707--720,
  \href{http://xxx.lanl.gov/abs/0905.3552}{{\normalfont
  [arXiv:astro-ph.CO/0905.3552]}}.
\newblock
  doi:{\changeurlcolor{black}\href{https://doi.org/10.1051/0004-6361/200912520}{\detokenize{10.1051/0004-6361/200912520}}}.

\bibitem[{Bonafede} \em{et~al.}(2010){Bonafede}, {Feretti}, {Murgia}, {Govoni},
  {Giovannini}, {Dallacasa}, {Dolag}, and {Taylor}]{bonaf10}
{Bonafede}, A.; {Feretti}, L.; {Murgia}, M.; {Govoni}, F.; {Giovannini}, G.;
  {Dallacasa}, D.; {Dolag}, K.; {Taylor}, G.B.
\newblock {The Coma cluster magnetic field from Faraday rotation measures}.
\newblock {\em \aap} {\bf 2010}, {\em 513},~A30,
  \href{http://xxx.lanl.gov/abs/1002.0594}{{\normalfont
  [arXiv:astro-ph.CO/1002.0594]}}.
\newblock
  doi:{\changeurlcolor{black}\href{https://doi.org/10.1051/0004-6361/200913696}{\detokenize{10.1051/0004-6361/200913696}}}.

\bibitem[{Kierdorf} \em{et~al.}(2017){Kierdorf}, {Beck}, {Hoeft}, {Klein}, {van
  Weeren}, {Forman}, and {Jones}]{Kierdorf+17}
{Kierdorf}, M.; {Beck}, R.; {Hoeft}, M.; {Klein}, U.; {van Weeren}, R.J.;
  {Forman}, W.R.; {Jones}, C.
\newblock {Relics in galaxy clusters at high radio frequencies}.
\newblock {\em \aap} {\bf 2017}, {\em 600},~A18,
  \href{http://xxx.lanl.gov/abs/1612.01764}{{\normalfont
  [arXiv:astro-ph.GA/1612.01764]}}.
\newblock
  doi:{\changeurlcolor{black}\href{https://doi.org/10.1051/0004-6361/201629570}{\detokenize{10.1051/0004-6361/201629570}}}.

\bibitem[{Subramanian} \em{et~al.}(2006){Subramanian}, {Shukurov}, and
  {Haugen}]{SSH06}
{Subramanian}, K.; {Shukurov}, A.; {Haugen}, N.E.L.
\newblock {Evolving turbulence and magnetic fields in galaxy clusters}.
\newblock {\em \mnras} {\bf 2006}, {\em 366},~1437--1454,
  \href{http://xxx.lanl.gov/abs/arXiv:astro-ph/0505144}{{\normalfont
  [arXiv:astro-ph/0505144]}}.
\newblock
  doi:{\changeurlcolor{black}\href{https://doi.org/10.1111/j.1365-2966.2006.09918.x}{\detokenize{10.1111/j.1365-2966.2006.09918.x}}}.

\bibitem[{Cho} and {Ryu}(2009)]{CR09}
{Cho}, J.; {Ryu}, D.
\newblock {Characteristic Lengths of Magnetic Field in Magnetohydrodynamic
  Turbulence}.
\newblock {\em \apjl} {\bf 2009}, {\em 705},~L90--L94,
  \href{http://xxx.lanl.gov/abs/0908.0610}{{\normalfont
  [arXiv:astro-ph.CO/0908.0610]}}.
\newblock
  doi:{\changeurlcolor{black}\href{https://doi.org/10.1088/0004-637X/705/1/L90}{\detokenize{10.1088/0004-637X/705/1/L90}}}.

\bibitem[{Bhat} and {Subramanian}(2013)]{BS13}
{Bhat}, P.; {Subramanian}, K.
\newblock {Fluctuation dynamos and their Faraday rotation signatures}.
\newblock {\em \mnras} {\bf 2013}, {\em 429},~2469--2481,
  \href{http://xxx.lanl.gov/abs/1210.3243}{{\normalfont
  [arXiv:astro-ph.CO/1210.3243]}}.
\newblock
  doi:{\changeurlcolor{black}\href{https://doi.org/10.1093/mnras/sts516}{\detokenize{10.1093/mnras/sts516}}}.

\bibitem[{Porter} \em{et~al.}(2015){Porter}, {Jones}, and {Ryu}]{PJR15}
{Porter}, D.H.; {Jones}, T.W.; {Ryu}, D.
\newblock {Vorticity, Shocks, and Magnetic Fields in Subsonic, ICM-like
  Turbulence}.
\newblock {\em \apj} {\bf 2015}, {\em 810},~93,
  \href{http://xxx.lanl.gov/abs/1507.08737}{{\normalfont
  [arXiv:astro-ph.CO/1507.08737]}}.
\newblock
  doi:{\changeurlcolor{black}\href{https://doi.org/10.1088/0004-637X/810/2/93}{\detokenize{10.1088/0004-637X/810/2/93}}}.

\bibitem[{Sur} \em{et~al.}(2018){Sur}, {Bhat}, and {Subramanian}]{SBS18}
{Sur}, S.; {Bhat}, P.; {Subramanian}, K.
\newblock {Faraday rotation signatures of fluctuation dynamos in young
  galaxies}.
\newblock {\em \mnras} {\bf 2018}, {\em 475},~L72--L76,
  \href{http://xxx.lanl.gov/abs/1711.08865}{{\normalfont
  [arXiv:astro-ph.GA/1711.08865]}}.
\newblock
  doi:{\changeurlcolor{black}\href{https://doi.org/10.1093/mnrasl/sly007}{\detokenize{10.1093/mnrasl/sly007}}}.

\bibitem[{Sur}(2019)]{Sur19}
{Sur}, S.
\newblock {Decaying turbulence and magnetic fields in galaxy clusters}.
\newblock {\em \mnras} {\bf 2019}, {\em 488},~3439--3445,
  \href{http://xxx.lanl.gov/abs/1907.04488}{{\normalfont
  [arXiv:astro-ph.CO/1907.04488]}}.
\newblock
  doi:{\changeurlcolor{black}\href{https://doi.org/10.1093/mnras/stz1918}{\detokenize{10.1093/mnras/stz1918}}}.

\bibitem[{Brunetti} and {Jones}(2014)]{BJ14}
{Brunetti}, G.; {Jones}, T.W.
\newblock {Cosmic Rays in Galaxy Clusters and Their Nonthermal Emission}.
\newblock {\em International Journal of Modern Physics D} {\bf 2014}, {\em
  23},~1430007--98,  \href{http://xxx.lanl.gov/abs/1401.7519}{{\normalfont
  [arXiv:astro-ph.CO/1401.7519]}}.
\newblock
  doi:{\changeurlcolor{black}\href{https://doi.org/10.1142/S0218271814300079}{\detokenize{10.1142/S0218271814300079}}}.

\bibitem[{Kunz} \em{et~al.}(2011){Kunz}, {Schekochihin}, {Cowley}, {Binney},
  and {Sanders}]{Kunz+11}
{Kunz}, M.W.; {Schekochihin}, A.A.; {Cowley}, S.C.; {Binney}, J.J.; {Sanders},
  J.S.
\newblock {A thermally stable heating mechanism for the intracluster medium:
  turbulence, magnetic fields and plasma instabilities}.
\newblock {\em \mnras} {\bf 2011}, {\em 410},~2446--2457,
  \href{http://xxx.lanl.gov/abs/1003.2719}{{\normalfont
  [arXiv:astro-ph.CO/1003.2719]}}.
\newblock
  doi:{\changeurlcolor{black}\href{https://doi.org/10.1111/j.1365-2966.2010.17621.x}{\detokenize{10.1111/j.1365-2966.2010.17621.x}}}.

\bibitem[{Komarov} \em{et~al.}(2014){Komarov}, {Churazov}, {Schekochihin}, and
  {ZuHone}]{Komarov+14}
{Komarov}, S.V.; {Churazov}, E.M.; {Schekochihin}, A.A.; {ZuHone}, J.A.
\newblock {Suppression of local heat flux in a turbulent magnetized
  intracluster medium}.
\newblock {\em \mnras} {\bf 2014}, {\em 440},~1153--1164,
  \href{http://xxx.lanl.gov/abs/1304.1857}{{\normalfont
  [arXiv:astro-ph.HE/1304.1857]}}.
\newblock
  doi:{\changeurlcolor{black}\href{https://doi.org/10.1093/mnras/stu281}{\detokenize{10.1093/mnras/stu281}}}.

\bibitem[{Roberg-Clark} \em{et~al.}(2016){Roberg-Clark}, {Drake}, {Reynolds},
  and {Swisdak}]{Roberg+16}
{Roberg-Clark}, G.T.; {Drake}, J.F.; {Reynolds}, C.S.; {Swisdak}, M.
\newblock {Suppression of Electron Thermal Conduction in the High
  {\ensuremath{\beta}} Intracluster Medium of Galaxy Clusters}.
\newblock {\em \apjl} {\bf 2016}, {\em 830},~L9,
  \href{http://xxx.lanl.gov/abs/1606.05261}{{\normalfont
  [arXiv:astro-ph.HE/1606.05261]}}.
\newblock
  doi:{\changeurlcolor{black}\href{https://doi.org/10.3847/2041-8205/830/1/L9}{\detokenize{10.3847/2041-8205/830/1/L9}}}.

\bibitem[{Bonafede} \em{et~al.}(2015){Bonafede}, {Vazza}, {Br{\"u}ggen},
  {Akahori}, {Carretti}, {Colafrancesco}, {Feretti}, {Ferrari}, {Giovannini},
  {Govoni}, {Johnston-Hollitt}, {Murgia}, {Scaife}, {Vacca}, {Govoni},
  {Rudnick}, and {Scaife}]{bonaf15}
{Bonafede}, A.; {Vazza}, F.; {Br{\"u}ggen}, M.; {Akahori}, T.; {Carretti}, E.;
  {Colafrancesco}, S.; {Feretti}, L.; {Ferrari}, C.; {Giovannini}, G.;
  {Govoni}, F.; {Johnston-Hollitt}, M.; {Murgia}, M.; {Scaife}, A.; {Vacca},
  V.; {Govoni}, F.; {Rudnick}, L.; {Scaife}, A.
\newblock {Unravelling the origin of large-scale magnetic fields in galaxy
  clusters and beyond through Faraday Rotation Measures with the SKA}.
\newblock  Advancing Astrophysics with the Square Kilometre Array (AASKA14),
  2015, p.~95,  \href{http://xxx.lanl.gov/abs/1501.00321}{{\normalfont
  [arXiv:astro-ph.CO/1501.00321]}}.

\bibitem[{Govoni} \em{et~al.}(2005){Govoni}, {Murgia}, {Feretti}, {Giovannini},
  {Dallacasa}, and {Taylor}]{Govoni+05}
{Govoni}, F.; {Murgia}, M.; {Feretti}, L.; {Giovannini}, G.; {Dallacasa}, D.;
  {Taylor}, G.B.
\newblock {A2255: The first detection of filamentary polarized emission in a
  radio halo}.
\newblock {\em \aap} {\bf 2005}, {\em 430},~L5--L8,
  \href{http://xxx.lanl.gov/abs/astro-ph/0411720}{{\normalfont
  [arXiv:astro-ph/astro-ph/0411720]}}.
\newblock
  doi:{\changeurlcolor{black}\href{https://doi.org/10.1051/0004-6361:200400113}{\detokenize{10.1051/0004-6361:200400113}}}.

\bibitem[{Girardi} \em{et~al.}(2016){Girardi}, {Boschin}, {Gastaldello},
  {Giovannini}, {Govoni}, {Murgia}, {Barrena}, {Ettori}, {Trasatti}, and
  {Vacca}]{girardi16}
{Girardi}, M.; {Boschin}, W.; {Gastaldello}, F.; {Giovannini}, G.; {Govoni},
  F.; {Murgia}, M.; {Barrena}, R.; {Ettori}, S.; {Trasatti}, M.; {Vacca}, V.
\newblock {A multiwavelength view of the galaxy cluster Abell 523 and its
  peculiar diffuse radio source}.
\newblock {\em \mnras} {\bf 2016}, {\em 456},~2829--2847,
  \href{http://xxx.lanl.gov/abs/1510.05951}{{\normalfont
  [arXiv:astro-ph.GA/1510.05951]}}.
\newblock
  doi:{\changeurlcolor{black}\href{https://doi.org/10.1093/mnras/stv2827}{\detokenize{10.1093/mnras/stv2827}}}.

\bibitem[{Pizzo} \em{et~al.}(2011){Pizzo}, {de Bruyn}, {Bernardi}, and
  {Brentjens}]{P+11}
{Pizzo}, R.F.; {de Bruyn}, A.G.; {Bernardi}, G.; {Brentjens}, M.A.
\newblock {Deep multi-frequency rotation measure tomography of the galaxy
  cluster A2255}.
\newblock {\em \aap} {\bf 2011}, {\em 525},~A104,
  \href{http://xxx.lanl.gov/abs/1008.3530}{{\normalfont
  [arXiv:astro-ph.CO/1008.3530]}}.
\newblock
  doi:{\changeurlcolor{black}\href{https://doi.org/10.1051/0004-6361/201014158}{\detokenize{10.1051/0004-6361/201014158}}}.

\bibitem[{Rajpurohit} \em{et~al.}(2021){Rajpurohit}, {Brunetti}, {Bonafede},
  {van Weeren}, {Botteon}, {Vazza}, {Hoeft}, {Riseley}, {Bonnassieux},
  {Brienza}, {Forman}, {R{\"o}ttgering}, {Rajpurohit}, {Locatelli}, {Shimwell},
  {Cassano}, {Di Gennaro}, {Br{\"u}ggen}, {Wittor}, {Drabent}, and
  {Ignesti}]{rajpurohit21}
{Rajpurohit}, K.; {Brunetti}, G.; {Bonafede}, A.; {van Weeren}, R.J.;
  {Botteon}, A.; {Vazza}, F.; {Hoeft}, M.; {Riseley}, C.J.; {Bonnassieux}, E.;
  {Brienza}, M.; {Forman}, W.R.; {R{\"o}ttgering}, H.J.A.; {Rajpurohit}, A.S.;
  {Locatelli}, N.; {Shimwell}, T.W.; {Cassano}, R.; {Di Gennaro}, G.;
  {Br{\"u}ggen}, M.; {Wittor}, D.; {Drabent}, A.; {Ignesti}, A.
\newblock {Physical insights from the spectrum of the radio halo in MACS
  J0717.5+3745}.
\newblock {\em \aap} {\bf 2021}, {\em 646},~A135,
  \href{http://xxx.lanl.gov/abs/2012.14373}{{\normalfont
  [arXiv:astro-ph.GA/2012.14373]}}.
\newblock
  doi:{\changeurlcolor{black}\href{https://doi.org/10.1051/0004-6361/202039591}{\detokenize{10.1051/0004-6361/202039591}}}.

\bibitem[{Wittor} \em{et~al.}(2019){Wittor}, {Hoeft}, {Vazza}, {Br{\"u}ggen},
  and {Dom{\'\i}nguez-Fern{\'a}ndez}]{wittor19}
{Wittor}, D.; {Hoeft}, M.; {Vazza}, F.; {Br{\"u}ggen}, M.;
  {Dom{\'\i}nguez-Fern{\'a}ndez}, P.
\newblock {Polarization of radio relics in galaxy clusters}.
\newblock {\em \mnras} {\bf 2019}, {\em 490},~3987--4006,
  \href{http://xxx.lanl.gov/abs/1909.11329}{{\normalfont
  [arXiv:astro-ph.CO/1909.11329]}}.
\newblock
  doi:{\changeurlcolor{black}\href{https://doi.org/10.1093/mnras/stz2715}{\detokenize{10.1093/mnras/stz2715}}}.

\bibitem[{B{\"o}hringer} \em{et~al.}(2016){B{\"o}hringer}, {Chon}, and
  {Kronberg}]{bohringer16}
{B{\"o}hringer}, H.; {Chon}, G.; {Kronberg}, P.P.
\newblock {The Cosmic Large-Scale Structure in X-rays (CLASSIX) Cluster Survey.
  I. Probing galaxy cluster magnetic fields with line of sight rotation
  measures}.
\newblock {\em \aap} {\bf 2016}, {\em 596},~A22,
  \href{http://xxx.lanl.gov/abs/1610.02887}{{\normalfont
  [arXiv:astro-ph.CO/1610.02887]}}.
\newblock
  doi:{\changeurlcolor{black}\href{https://doi.org/10.1051/0004-6361/201628873}{\detokenize{10.1051/0004-6361/201628873}}}.

\bibitem[{Vacca} \em{et~al.}(2010){Vacca}, {Murgia}, {Govoni}, {Feretti},
  {Giovannini}, {Orr{\`u}}, and {Bonafede}]{Vacca+10}
{Vacca}, V.; {Murgia}, M.; {Govoni}, F.; {Feretti}, L.; {Giovannini}, G.;
  {Orr{\`u}}, E.; {Bonafede}, A.
\newblock {The intracluster magnetic field power spectrum in Abell 665}.
\newblock {\em \aap} {\bf 2010}, {\em 514},~A71.
\newblock
  doi:{\changeurlcolor{black}\href{https://doi.org/10.1051/0004-6361/200913060}{\detokenize{10.1051/0004-6361/200913060}}}.

\bibitem[{Schekochihin} \em{et~al.}(2004){Schekochihin}, {Cowley}, {Taylor},
  {Maron}, and {McWilliams}]{Schek04}
{Schekochihin}, A.A.; {Cowley}, S.C.; {Taylor}, S.F.; {Maron}, J.L.;
  {McWilliams}, J.C.
\newblock {Simulations of the Small-Scale Turbulent Dynamo}.
\newblock {\em \apj} {\bf 2004}, {\em 612},~276--307,
  \href{http://xxx.lanl.gov/abs/astro-ph/0312046}{{\normalfont
  [arXiv:astro-ph/astro-ph/0312046]}}.
\newblock
  doi:{\changeurlcolor{black}\href{https://doi.org/10.1086/422547}{\detokenize{10.1086/422547}}}.

\bibitem[{Brandenburg} and {Subramanian}(2005)]{BS05}
{Brandenburg}, A.; {Subramanian}, K.
\newblock {Astrophysical magnetic fields and nonlinear dynamo theory}.
\newblock {\em \physrep} {\bf 2005}, {\em 417},~1--209,
  \href{http://xxx.lanl.gov/abs/astro-ph/0405052}{{\normalfont
  [arXiv:astro-ph/astro-ph/0405052]}}.
\newblock
  doi:{\changeurlcolor{black}\href{https://doi.org/10.1016/j.physrep.2005.06.005}{\detokenize{10.1016/j.physrep.2005.06.005}}}.

\bibitem[{Seta} \em{et~al.}(2020){Seta}, {Bushby}, {Shukurov}, and
  {Wood}]{Seta+20}
{Seta}, A.; {Bushby}, P.J.; {Shukurov}, A.; {Wood}, T.S.
\newblock {Saturation mechanism of the fluctuation dynamo at Pr$_{M}$
  {\ensuremath{\geq}} 1}.
\newblock {\em Physical Review Fluids} {\bf 2020}, {\em 5},~043702.
\newblock
  doi:{\changeurlcolor{black}\href{https://doi.org/10.1103/PhysRevFluids.5.043702}{\detokenize{10.1103/PhysRevFluids.5.043702}}}.

\bibitem[Basu \em{et~al.}(2019)Basu, Fletcher, Mao, Burkhart, Beck, and
  Schnitzeler]{basu19b}
Basu, A.; Fletcher, A.; Mao, S.A.; Burkhart, B.; Beck, R.; Schnitzeler, D.
\newblock An In-depth Investigation of Faraday Depth Spectrum Using Synthetic
  Observations of Turbulent MHD Simulations.
\newblock {\em Galaxies} {\bf 2019}, {\em 7},~89.
\newblock
  doi:{\changeurlcolor{black}\href{https://doi.org/10.3390/galaxies7040089}{\detokenize{10.3390/galaxies7040089}}}.

\bibitem[{Fryxell} \em{et~al.}(2000){Fryxell}, {Olson}, {Ricker}, {Timmes},
  {Zingale}, {Lamb}, {MacNeice}, {Rosner}, {Truran}, and {Tufo}]{Fry+00}
{Fryxell}, B.; {Olson}, K.; {Ricker}, P.; {Timmes}, F.X.; {Zingale}, M.;
  {Lamb}, D.Q.; {MacNeice}, P.; {Rosner}, R.; {Truran}, J.W.; {Tufo}, H.
\newblock {FLASH: An Adaptive Mesh Hydrodynamics Code for Modeling
  Astrophysical Thermonuclear Flashes}.
\newblock {\em \apjs} {\bf 2000}, {\em 131},~273--334.
\newblock
  doi:{\changeurlcolor{black}\href{https://doi.org/10.1086/317361}{\detokenize{10.1086/317361}}}.

\bibitem[{Miniati}(2015)]{Miniati15}
{Miniati}, F.
\newblock {The Matryoshka Run. II. Time-dependent Turbulence Statistics,
  Stochastic Particle Acceleration, and Microphysics Impact in a Massive Galaxy
  Cluster}.
\newblock {\em \apj} {\bf 2015}, {\em 800},~60,
  \href{http://xxx.lanl.gov/abs/1409.3576}{{\normalfont
  [arXiv:astro-ph.CO/1409.3576]}}.
\newblock
  doi:{\changeurlcolor{black}\href{https://doi.org/10.1088/0004-637X/800/1/60}{\detokenize{10.1088/0004-637X/800/1/60}}}.

\bibitem[{Vazza} \em{et~al.}(2017){Vazza}, {Jones}, {Br{\"u}ggen}, {Brunetti},
  {Gheller}, {Porter}, and {Ryu}]{Vazza+17}
{Vazza}, F.; {Jones}, T.W.; {Br{\"u}ggen}, M.; {Brunetti}, G.; {Gheller}, C.;
  {Porter}, D.; {Ryu}, D.
\newblock {Turbulence and vorticity in Galaxy clusters generated by structure
  formation}.
\newblock {\em \mnras} {\bf 2017}, {\em 464},~210--230,
  \href{http://xxx.lanl.gov/abs/1609.03558}{{\normalfont
  [arXiv:astro-ph.CO/1609.03558]}}.
\newblock
  doi:{\changeurlcolor{black}\href{https://doi.org/10.1093/mnras/stw2351}{\detokenize{10.1093/mnras/stw2351}}}.

\bibitem[{Wittor} \em{et~al.}(2017){Wittor}, {Jones}, {Vazza}, and
  {Br{\"u}ggen}]{Wittor+17}
{Wittor}, D.; {Jones}, T.; {Vazza}, F.; {Br{\"u}ggen}, M.
\newblock {Evolution of vorticity and enstrophy in the intracluster medium}.
\newblock {\em \mnras} {\bf 2017}, {\em 471},~3212--3225,
  \href{http://xxx.lanl.gov/abs/1706.02315}{{\normalfont
  [arXiv:astro-ph.CO/1706.02315]}}.
\newblock
  doi:{\changeurlcolor{black}\href{https://doi.org/10.1093/mnras/stx1769}{\detokenize{10.1093/mnras/stx1769}}}.

\bibitem[{Vall{\'e}s-P{\'e}rez} \em{et~al.}(2021){Vall{\'e}s-P{\'e}rez},
  {Planelles}, and {Quilis}]{PPQ21}
{Vall{\'e}s-P{\'e}rez}, D.; {Planelles}, S.; {Quilis}, V.
\newblock {Troubled cosmic flows: turbulence, enstrophy, and helicity from the
  assembly history of the intracluster medium}.
\newblock {\em \mnras} {\bf 2021}, {\em 504},~510--527,
  \href{http://xxx.lanl.gov/abs/2103.13449}{{\normalfont
  [arXiv:astro-ph.CO/2103.13449]}}.
\newblock
  doi:{\changeurlcolor{black}\href{https://doi.org/10.1093/mnras/stab880}{\detokenize{10.1093/mnras/stab880}}}.

\bibitem[{\textit{Hitomi} Collaboration: Aharonian}
  \em{et~al.}(2018){\textit{Hitomi} Collaboration: Aharonian}, {Akamatsu},
  {Akimoto}, {Allen}, {Angelini}, {Audard}, {Awaki}, {Axelsson}, {Bamba},
  {Bautz}, {Blandford}, {Brenneman}, {Brown}, {Bulbul}, {Cackett}, {Canning},
  {Chernyakova}, {Chiao}, {Coppi}, {Costantini}, {de Plaa}, {de Vries}, {den
  Herder}, {Done}, {Dotani}, {Ebisawa}, {Eckart}, {Enoto}, {Ezoe}, {Fabian},
  {Ferrigno}, {Foster}, {Fujimoto}, {Fukazawa}, {Furuzawa}, {Galeazzi},
  {Gallo}, {Gandhi}, {Giustini}, {Goldwurm}, {Gu}, {Guainazzi}, {Haba},
  {Hagino}, {Hamaguchi}, {Harrus}, {Hatsukade}, {Hayashi}, {Hayashi},
  {Hayashi}, {Hayashida}, {Hiraga}, {Hornschemeier}, {Hoshino}, {Hughes},
  {Ichinohe}, {Iizuka}, {Inoue}, {Inoue}, {Inoue}, {Ishida}, {Ishikawa},
  {Ishisaki}, {Iwai}, {Kaastra}, {Kallman}, {Kamae}, {Kataoka}, {Katsuda},
  {Kawai}, {Kelley}, {Kilbourne}, {Kitaguchi}, {Kitamoto}, {Kitayama},
  {Kohmura}, {Kokubun}, {Koyama}, {Koyama}, {Kretschmar}, {Krimm}, {Kubota},
  {Kunieda}, {Laurent}, {Lee}, {Leutenegger}, {Limousin}, {Loewenstein},
  {Long}, {Lumb}, {Madejski}, {Maeda}, {Maier}, {Makishima}, {Markevitch},
  {Matsumoto}, {Matsushita}, {McCammon}, {McNamara}, {Mehdipour}, {Miller},
  {Miller}, {Mineshige}, {Mitsuda}, {Mitsuishi}, {Miyazawa}, {Mizuno}, {Mori},
  {Mori}, {Mukai}, {Murakami}, {Mushotzky}, {Nakagawa}, {Nakajima}, {Nakamori},
  {Nakashima}, {Nakazawa}, {Nobukawa}, {Nobukawa}, {Noda}, {Odaka}, {Ohashi},
  {Ohno}, {Okajima}, {Ota}, {Ozaki}, {Paerels}, {Paltani}, {Petre}, {Pinto},
  {Porter}, {Pottschmidt}, {Reynolds}, {Safi-Harb}, {Saito}, {Sakai}, {Sasaki},
  {Sato}, {Sato}, {Sato}, {Sawada}, {Schartel}, {Serlemtsos}, {Seta},
  {Shidatsu}, {Simionescu}, {Smith}, {Soong}, {Stawarz}, {Sugawara}, {Sugita},
  {Szymkowiak}, {Tajima}, {Takahashi}, {Takahashi}, {Takeda}, {Takei},
  {Tamagawa}, {Tamura}, {Tanaka}, {Tanaka}, {Tanaka}, {Tanaka}, {Tashiro},
  {Tawara}, {Terada}, {Terashima}, {Tombesi}, {Tomida}, {Tsuboi}, {Tsujimoto},
  {Tsunemi}, {Tsuru}, {Uchida}, {Uchiyama}, {Uchiyama}, {Ueda}, {Ueda}, {Uno},
  {Urry}, {Ursino}, {Wang}, {Watanabe}, {Werner}, {Wilkins}, {Williams},
  {Yamada}, {Yamaguchi}, {Yamaoka}, {Yamasaki}, {Yamauchi}, {Yamauchi},
  {Yaqoob}, {Yatsu}, {Yonetoku}, {Zhuravleva}, and {Zoghbi}]{Hitomi18a}
{\textit{Hitomi} Collaboration.}
\newblock {Atmospheric gas dynamics in the Perseus cluster observed with
  \textit{Hitomi}}.
\newblock {\em \pasj} {\bf 2018}, {\em 70},~9,
  \href{http://xxx.lanl.gov/abs/1711.00240}{{\normalfont
  [arXiv:astro-ph.HE/1711.00240]}}.
\newblock
  doi:{\changeurlcolor{black}\href{https://doi.org/10.1093/pasj/psx138}{\detokenize{10.1093/pasj/psx138}}}.

\bibitem[{Churazov} \em{et~al.}(2012){Churazov}, {Vikhlinin}, {Zhuravleva},
  {Schekochihin}, {Parrish}, {Sunyaev}, {Forman}, {B{\"o}hringer}, and
  {Randall}]{Churazov+12}
{Churazov}, E.; {Vikhlinin}, A.; {Zhuravleva}, I.; {Schekochihin}, A.;
  {Parrish}, I.; {Sunyaev}, R.; {Forman}, W.; {B{\"o}hringer}, H.; {Randall},
  S.
\newblock {X-ray surface brightness and gas density fluctuations in the Coma
  cluster}.
\newblock {\em \mnras} {\bf 2012}, {\em 421},~1123--1135,
  \href{http://xxx.lanl.gov/abs/1110.5875}{{\normalfont
  [arXiv:astro-ph.CO/1110.5875]}}.
\newblock
  doi:{\changeurlcolor{black}\href{https://doi.org/10.1111/j.1365-2966.2011.20372.x}{\detokenize{10.1111/j.1365-2966.2011.20372.x}}}.

\bibitem[{Zhuravleva} \em{et~al.}(2019){Zhuravleva}, {Churazov},
  {Schekochihin}, {Allen}, {Vikhlinin}, and {Werner}]{Zhurav+19}
{Zhuravleva}, I.; {Churazov}, E.; {Schekochihin}, A.A.; {Allen}, S.W.;
  {Vikhlinin}, A.; {Werner}, N.
\newblock {Suppressed effective viscosity in the bulk intergalactic plasma}.
\newblock {\em Nature Astronomy} {\bf 2019}, {\em 3},~832--837,
  \href{http://xxx.lanl.gov/abs/1906.06346}{{\normalfont
  [arXiv:astro-ph.HE/1906.06346]}}.
\newblock
  doi:{\changeurlcolor{black}\href{https://doi.org/10.1038/s41550-019-0794-z}{\detokenize{10.1038/s41550-019-0794-z}}}.

\bibitem[{Sarazin}(1988)]{Sar88}
{Sarazin}, C.L.
\newblock {\em {X-Ray Emission From Clusters Of Galaxies}}; Cambridge Univ.\
  Press: Cambridge,  1988.

\bibitem[{Brentjens} and {de Bruyn}(2005)]{brent05}
{Brentjens}, M.A.; {de Bruyn}, A.G.
\newblock {Faraday rotation measure synthesis}.
\newblock {\em \aap} {\bf 2005}, {\em 441},~1217--1228,
  \href{http://xxx.lanl.gov/abs/astro-ph/0507349}{{\normalfont
  [astro-ph/0507349]}}.
\newblock
  doi:{\changeurlcolor{black}\href{https://doi.org/10.1051/0004-6361:20052990}{\detokenize{10.1051/0004-6361:20052990}}}.

\bibitem[{Heald} \em{et~al.}(2009){Heald}, {Braun}, and {Edmonds}]{heald09}
{Heald}, G.; {Braun}, R.; {Edmonds}, R.
\newblock {The Westerbork SINGS survey. II Polarization, Faraday rotation, and
  magnetic fields}.
\newblock {\em \aap} {\bf 2009}, {\em 503},~409--435,
  \href{http://xxx.lanl.gov/abs/0905.3995}{{\normalfont [0905.3995]}}.
\newblock
  doi:{\changeurlcolor{black}\href{https://doi.org/10.1051/0004-6361/200912240}{\detokenize{10.1051/0004-6361/200912240}}}.

\bibitem[{Cuciti} \em{et~al.}(2021){Cuciti}, {Cassano}, {Brunetti},
  {Dallacasa}, {van Weeren}, {Giacintucci}, {Bonafede}, {de Gasperin},
  {Ettori}, {Kale}, {Pratt}, and {Venturi}]{cuciti21}
{Cuciti}, V.; {Cassano}, R.; {Brunetti}, G.; {Dallacasa}, D.; {van Weeren},
  R.J.; {Giacintucci}, S.; {Bonafede}, A.; {de Gasperin}, F.; {Ettori}, S.;
  {Kale}, R.; {Pratt}, G.W.; {Venturi}, T.
\newblock {Radio halos in a mass-selected sample of 75 galaxy clusters. I.
  Sample selection and data analysis}.
\newblock {\em \aap} {\bf 2021}, {\em 647},~A50,
  \href{http://xxx.lanl.gov/abs/2101.01640}{{\normalfont
  [arXiv:astro-ph.CO/2101.01640]}}.
\newblock
  doi:{\changeurlcolor{black}\href{https://doi.org/10.1051/0004-6361/202039206}{\detokenize{10.1051/0004-6361/202039206}}}.

\bibitem[{Govoni} and {Feretti}(2004)]{GF04}
{Govoni}, F.; {Feretti}, L.
\newblock {Magnetic Fields in Clusters of Galaxies}.
\newblock {\em International Journal of Modern Physics D} {\bf 2004}, {\em
  13},~1549--1594,
  \href{http://xxx.lanl.gov/abs/astro-ph/0410182}{{\normalfont
  [arXiv:astro-ph/astro-ph/0410182]}}.
\newblock
  doi:{\changeurlcolor{black}\href{https://doi.org/10.1142/S0218271804005080}{\detokenize{10.1142/S0218271804005080}}}.

\bibitem[{Thierbach} \em{et~al.}(2003){Thierbach}, {Klein}, and
  {Wielebinski}]{TKW03}
{Thierbach}, M.; {Klein}, U.; {Wielebinski}, R.
\newblock {The diffuse radio emission from the Coma cluster at 2.675 GHz and
  4.85 GHz}.
\newblock {\em \aap} {\bf 2003}, {\em 397},~53--61,
  \href{http://xxx.lanl.gov/abs/astro-ph/0210147}{{\normalfont
  [arXiv:astro-ph/astro-ph/0210147]}}.
\newblock
  doi:{\changeurlcolor{black}\href{https://doi.org/10.1051/0004-6361:20021474}{\detokenize{10.1051/0004-6361:20021474}}}.

\bibitem[Sokoloff \em{et~al.}(1998)Sokoloff, Bykov, Shukurov, Berkhuijsen,
  Beck, and Poezd]{sokol98}
Sokoloff, D.; Bykov, A.; Shukurov, A.; Berkhuijsen, E.; Beck, R.; Poezd, A.
\newblock {Depolarization and Faraday effects in galaxies}.
\newblock {\em \mnras} {\bf 1998}, {\em 299},~189--206.
\newblock
  doi:{\changeurlcolor{black}\href{https://doi.org/10.1046/j.1365-8711.1998.01782.x}{\detokenize{10.1046/j.1365-8711.1998.01782.x}}}.

\bibitem[O'Sullivan \em{et~al.}(2012)O'Sullivan, Brown, Robishaw, Schnitzeler,
  McClure-Griffiths, Feain, Taylor, Gaensler, Landecker, Harvey-Smith, and
  Carretti]{sulli12}
O'Sullivan, S.; Brown, S.; Robishaw, T.; Schnitzeler, D.; McClure-Griffiths,
  N.; Feain, I.; Taylor, A.; Gaensler, B.; Landecker, T.; Harvey-Smith, L.;
  Carretti, E.
\newblock {Complex Faraday depth structure of active galactic nuclei as
  revealed by broad-band radio polarimetry}.
\newblock {\em \mnras} {\bf 2012}, {\em 421},~3300--3315,
  \href{http://xxx.lanl.gov/abs/1201.3161}{{\normalfont [1201.3161]}}.
\newblock
  doi:{\changeurlcolor{black}\href{https://doi.org/10.1111/j.1365-2966.2012.20554.x}{\detokenize{10.1111/j.1365-2966.2012.20554.x}}}.

\bibitem[{O'Sullivan} \em{et~al.}(2015){O'Sullivan}, {Gaensler},
  {Lara-L{\'o}pez}, {van Velzen}, {Banfield}, and {Farnes}]{sulli15}
{O'Sullivan}, S.P.; {Gaensler}, B.M.; {Lara-L{\'o}pez}, M.A.; {van Velzen}, S.;
  {Banfield}, J.K.; {Farnes}, J.S.
\newblock {The Magnetic Field and Polarization Properties of Radio Galaxies in
  Different Accretion States}.
\newblock {\em \apj} {\bf 2015}, {\em 806},~83,
  \href{http://xxx.lanl.gov/abs/1504.06679}{{\normalfont [1504.06679]}}.
\newblock
  doi:{\changeurlcolor{black}\href{https://doi.org/10.1088/0004-637X/806/1/83}{\detokenize{10.1088/0004-637X/806/1/83}}}.

\bibitem[{Akahori} \em{et~al.}(2016){Akahori}, {Ryu}, and {Gaensler}]{akaho16}
{Akahori}, T.; {Ryu}, D.; {Gaensler}, B.M.
\newblock {Fast Radio Bursts as Probes of Magnetic Fields in the Intergalactic
  Medium}.
\newblock {\em \apj} {\bf 2016}, {\em 824},~105,
  \href{http://xxx.lanl.gov/abs/1602.03235}{{\normalfont [1602.03235]}}.
\newblock
  doi:{\changeurlcolor{black}\href{https://doi.org/10.3847/0004-637X/824/2/105}{\detokenize{10.3847/0004-637X/824/2/105}}}.

\bibitem[{Vazza} \em{et~al.}(2014){Vazza}, {Br{\"u}ggen}, {Gheller}, and
  {Wang}]{vazza14}
{Vazza}, F.; {Br{\"u}ggen}, M.; {Gheller}, C.; {Wang}, P.
\newblock {On the amplification of magnetic fields in cosmic filaments and
  galaxy clusters}.
\newblock {\em \mnras} {\bf 2014}, {\em 445},~3706--3722,
  \href{http://xxx.lanl.gov/abs/1409.2640}{{\normalfont
  [arXiv:astro-ph.CO/1409.2640]}}.
\newblock
  doi:{\changeurlcolor{black}\href{https://doi.org/10.1093/mnras/stu1896}{\detokenize{10.1093/mnras/stu1896}}}.

\bibitem[{O'Sullivan} \em{et~al.}(2019){O'Sullivan}, {Machalski}, {Van Eck},
  {Heald}, {Br{\"u}ggen}, {Fynbo}, {Heintz}, {Lara-Lopez}, {Vacca},
  {Hardcastle}, {Shimwell}, {Tasse}, {Vazza}, {Andernach}, {Birkinshaw},
  {Haverkorn}, {Horellou}, {Williams}, {Harwood}, {Brunetti}, {Anderson},
  {Mao}, {Nikiel-Wroczy{\'n}ski}, {Takahashi}, {Carretti}, {Vernstrom}, {van
  Weeren}, {Orr{\'u}}, {Morabito}, and {Callingham}]{sulli19}
{O'Sullivan}, S.P.; {Machalski}, J.; {Van Eck}, C.L.; {Heald}, G.;
  {Br{\"u}ggen}, M.; {Fynbo}, J.P.U.; {Heintz}, K.E.; {Lara-Lopez}, M.A.;
  {Vacca}, V.; {Hardcastle}, M.J.; {Shimwell}, T.W.; {Tasse}, C.; {Vazza}, F.;
  {Andernach}, H.; {Birkinshaw}, M.; {Haverkorn}, M.; {Horellou}, C.;
  {Williams}, W.L.; {Harwood}, J.J.; {Brunetti}, G.; {Anderson}, J.M.; {Mao},
  S.A.; {Nikiel-Wroczy{\'n}ski}, B.; {Takahashi}, K.; {Carretti}, E.;
  {Vernstrom}, T.; {van Weeren}, R.J.; {Orr{\'u}}, E.; {Morabito}, L.K.;
  {Callingham}, J.R.
\newblock {The intergalactic magnetic field probed by a giant radio galaxy}.
\newblock {\em \aap} {\bf 2019}, {\em 622},~A16,
  \href{http://xxx.lanl.gov/abs/1811.07934}{{\normalfont
  [arXiv:astro-ph.HE/1811.07934]}}.
\newblock
  doi:{\changeurlcolor{black}\href{https://doi.org/10.1051/0004-6361/201833832}{\detokenize{10.1051/0004-6361/201833832}}}.

\bibitem[{Briel} \em{et~al.}(1992){Briel}, {Henry}, and {Boehringer}]{BHB92}
{Briel}, U.G.; {Henry}, J.P.; {Boehringer}, H.
\newblock {Observation of the Coma cluster of galaxies with ROSAT during the
  all-sky-survey.}
\newblock {\em \aap} {\bf 1992}, {\em 259},~L31--L34.

\bibitem[{Bonafede} \em{et~al.}(2010){Bonafede}, {Feretti}, {Murgia}, {Govoni},
  {Giovannini}, {Dallacasa}, {Dolag}, and {Taylor}]{B+10}
{Bonafede}, A.; {Feretti}, L.; {Murgia}, M.; {Govoni}, F.; {Giovannini}, G.;
  {Dallacasa}, D.; {Dolag}, K.; {Taylor}, G.B.
\newblock {The Coma cluster magnetic field from Faraday rotation measures}.
\newblock {\em \aap} {\bf 2010}, {\em 513},~A30,
  \href{http://xxx.lanl.gov/abs/1002.0594}{{\normalfont [1002.0594]}}.
\newblock
  doi:{\changeurlcolor{black}\href{https://doi.org/10.1051/0004-6361/200913696}{\detokenize{10.1051/0004-6361/200913696}}}.

\bibitem[{Norman} and {Bryan}(1999)]{NB99}
{Norman}, M.L.; {Bryan}, G.L., {Cluster Turbulence}.
\newblock In {\em The Radio Galaxy Messier 87}; {R{\"o}ser}, H.J.;
  {Meisenheimer}, K., Eds.;  1999; Vol. 530, p. 106.
\newblock
  doi:{\changeurlcolor{black}\href{https://doi.org/10.1007/BFb0106425}{\detokenize{10.1007/BFb0106425}}}.

\bibitem[{Ryu} \em{et~al.}(2008){Ryu}, {Kang}, {Cho}, and {Das}]{RKCD08}
{Ryu}, D.; {Kang}, H.; {Cho}, J.; {Das}, S.
\newblock {Turbulence and Magnetic Fields in the Large-Scale Structure of the
  Universe}.
\newblock {\em Science} {\bf 2008}, {\em 320},~909,
  \href{http://xxx.lanl.gov/abs/0805.2466}{{\normalfont
  [arXiv:astro-ph/0805.2466]}}.
\newblock
  doi:{\changeurlcolor{black}\href{https://doi.org/10.1126/science.1154923}{\detokenize{10.1126/science.1154923}}}.

\bibitem[{Xu} \em{et~al.}(2012){Xu}, {Govoni}, {Murgia}, {Li}, {Collins},
  {Norman}, {Cen}, {Feretti}, and {Giovannini}]{Xu+12}
{Xu}, H.; {Govoni}, F.; {Murgia}, M.; {Li}, H.; {Collins}, D.C.; {Norman},
  M.L.; {Cen}, R.; {Feretti}, L.; {Giovannini}, G.
\newblock {Comparisons of Cosmological Magnetohydrodynamic Galaxy Cluster
  Simulations to Radio Observations}.
\newblock {\em \apj} {\bf 2012}, {\em 759},~40,
  \href{http://xxx.lanl.gov/abs/1209.2737}{{\normalfont
  [arXiv:astro-ph.CO/1209.2737]}}.
\newblock
  doi:{\changeurlcolor{black}\href{https://doi.org/10.1088/0004-637X/759/1/40}{\detokenize{10.1088/0004-637X/759/1/40}}}.

\bibitem[{Donnert} \em{et~al.}(2009){Donnert}, {Dolag}, {Lesch}, and
  {M{\"u}ller}]{donnert2009}
{Donnert}, J.; {Dolag}, K.; {Lesch}, H.; {M{\"u}ller}, E.
\newblock {Cluster magnetic fields from galactic outflows}.
\newblock {\em \mnras} {\bf 2009}, {\em 392},~1008--1021,
  \href{http://xxx.lanl.gov/abs/0808.0919}{{\normalfont
  [arXiv:astro-ph/0808.0919]}}.
\newblock
  doi:{\changeurlcolor{black}\href{https://doi.org/10.1111/j.1365-2966.2008.14132.x}{\detokenize{10.1111/j.1365-2966.2008.14132.x}}}.

\bibitem[Dubois \em{et~al.}(2012)Dubois, Devriendt, Slyz, and
  Teyssier]{dubois2012}
Dubois, Y.; Devriendt, J.; Slyz, A.; Teyssier, R.
\newblock {Self-regulated growth of supermassive black holes by a dual
  jet–heating active galactic nucleus feedback mechanism: methods, tests and
  implications for cosmological simulations}.
\newblock {\em \mnras} {\bf 2012}, {\em 420},~2662--2683,
  \href{http://xxx.lanl.gov/abs/https://academic.oup.com/mnras/article-pdf/420/3/2662/3028346/mnras0420-2662.pdf}{{\normalfont
  [https://academic.oup.com/mnras/article-pdf/420/3/2662/3028346/mnras0420-2662.pdf]}}.
\newblock
  doi:{\changeurlcolor{black}\href{https://doi.org/10.1111/j.1365-2966.2011.20236.x}{\detokenize{10.1111/j.1365-2966.2011.20236.x}}}.

\bibitem[{Pakmor} \em{et~al.}(2016){Pakmor}, {Pfrommer}, {Simpson}, and
  {Springel}]{pakmor2016}
{Pakmor}, R.; {Pfrommer}, C.; {Simpson}, C.M.; {Springel}, V.
\newblock {Galactic Winds Driven by Isotropic and Anisotropic Cosmic-Ray
  Diffusion in Disk Galaxies}.
\newblock {\em \apjl} {\bf 2016}, {\em 824},~L30,
  \href{http://xxx.lanl.gov/abs/1605.00643}{{\normalfont
  [arXiv:astro-ph.GA/1605.00643]}}.
\newblock
  doi:{\changeurlcolor{black}\href{https://doi.org/10.3847/2041-8205/824/2/L30}{\detokenize{10.3847/2041-8205/824/2/L30}}}.

\bibitem[{Wiener} \em{et~al.}(2017){Wiener}, {Pfrommer}, and {Peng
  Oh}]{wiene17}
{Wiener}, J.; {Pfrommer}, C.; {Peng Oh}, S.
\newblock {Cosmic ray-driven galactic winds: streaming or diffusion?}
\newblock {\em \mnras} {\bf 2017}, {\em 467},~906--921,
  \href{http://xxx.lanl.gov/abs/1608.02585}{{\normalfont [1608.02585]}}.
\newblock
  doi:{\changeurlcolor{black}\href{https://doi.org/10.1093/mnras/stx127}{\detokenize{10.1093/mnras/stx127}}}.

\bibitem[{Fabian}(2012)]{fabian2012}
{Fabian}, A.C.
\newblock {Observational Evidence of Active Galactic Nuclei Feedback}.
\newblock {\em \araa} {\bf 2012}, {\em 50},~455--489,
  \href{http://xxx.lanl.gov/abs/1204.4114}{{\normalfont
  [arXiv:astro-ph.CO/1204.4114]}}.
\newblock
  doi:{\changeurlcolor{black}\href{https://doi.org/10.1146/annurev-astro-081811-125521}{\detokenize{10.1146/annurev-astro-081811-125521}}}.

\bibitem[{Bourne} and {Sijacki}(2017)]{bourne2017}
{Bourne}, M.A.; {Sijacki}, D.
\newblock {AGN jet feedback on a moving mesh: cocoon inflation, gas flows and
  turbulence}.
\newblock {\em \mnras} {\bf 2017}, {\em 472},~4707--4735,
  \href{http://xxx.lanl.gov/abs/1705.07900}{{\normalfont
  [arXiv:astro-ph.GA/1705.07900]}}.
\newblock
  doi:{\changeurlcolor{black}\href{https://doi.org/10.1093/mnras/stx2269}{\detokenize{10.1093/mnras/stx2269}}}.

\bibitem[{Ehlert} \em{et~al.}(2021){Ehlert}, {Weinberger}, {Pfrommer}, and
  {Springel}]{ehlert2021}
{Ehlert}, K.; {Weinberger}, R.; {Pfrommer}, C.; {Springel}, V.
\newblock {Connecting turbulent velocities and magnetic fields in galaxy
  cluster simulations with active galactic nuclei jets}.
\newblock {\em \mnras} {\bf 2021}, {\em 503},~1327--1344,
  \href{http://xxx.lanl.gov/abs/2011.13964}{{\normalfont
  [arXiv:astro-ph.GA/2011.13964]}}.
\newblock
  doi:{\changeurlcolor{black}\href{https://doi.org/10.1093/mnras/stab551}{\detokenize{10.1093/mnras/stab551}}}.

\bibitem[{Braun} \em{et~al.}(2019){Braun}, {Bonaldi}, {Bourke}, {Keane}, and
  {Wagg}]{braun2019}
{Braun}, R.; {Bonaldi}, A.; {Bourke}, T.; {Keane}, E.; {Wagg}, J.
\newblock {Anticipated Performance of the Square Kilometre Array -- Phase 1
  (SKA1)}.
\newblock {\em arXiv e-prints} {\bf 2019}, p. arXiv:1912.12699,
  \href{http://xxx.lanl.gov/abs/1912.12699}{{\normalfont
  [arXiv:astro-ph.IM/1912.12699]}}.

\bibitem[{Yuan} \em{et~al.}(2015){Yuan}, {Han}, and {Wen}]{yuan2015}
{Yuan}, Z.S.; {Han}, J.L.; {Wen}, Z.L.
\newblock {The Scaling Relations and the Fundamental Plane for Radio Halos and
  Relics of Galaxy Clusters}.
\newblock {\em \apj} {\bf 2015}, {\em 813},~77,
  \href{http://xxx.lanl.gov/abs/1510.04980}{{\normalfont
  [arXiv:astro-ph.CO/1510.04980]}}.
\newblock
  doi:{\changeurlcolor{black}\href{https://doi.org/10.1088/0004-637X/813/1/77}{\detokenize{10.1088/0004-637X/813/1/77}}}.

\bibitem[{Astropy Collaboration} \em{et~al.}(2013){Astropy Collaboration},
  {Robitaille}, {Tollerud}, {Greenfield}, {Droettboom}, {Bray}, {Aldcroft},
  {Davis}, {Ginsburg}, {Price-Whelan}, {Kerzendorf}, {Conley}, {Crighton},
  {Barbary}, {Muna}, {Ferguson}, {Grollier}, {Parikh}, {Nair}, {Unther},
  {Deil}, {Woillez}, {Conseil}, {Kramer}, {Turner}, {Singer}, {Fox}, {Weaver},
  {Zabalza}, {Edwards}, {Azalee Bostroem}, {Burke}, {Casey}, {Crawford},
  {Dencheva}, {Ely}, {Jenness}, {Labrie}, {Lim}, {Pierfederici}, {Pontzen},
  {Ptak}, {Refsdal}, {Servillat}, and {Streicher}]{astropy:2013}
{Astropy Collaboration}.
\newblock {Astropy: A community Python package for astronomy}.
\newblock {\em \aap} {\bf 2013}, {\em 558},~A33,
  \href{http://xxx.lanl.gov/abs/1307.6212}{{\normalfont
  [arXiv:astro-ph.IM/1307.6212]}}.
\newblock
  doi:{\changeurlcolor{black}\href{https://doi.org/10.1051/0004-6361/201322068}{\detokenize{10.1051/0004-6361/201322068}}}.

\bibitem[{Price-Whelan} \em{et~al.}(2018){Price-Whelan}, {Sip{\H{o}}cz},
  {G{\"u}nther}, {Lim}, {Crawford}, {Conseil}, {Shupe}, {Craig}, {Dencheva},
  {Ginsburg}, {VanderPlas}, {Bradley}, {P{\'e}rez-Su{\'a}rez}, {de Val-Borro},
  {Paper Contributors}, {Aldcroft}, {Cruz}, {Robitaille}, {Tollerud},
  {Coordination Committee}, {Ardelean}, {Babej}, {Bach}, {Bachetti}, {Bakanov},
  {Bamford}, {Barentsen}, {Barmby}, {Baumbach}, {Berry}, {Biscani}, {Boquien},
  {Bostroem}, {Bouma}, {Brammer}, {Bray}, {Breytenbach}, {Buddelmeijer},
  {Burke}, {Calderone}, {Cano Rodr{\'\i}guez}, {Cara}, {Cardoso}, {Cheedella},
  {Copin}, {Corrales}, {Crichton}, {D{\textquoteright}Avella}, {Deil},
  {Depagne}, {Dietrich}, {Donath}, {Droettboom}, {Earl}, {Erben}, {Fabbro},
  {Ferreira}, {Finethy}, {Fox}, {Garrison}, {Gibbons}, {Goldstein}, {Gommers},
  {Greco}, {Greenfield}, {Groener}, {Grollier}, {Hagen}, {Hirst}, {Homeier},
  {Horton}, {Hosseinzadeh}, {Hu}, {Hunkeler}, {Ivezi{\'c}}, {Jain}, {Jenness},
  {Kanarek}, {Kendrew}, {Kern}, {Kerzendorf}, {Khvalko}, {King}, {Kirkby},
  {Kulkarni}, {Kumar}, {Lee}, {Lenz}, {Littlefair}, {Ma}, {Macleod},
  {Mastropietro}, {McCully}, {Montagnac}, {Morris}, {Mueller}, {Mumford},
  {Muna}, {Murphy}, {Nelson}, {Nguyen}, {Ninan}, {N{\"o}the}, {Ogaz}, {Oh},
  {Parejko}, {Parley}, {Pascual}, {Patil}, {Patil}, {Plunkett}, {Prochaska},
  {Rastogi}, {Reddy Janga}, {Sabater}, {Sakurikar}, {Seifert}, {Sherbert},
  {Sherwood-Taylor}, {Shih}, {Sick}, {Silbiger}, {Singanamalla}, {Singer},
  {Sladen}, {Sooley}, {Sornarajah}, {Streicher}, {Teuben}, {Thomas},
  {Tremblay}, {Turner}, {Terr{\'o}n}, {van Kerkwijk}, {de la Vega}, {Watkins},
  {Weaver}, {Whitmore}, {Woillez}, {Zabalza}, and {Contributors}]{astropy:2018}
{Price-Whelan}, A.M.; {Sip{\H{o}}cz}, B.M.; {G{\"u}nther}, H.M.; {Lim}, P.L.;
  {Crawford}, S.M.; {Conseil}, S.; {Shupe}, D.L.; {Craig}, M.W.; {Dencheva},
  N.; {Ginsburg}, A.;  \textit{et al.}
\newblock {The Astropy Project: Building an Open-science Project and Status of
  the v2.0 Core Package}.
\newblock {\em \aj} {\bf 2018}, {\em 156},~123.
\newblock
  doi:{\changeurlcolor{black}\href{https://doi.org/10.3847/1538-3881/aabc4f}{\detokenize{10.3847/1538-3881/aabc4f}}}.

\bibitem[{van der Walt} \em{et~al.}(2011){van der Walt}, {Colbert}, and
  {Varoquaux}]{numpy11}
{van der Walt}, S.; {Colbert}, S.C.; {Varoquaux}, G.
\newblock The NumPy Array: A Structure for Efficient Numerical Computation.
\newblock {\em Computing in Science Engineering} {\bf 2011}, {\em 13},~22--30.

\bibitem[Hunter(2007)]{matplotlib07}
Hunter, J.D.
\newblock Matplotlib: A 2D graphics environment.
\newblock {\em Computing in Science \& Engineering} {\bf 2007}, {\em
  9},~90--95.
\newblock
  doi:{\changeurlcolor{black}\href{https://doi.org/10.1109/MCSE.2007.55}{\detokenize{10.1109/MCSE.2007.55}}}.

\end{thebibliography}

\end{document}